\documentclass[10pt]{iopart}
\usepackage{bm}              
\usepackage{graphicx}
\usepackage{amssymb}

\newcommand{\Ci}{{\rm Ci}}
\newcommand{\veck}{{\rm\bf k}}

\newcommand{\vecR}{{\rm\bf R}}
\newcommand{\vecS}{{\rm\bf S}}
\newcommand{\vecRprime}{{\rm\bf R}^{\prime}}
\newcommand{\vecQ}{{\rm\bf Q}}
\newcommand{\vecG}{{\rm\bf G}}
\newcommand{\vecs}{{\rm\bf s}}
\newcommand{\veclambda}{{\bm\lambda}}
\newcommand{\veczero}{{\rm\bf 0}}
\newcommand{\Cot}{{\rm Cot}}
\newcommand{\openone}{\leavevmode\hbox{\small1\normalsize\kern-.33em1}}

\newcommand{\mytoday}{July 18, 2017} 

\begin{document}

\title{Gutzwiller variational approach to the two-impurity Anderson model
at particle-hole symmetry}

\author{Thorben Linneweber$^{1}$, J\"org B\"unemann$^{2,3}$,
Zakaria M.M.\ Mahmoud$^{1,4}$, and Florian Gebhard$^{1}$}


\address{$^{1}$ Fachbereich Physik, Philipps-Universit\"at Marburg,
D-35032 Marburg, Germany}
\address{$^{2}$ Institut f\"ur Physik, 
BTU Cottbus-Senftenberg, D-03013 Cottbus, Germany}
\address{$^{3}$ Fakult\"at Physik, 
TU Dortmund, D-44221 Dortmund, Germany}
\address{$^{4}$ Department of Physics, New-Valley Faculty of Science, 
El-Kharga, Assiut University, Egypt}


\date{\mytoday}

\begin{abstract}
We study Gutzwiller-correlated wave functions as variational
ground states for the two-impurity Anderson model (TIAM) at particle-hole symmetry
as a function of the impurity separation $\vecR$.
Our variational state is obtained by applying the Gutzwiller many-particle correlator
to a single-particle product state.
We determine the optimal single-particle product state fully variationally
from an effective non-interacting TIAM that contains 
a direct electron transfer between the impurities as variational degree of freedom.
For a large Hubbard interaction~$U$ between the
electrons on the impurities, 
the impurity spins experience a Heisenberg coupling proportional to $V^2/U$
where $V$ parameterizes the strength of the on-site hybridization.
For small Hubbard interactions we observe weakly coupled impurities.
In general, for a three-dimensional simple cubic lattice
we find discontinuous quantum phase transitions that separate 
weakly interacting impurities for small interactions from 
singlet pairs for large interactions.
\end{abstract}


\submitto{\JPCM on \mytoday}
\ioptwocol

\section{Introduction}

The description of impurities in a metallic host poses a
fundamental problem in solid-state theory.
The single-impurity Anderson model, the $s$-$d$ (or `Kondo') model,
and other single-impurity Hamiltonians are among the most studied 
many-particle problems because they can be treated analytically 
and numerically with a variety of methods and some of them can even be solved
exactly; for an overview, see 
Refs.~\cite{Hewson,0953-8984-10-12-009,RevModPhys.80.395,RevModPhys.83.349},
and references therein. Nowadays, the `Kondo effect' 
is well understood:
the host electrons build a `Kondo cloud' around the impurity 
so that the impurity spin-1/2 is screened into a `Kondo spin'.
At zero temperature, the host electrons and the impurity spin eventually
form a `Kondo singlet', and the ground state of the system is non-degenerate.

When there are two (magnetic) impurities present in the system, 
they interact via the RKKY mechanism,
named after Ruderman and Kittel~\cite{PhysRev.96.99},
Kasuya~\cite{Kasuya01071956}, and Yosida~\cite{PhysRev.106.893}.
The electrons scatter off both impurities and thereby mediate
an effective interaction between the impurities.
For large enough couplings, two im\-purity spins can bind
into a singlet, and the ground state is also non-degenerate.
Apparently, the RKKY and Kondo mechanisms for singlet formation 
compete with each other.
Consequently, as pointed out by 
Jones, Varma, and Wilkins~\cite{VarmaJonesPRL} and 
by Jones and Varma~\cite{VarmaJonesPRB}, 
a quantum phase transition between Kondo singlet and spin-pair phases might occur, 
depending on the ratio of Kondo and RKKY couplings in the two-impurity Kondo model.
This proposition was sup\-ported
by a slave-boson mean-field study~\cite{JonesKotliarMillis}.

Subsequent numerical~\cite{RMFye} and 
variational studies~\cite{PhysRevB.48.7322,PhysRevB.44.450,Yanagisawa} 
questioned the existence of a quantum phase transition 
in the two-impurity Kondo and Anderson models.
It was later shown analytically~\cite{PhysRevLett.68.1046,AffleckLudwigJones} 
that the appearance of a quantum phase transition
in the two-impurity Kondo model at particle-hole symmetry
depends on the impurities' lattice positions. 
To complicate matters, for impurities in the same bath of host electrons,
the two competing energy scales 
prevent a straightforward mapping of the two-impurity
Anderson model to the two-impurity Kondo model employing
the Schrieffer-Wolff transformation~\cite{Ong}.
Therefore, it is not obvious that the two models belong to the same universality class,
and it remains interesting to investigate the competition between
Kondo and RKKY interactions for the two-impurity Anderson model.

In this work, we investigate the ground state of the
particle-hole symmetric two-impurity Ander\-son model at 
half band-filling~\cite{AlexanderAnderson}.
In the Gutzwiller variational ground state,
the Gutzwiller many-particle correlator
is applied to a single-particle product state~\cite{Gutzwiller1963}
that can be viewed as the ground state of an effective single-particle Hamiltonian.
Since only the impurity electrons
are correlated, the wave function can be evaluated without further ap\-proxi\-ma\-tions.
Therefore, we derive upper bounds to the exact ground-state energy.

In contrast to previous variational 
studies~\cite{PhysRevB.48.7322,PhysRevB.44.450,Yanagisawa},
our single-particle product state for the Gutzwiller wave function
is determined fully variationally as the optimal
ground state of an effective 
non-in\-ter\-ac\-ting two-impurity Anderson model.
In a previous article~\cite{nonintTIAM}, referred to as~MBG,
we studied the non-in\-ter\-ac\-ting Hamiltonian.
The solution in~MBG parametrically depends 
on the effective electron trans\-fer between the two impurities.
The variational free\-dom 
to generate an inter-impurity electron transfer 
is decisive for the Gutzwiller ground-state phase diagram
for the interacting two-impurity Anderson model.
We find a (generically discontinuous) phase transition
as a function of the Hubbard interaction between two phases,
(i), weakly interacting impurities for small interactions 
and, (ii), spin-singlets formed by the impurity spins at large interactions.
The transition appears generically
in a parameter range where the two-impurity Anderson model
cannot be described faithfully by an effective spin model.

Our work is organized as follows. 
In Sect.~\ref{sec:TIAMdef}, we define the two-impurity Anderson model 
and rewrite it in the form of a two-orbital model.
In Sect.~\ref{sec:phsymm} we re\-call the conditions
for particle-hole symmetry at half band-filling, we
define the various parameter limits of interest
(atomic, spin-model, Kondo, itinerant), and 
we discuss the two phases that we expect to find.
Next, in Sect.~\ref{sec:GWF}, we introduce and evaluate the Gutzwiller variational 
ground state. In particular, we identify the effective non-interacting two-impurity 
model and recover the exact results for the atomic limit.
In Sect.~\ref{sec:tb-simplecubiclattice}
we investigate host electrons with
nearest-neighbor trans\-fers on a simple-cubic lattice and provide explicit expressions 
for the single-particle contribution to the variational
ground-state energy for small hybridizations.
In Sect.~\ref{subsec:optimizationofGutzwparameters}
we study the spin-model and Kondo limits 
where the Hubbard interaction is large.
In this limit, the optimization of the Gutzwiller variational parameters
can be done analytically to a far extent. This provides 
useful in\-sights into the quantum phase transition
from weakly coupled impurities to singlet pairs.
In Sect.~\ref{numminHeislimit} we discuss the numerical results
for the quantum phase transitions in the whole parameter space.
Short conclusions, Sect.~\ref{sec:conclusions}, close our presentation.
We defer technical details to six extensive appendices.

\section{Two-impurity Anderson model}
\label{sec:TIAMdef}

We start our investigation with the definition of the Hamiltonian. 
Then, we rephrase the problem in terms of a two-orbital model.

\subsection{Hamiltonian}

Two impurities in a metallic host on a lattice 
are modeled by the Hamiltonian~\cite{AlexanderAnderson}
\begin{equation}
\hat{H}=\hat{T}+\hat{T}_d+\hat{V}+\hat{H}_{\rm int} 
\equiv \hat{H}_0+\hat{H}_{\rm int}
\; .
\label{eq:defH}
\end{equation}
Here, $\hat{T}$ is the kinetic energy of the non-interacting spin-1/2
host electrons ($\sigma=\uparrow,\downarrow$), 
\begin{equation}
\hat{T}=  \sum_{\vecR,\vecRprime,\sigma}t(\vecR-\vecRprime)
\hat{c}_{\vecR,\sigma}^+\hat{c}_{\vecRprime,\sigma}^{\vphantom{+}}\; ,
\label{eq:defT}
\end{equation}
where the electrons tunnel between the sites $\vecR$ and $\vecRprime$
of the lattice with amplitude $t(\vecR-\vecRprime)$.
The kinetic energy is diagonal in Fourier space.
For $\veck$ from the first Brillouin zone we define
\begin{eqnarray}
\hat{c}_{\veck,\sigma}^{\vphantom{+}}&=&
\frac{1}{\sqrt{L}} \sum_{\vecR} e^{-\rmi \veck \cdot \vecR}
\hat{c}_{\vecR,\sigma}^{\vphantom{+}}\; , \nonumber \\
\hat{c}_{\vecR,\sigma}^{\vphantom{+}}&=&
\frac{1}{\sqrt{L}} \sum_{\veck} e^{\rmi \veck \cdot \vecR}
\hat{c}_{\veck,\sigma}^{\vphantom{+}}\; ,
\end{eqnarray}
where $L$ is the (even) number of lattice sites. With
\begin{eqnarray}
t(\vecR) &= &\frac{1}{L} \sum_{\veck} e^{\rmi \veck\cdot \vecR} \epsilon(\veck)
\; , \nonumber \\
\epsilon(\veck)&=&\sum_{\vecR}t(\vecR)e^{-\rmi \veck\cdot \vecR}\; ,
\end{eqnarray}
the host electron kinetic energy becomes diagonal, 
\begin{equation}
\hat{T}= \sum_{\veck,\sigma} \epsilon(\veck) 
\hat{c}_{\veck,\sigma}^+\hat{c}_{\veck,\sigma}^{\vphantom{+}} \;,
\end{equation}
where $\epsilon(\veck)$ is the dispersion relation.

With $\hat{T}_d$ we also permit a direct electron transfer 
with amplitude $t_{12}$ between the impurity orbitals at sites $\vecR_1$ 
and $\vecR_2$,
\begin{equation}
\hat{T}_{d}=
\sum_{\sigma}t_{12}\hat{d}_{1,\sigma}^+\hat{d}_{2,\sigma}^{\vphantom{+}}
+ t_{12}^*\hat{d}_{2,\sigma}^+\hat{d}_{1,\sigma}^{\vphantom{+}}
\; .
\label{eq:defTd}
\end{equation}
For most of the paper, however, 
we shall focus on the case of vanishingly small direct electron 
transfer between the impurities, $|t_{12}|\to 0$.

Next, $\hat{V}$ describes the hybridization between im\-puri\-ty and host electron
states ($b=1,2$),
\begin{eqnarray}
\hat{V}&=&  
\sum_{\vecR,b,\sigma}
V(\vecR-\vecR_b)
\hat{c}_{\vecR,\sigma}^+\hat{d}_{b,\sigma}^{\vphantom{+}}
+ V^*(\vecR-\vecR_b) \hat{d}_{b,\sigma}^+\hat{c}_{\vecR,\sigma}^{\vphantom{+}}
\nonumber \\
&=&
\frac{1}{\sqrt{L}}\sum_{\veck,b,\sigma}
V_{\veck}e^{-\rmi \veck \cdot \vecR_b}
\hat{c}_{\veck,\sigma}^+\hat{d}_{b,\sigma}^{\vphantom{+}}
+ V_{\veck}^* e^{\rmi \veck\cdot \vecR_b} 
\hat{d}_{b,\sigma}^+\hat{c}_{\veck,\sigma}^{\vphantom{+}}\; ,\nonumber \\
V_{\veck}&=& \sum_{\vecR}V(\vecR)e^{-\rmi \veck\cdot \vecR} \,, \, 
V(\vecR)=\frac{1}{L}\sum_{\veck}e^{\rmi \veck \cdot \vecR}V_{\veck} \; .
\label{eq:defV}
\end{eqnarray}
In Sects.~\ref{sec:tb-simplecubiclattice}--
\ref{numminHeislimit}, 
we shall employ a local hybridization, $V_{\veck}\equiv V$.
The non-interacting two-impurity Anderson model
$\hat{H}_0=\hat{T}+\hat{T}_d+\hat{V}$ can be solved exactly using
the equation-of-motion method~\cite{AlexanderAnderson}, see~MBG.

Last, $\hat{H}_{\rm int}$ represents the Hubbard interaction
to model the Coulomb repulsion on the impurities,
\begin{equation}
\hat{H}_{\rm int}= U \sum_{b=1}^2 
(\hat{n}_{b,\uparrow}^d-1/2)
(\hat{n}_{b,\downarrow}^d-1/2)\; ,
\label{eq:defHint}
\end{equation}
where 
$\hat{n}_{b,\sigma}^d
=\hat{d}_{b,\sigma}^+\hat{d}_{b,\sigma}^{\vphantom{+}}$
counts the number of impurity electrons.
The two-impurity Anderson model $\hat{H}=\hat{H}_0+\hat{H}_{\rm int}$ 
in eq.~(\ref{eq:defH}) 
poses a difficult many-particle problem
that cannot be solved in general.

\subsection{Single-site two-orbital model}

As a second step,
we map the two-impurity model onto an asymmetric two-orbital model.
This step is equivalent to the introduction of even and odd parity channels.

\subsubsection{Kinetic energy of d-electrons}

We introduce the new `$h$-basis' 
for the impurity electrons using the unitary transformation
\begin{eqnarray}
\hat{d}_{1,\sigma}^+&=& \frac{1}{\sqrt{2}} \Bigl( \hat{h}_{1,\sigma}^+
+\alpha_{12} \hat{h}_{2,\sigma}^+\Bigr)  \; ,\nonumber \\
\hat{d}_{2,\sigma}^+&=& \frac{1}{\sqrt{2}} \Bigl(-\alpha_{12}^* \hat{h}_{1,\sigma}^+
+\hat{h}_{2,\sigma}^+\Bigr)  \; ,
\end{eqnarray}
where
\begin{equation}
\alpha_{12}=
\frac{t_{12}^*}{|t_{12}|} 
\quad, \quad  
\alpha_{12}^2=
\frac{t_{12}^*}{t_{12}}\; .
\label{eq:defalpha12}
\end{equation}
The inverse transformation reads
\begin{eqnarray}
\hat{h}_{1,\sigma}^+&=& \frac{1}{\sqrt{2}} \Bigl( \hat{d}_{1,\sigma}^+
-\alpha_{12} \hat{d}_{2,\sigma}^+\Bigr)
\nonumber \; , \\
\hat{h}_{2,\sigma}^+&=& 
\frac{1}{\sqrt{2}} \Bigl( \alpha_{12}^* \hat{d}_{1,\sigma}^+
+\hat{d}_{2,\sigma}^+\Bigr) \; . 
\end{eqnarray}
For a unitary transformation we have 
($\hat{n}_{b,\sigma}=\hat{h}_{b,\sigma}^+\hat{h}_{b,\sigma}^{\vphantom{+}})$
\begin{equation}
\hat{d}_{1,\sigma}^+\hat{d}_{1,\sigma}^{\vphantom{+}} 
+ 
\hat{d}_{2,\sigma}^+\hat{d}_{2,\sigma}^{\vphantom{+}} 
= 
\hat{n}_{1,\sigma}+ \hat{n}_{2,\sigma}\; ,
\label{eq:traceinvariant}
\end{equation}
and the average number of $d_{\sigma}$-electrons obviously equals the 
average number of $h_{\sigma}$-electrons.

We introduced the $h$-basis because it diagonalizes $\hat{T}_d$,
eq.~(\ref{eq:defTd}), 
\begin{equation}
\hat{T}_d= |t_{12}| \sum_{\sigma} \Bigl( 
\hat{h}_{2,\sigma}^+\hat{h}_{2,\sigma}^{\vphantom{+}}
-\hat{h}_{1,\sigma}^+\hat{h}_{1,\sigma}^{\vphantom{+}} 
\Bigr) \; .
\end{equation}
In the $h$-basis representation, $\hat{T}_d$ has the form of a splitting 
of the  two impurity levels.
For $|t_{12}|\to 0$, the occupancies of both orbitals
should be the same. A broken $h$-orbital symmetry in the ground state $|\Psi_0\rangle$,
$\langle \Psi_0 |\hat{n}_{1,\sigma}|\Psi_0\rangle \neq 
\langle \Psi_0 |\hat{n}_{2,\sigma}|\Psi_0\rangle$,
indicates that a finite electron transfer between the impurities
is enhanced by the interplay of the electrons' kinetic energy
and their Coulomb interactions, $\langle \Psi_0 |\hat{d}_{1,\sigma}^+
\hat{d}_{2,\sigma}^{\vphantom{+}}| \Psi_0 \rangle \neq 0$.

\begin{table}[b]
{\scriptsize \tabcolsep=2pt\begin{tabular}{|c|l|c|c|}
\hline
\vphantom{\large A}$\,n\,$ & \multicolumn{1}{c|}{$|\Gamma\rangle$}  & 
$E_{\Gamma}$ &$E_d$\\[1pt]
\hline
\vphantom{\large A}0 & 
$|1\rangle=|\emptyset,\emptyset\rangle\equiv |{\rm vac}\rangle$ & $U/2$ &0\\[1pt]
\hline
\vphantom{\large A}1 & \begin{tabular}[t]{@{}ll@{}}
$|2\rangle$& $=\hat{h}_{1,\uparrow}^+|{\rm vac}\rangle$\\
&$=(\hat{d}_{1,\uparrow}^+-\alpha\hat{d}_{2,\uparrow}^+)/\sqrt{2}|{\rm vac}\rangle$ \\
& $=(|\uparrow,\emptyset\rangle
-\alpha|\emptyset,\uparrow\rangle)/\sqrt{2}$ 
\end{tabular}
&0 & $-|t_{12}|$\\
& \begin{tabular}[t]{@{}ll@{}}
$|3\rangle$ &$=\hat{h}_{1,\downarrow}^+|{\rm vac}\rangle$\\
&$=(\hat{d}_{1,\downarrow}^+-\alpha\hat{d}_{2,\downarrow}^+)/\sqrt{2}
|{\rm vac}\rangle$ \\
& $=(|\downarrow,\emptyset\rangle
-\alpha|\emptyset,\downarrow\rangle)/\sqrt{2}$ 
\end{tabular}
&0&$-|t_{12}|$\\
 & \begin{tabular}[t]{@{}ll@{}}
$|4\rangle$ &$=\hat{h}_{2,\uparrow}^+|{\rm vac}\rangle$\\
&$=(\alpha^*\hat{d}_{1,\uparrow}^++\hat{d}_{2,\uparrow}^+)/\sqrt{2}|{\rm vac}\rangle$\\
& $=(\alpha^*|\uparrow,\emptyset\rangle
+|\emptyset,\uparrow\rangle)/\sqrt{2}$ 
\end{tabular}
&0 &$|t_{12}|$\\
& \begin{tabular}[t]{@{}ll@{}}
$|5\rangle$ &$=\hat{h}_{2,\downarrow}^+|{\rm vac}\rangle$\\
&$=(\alpha^*\hat{d}_{1,\downarrow}^++\hat{d}_{2,\downarrow}^+)/\sqrt{2}
|{\rm vac}\rangle$ \\
& $=(\alpha^*|\downarrow,\emptyset\rangle
+|\emptyset,\downarrow\rangle)/\sqrt{2}$ \\[1pt]
\end{tabular}
&0 &$|t_{12}|$\\
\hline
\vphantom{\large A}2 & $|6\rangle 
=\hat{h}_{1,\uparrow}^+\hat{h}_{2,\uparrow}^+|{\rm vac}\rangle
=\hat{d}_{1,\uparrow}^+\hat{d}_{2,\uparrow}^+
|{\rm vac}\rangle =|\uparrow,\uparrow\rangle$
& $-U/2$ &0\\
&\begin{tabular}[t]{@{}ll@{}}
$|7\rangle$
&$=(\hat{h}_{1,\uparrow}^+\hat{h}_{2,\downarrow}^+
+\hat{h}_{1,\downarrow}^+\hat{h}_{2,\uparrow}^+)/\sqrt{2}|{\rm vac}\rangle$ \\
&$=(\hat{d}_{1,\uparrow}^+\hat{d}_{2,\downarrow}^+
+\hat{d}_{1,\downarrow}^+\hat{d}_{2,\uparrow}^+)/\sqrt{2}|{\rm vac}\rangle$ \\
& $=(|\uparrow,\downarrow\rangle
+|\downarrow,\uparrow\rangle)/\sqrt{2}$ 
\end{tabular}
& $-U/2$ &0 \\
& $|8\rangle =\hat{h}_{1,\downarrow}^+\hat{h}_{2,\downarrow}^+|{\rm vac}\rangle
=\hat{d}_{1,\downarrow}^+\hat{d}_{2,\downarrow}^+|{\rm vac}\rangle
=|\downarrow,\downarrow\rangle$
& $-U/2$ &0\\
&\begin{tabular}[t]{@{}ll@{}}
$|9\rangle$ 
&$=(\hat{h}_{1,\uparrow}^+\hat{h}_{2,\downarrow}^+
-\hat{h}_{1,\downarrow}^+\hat{h}_{2,\uparrow}^+)/\sqrt{2}|{\rm vac}\rangle$ 
\\
&$=(\alpha^*\hat{d}_{1,\uparrow}^+\hat{d}_{1,\downarrow}^+
-\alpha\hat{d}_{2,\uparrow}^+\hat{d}_{2,\downarrow}^+)/\sqrt{2}|{\rm vac}\rangle$ \\
& $=(\alpha^*|\uparrow\downarrow,\emptyset\rangle
-\alpha|\emptyset,\uparrow\downarrow\rangle)/\sqrt{2}$ 
\end{tabular}
& $U/2$ &0\\
&\begin{tabular}[t]{@{}ll@{}}
$|10\rangle$ &$=(\alpha^*\hat{h}_{1,\uparrow}^+\hat{h}_{1,\downarrow}^+
-\alpha\hat{h}_{2,\uparrow}^+\hat{h}_{2,\downarrow}^+)/\sqrt{2}|{\rm vac}\rangle$\\
&$=(-\hat{d}_{1,\uparrow}^+\hat{d}_{2,\downarrow}^+
+\hat{d}_{1,\downarrow}^+\hat{d}_{2,\uparrow}^+)/\sqrt{2}|{\rm vac}\rangle$ \\
& $=(-|\uparrow,\downarrow\rangle
+|\downarrow,\uparrow\rangle)/\sqrt{2}$ 
\end{tabular}
& $-U/2$ & -- \\
&\begin{tabular}[t]{@{}ll@{}}
$|11\rangle$ 
&$=(\alpha^*\hat{h}_{1,\uparrow}^+\hat{h}_{1,\downarrow}^+
+\alpha\hat{h}_{2,\uparrow}^+\hat{h}_{2,\downarrow}^+)/\sqrt{2}|{\rm vac}\rangle$\\
&$=(\alpha^*\hat{d}_{1,\uparrow}^+\hat{d}_{1,\downarrow}^+
+\alpha\hat{d}_{2,\uparrow}^+\hat{d}_{2,\downarrow}^+)/\sqrt{2}|{\rm vac}\rangle$\\ 
& $=(\alpha^*|\uparrow\downarrow,\emptyset\rangle
+\alpha|\uparrow\downarrow,\emptyset\rangle)/\sqrt{2}$ \\[1pt]
\end{tabular}
& $U/2$ & -- \\
\hline
\vphantom{\large A}3 & \begin{tabular}[t]{@{}ll@{}}
$|12\rangle$ 
&$=\hat{h}_{1,\uparrow}^+\hat{h}_{2,\uparrow}^+\hat{h}_{2,\downarrow}^+
|{\rm vac}\rangle$\\
&$=(\hat{d}_{1,\uparrow}^+\hat{d}_{2,\uparrow}^+\hat{d}_{2,\downarrow}^+
-\alpha^*\hat{d}_{1,\uparrow}^+\hat{d}_{1,\downarrow}^+
\hat{d}_{2,\uparrow}^+)/\sqrt{2}|{\rm vac}\rangle$ \\
& $=(|\uparrow,\uparrow\downarrow\rangle
-\alpha^*|\uparrow\downarrow,\uparrow\rangle)/\sqrt{2}$
\end{tabular}
& 0 &$|t_{12}|$\\
& \begin{tabular}[t]{@{}ll@{}}
$|13\rangle$ 
&$=\hat{h}_{1,\downarrow}^+\hat{h}_{2,\uparrow}^+\hat{h}_{2,\downarrow}^+
|{\rm vac}\rangle$\\
&$=(\hat{d}_{1,\downarrow}^+\hat{d}_{2,\uparrow}^+\hat{d}_{2,\downarrow}^+
-\alpha^*\hat{d}_{1,\uparrow}^+\hat{d}_{1,\downarrow}^+
\hat{d}_{2,\downarrow}^+)/\sqrt{2}|{\rm vac}\rangle$ \\
& $=(|\downarrow,\uparrow\downarrow\rangle
-\alpha^*|\uparrow\downarrow,\downarrow\rangle)/\sqrt{2}$
\end{tabular}
& 0 &$|t_{12}|$\\
& \begin{tabular}[t]{@{}ll@{}}
$|14\rangle$ 
&$=\hat{h}_{1,\uparrow}^+\hat{h}_{1,\downarrow}^+\hat{h}_{2,\uparrow}^+
|{\rm vac}\rangle$\\
&$=(\alpha\hat{d}_{1,\uparrow}^+\hat{d}_{2,\uparrow}^+\hat{d}_{2,\downarrow}^+
+\hat{d}_{1,\uparrow}^+\hat{d}_{1,\downarrow}^+
\hat{d}_{2,\uparrow}^+)/\sqrt{2}|{\rm vac}\rangle$ \\
& $=(\alpha|\uparrow,\uparrow\downarrow\rangle
+|\uparrow\downarrow,\uparrow\rangle)/\sqrt{2}$
\end{tabular}
& 0 &$-|t_{12}|$\\
& \begin{tabular}[t]{@{}ll@{}}
$|15\rangle$ 
&$=\hat{h}_{1,\uparrow}^+\hat{h}_{1,\downarrow}^+\hat{h}_{2,\downarrow}^+
|{\rm vac}\rangle$\\
&$=(\alpha\hat{d}_{1,\downarrow}^+\hat{d}_{2,\uparrow}^+\hat{d}_{2,\downarrow}^+
+\hat{d}_{1,\uparrow}^+\hat{d}_{1,\downarrow}^+
\hat{d}_{2,\downarrow}^+)/\sqrt{2}|{\rm vac}\rangle$ \\
& $=(\alpha|\downarrow,\uparrow\downarrow\rangle
+|\uparrow\downarrow,\downarrow\rangle)/\sqrt{2}$\\[1pt]
\end{tabular}
& 0 &$-|t_{12}|$\\
\hline
\vphantom{\large A}4 & \begin{tabular}[t]{@{}ll@{}}
$|16\rangle$ 
& $=\hat{h}_{1,\uparrow}^+\hat{h}_{1,\downarrow}^+
\hat{h}_{2,\uparrow}^+\hat{h}_{2,\downarrow}^+|{\rm vac}\rangle$\\
& $= \hat{d}_{1,\uparrow}^+\hat{d}_{1,\downarrow}^+
\hat{d}_{2,\uparrow}^+\hat{d}_{2,\downarrow}^+|{\rm vac}\rangle=
|\uparrow\downarrow,\uparrow\downarrow\rangle$ \\[2pt]
\end{tabular}
& $U/2$ &0\\
\hline
\end{tabular}}
\caption{Atomic eigenstates $|\Gamma\rangle$ of $\hat{H}_{\rm int}$ 
with energy $E_{\Gamma}$; all states apart from $|10\rangle$  and $|11\rangle$ 
are also eigenstates of $\hat{T}_d$ with energy $E_d$;
$\alpha\equiv \alpha_{12}$.\label{tab:Gammas}}
\end{table}

\subsubsection{Hybridization}

In the $h$-basis, the hybridization $\hat{V}$, see eq.~(\ref{eq:defV}), takes the form
\begin{equation}
\hat{V}=  \frac{1}{\sqrt{L}}\sum_{\veck,b,\sigma}
V_{\veck,b}\hat{c}_{\veck,\sigma}^+\hat{h}_{b,\sigma}^{\vphantom{+}}
+ V_{\veck,b}^* 
\hat{h}_{b,\sigma}^+\hat{c}_{\veck,\sigma}^{\vphantom{+}}\; .
\label{eq:defVdiagonal}
\end{equation}
The two impurity levels hybridize with the host elec\-trons with 
the matrix elements
\begin{eqnarray}
V_{\veck,1}\equiv V_{\veck,1}(\vecR_1,\vecR_2)
&=&\frac{V_{\veck}}{\sqrt{2}}
( e^{-\rmi \veck\cdot \vecR_1}
-\alpha_{12} e^{-\rmi \veck\cdot \vecR_2}) \, , \nonumber \\
V_{\veck,2}\equiv V_{\veck,2}(\vecR_1,\vecR_2)&=&\frac{V_{\veck}}{\sqrt{2}}
( \alpha_{12}^* e^{-\rmi \veck\cdot \vecR_1}
+ e^{-\rmi \veck\cdot \vecR_2}) \,. 
\label{eq:hybridizationsforh}
\end{eqnarray}
In the absence of a direct electron transfer between the impurities, $t_{12}=0$,
we may set $\alpha_{12}=1$ for all $\vecR_1,\vecR_2$.
Then, 
$V_{\veck,1}$ ($V_{\veck,2}$) describes the hybridization
in the odd (even) parity channel,
$V_{\veck,1}(\vecR_2,\vecR_1)=-V_{\veck,1}(\vecR_1,\vecR_2)$
[$V_{\veck,2}(\vecR_2,\vecR_1)=V_{\veck,2}(\vecR_1,\vecR_2)$].
For our study, we keep $|t_{12}|$ infinitesimally small so that
$\alpha_{12}$ remains well defined by eq.~(\ref{eq:defalpha12}).

\subsubsection{Interaction}

We write the interaction term in its eigenbasis~$|\Gamma\rangle$,
\begin{equation}
\hat{H}_{\rm int}= \sum_{\Gamma} E_{\Gamma} \hat{m}_{\Gamma} 
\quad , \quad \hat{m}_{\Gamma} =|\Gamma\rangle\langle \Gamma | \; ,
\end{equation}
where $\Gamma=1,\ldots,16$ labels the 16 possible configurations
on the two impurity sites. 
They are listed in the local $h$-basis in table~\ref{tab:Gammas}, 
together with the atomic spectrum.
The operator $\hat{T}_d$ mixes the states
$|10\rangle$ and $|11\rangle$. All other states
$|\Gamma\rangle$ in table~\ref{tab:Gammas} are also eigenstates of $\hat{T}_d$
with energy $E_d=0,\pm|t_{12}|$.

\section{Particle-hole symmetry at half band-filling}
\label{sec:phsymm}

We are interested in the case where there is on average one electron on each 
of the impurities. This can be assured for a particle-hole symmetric 
Hamiltonian~(\ref{eq:defH}) at half band-filling. 

\subsection{Conditions}

We consider a bipartite lattice and assume that there exists half a reciprocal
lattice vector~$\vecQ=\vecG/2$ for which
\begin{eqnarray}
\epsilon(\veck\pm \vecQ) = -\epsilon(\veck)
&\, , \,& e^{\rmi \vecQ\cdot \vecR}= \left\{
{\arraycolsep=1pt\begin{array}{@{}rcl@{}}
1 & \hbox{if} &\vecR \in \hbox{$A$-lattice} \\
-1 & \hbox{if} &\vecR \in \hbox{$B$-lattice}
\end{array}}
\right.  .\label{eq:phdemandepsk}
\end{eqnarray}
We also assume inversion symmetry, $\epsilon(-\veck)= \epsilon(\veck)$;
recall that $\epsilon(\veck + \vecG)=\epsilon(\veck)$.
Note that the transfer matrix elements between sites on different sublattices 
are real and those between
sites on the same sublattice are purely imaginary, see~MBG.
The same applies to the impurity transfer matrix element $t_{12}$.
In the main text we focus on the case that the two impurities are on different sublattices
and consider the other case in the appendix.

Moreover, we demand that
\begin{equation}
V_{\veck}= V_{\vecQ-\veck}^* \; .
\label{eq:phdemandVk}
\end{equation}
Note that a $\veck$-independent hybridization,
$V_{\veck}\equiv V$, must necessarily be real. 
The conditions~(\ref{eq:phdemandepsk})
and~(\ref{eq:phdemandVk}) make the Hamiltonian
invariant under particle-hole trans\-formation, 
$\hat{H}=\hat{\tau}_{\rm ph}^+\hat{H}\hat{\tau}_{\rm ph}^{\vphantom{+}}$,
see~MBG.

\subsection{Half-filled bands}

In the following we consider paramagnetism at half band-filling
where the number of electrons $N=N_{\uparrow}+N_{\downarrow}$ 
equals the (even) number of orbitals, $N=L+2$,
and $N_{\uparrow}=N_{\downarrow}=L/2+1$. Note that there are $L$ lattices
sites for the host electrons and two additional impurity orbitals
on the lattice sites $\vecR_1$ and $\vecR_2$.

At half band-filling, the non-degenerate ground state $|\Psi_0\rangle$
maps onto itself under the particle-hole transformation~$\tau_{\rm ph}$.
Therefore, we find that
each im\-purity level is exactly half filled for all
hybridizations and interaction strengths,
\begin{equation}
\langle \Psi_0 | \hat{d}_{1,\sigma}^+\hat{d}_{1,\sigma}^{\vphantom{+}}  | \Psi_0 \rangle 
=\langle \Psi_0 | \hat{d}_{2,\sigma}^+\hat{d}_{2,\sigma}^{\vphantom{+}}  | \Psi_0 \rangle 
=1/2 \; .
\label{eq:dsarehalffilled}
\end{equation}
Moreover, it is readily shown that the bare density of states
is symmetric,
\begin{equation}
D_{\sigma,0}(\epsilon) = \frac{1}{L} \sum_{\veck} \delta(\epsilon-\epsilon(\veck))
=D_{\sigma,0}(-\epsilon) \; ,
\end{equation}
so that the Fermi energy is $E_{\rm F}=0$ at half band-filling.

At half band-filling, eq.~(\ref{eq:traceinvariant}) implies
\begin{equation}
\langle \Psi_0 | 
\hat{n}_{1,\sigma}+\hat{n}_{2,\sigma}|\Psi_0\rangle=1 \; .
\label{eq:btwoandbonedensities}
\end{equation}
Moreover, as shown in~MBG, particle-hole symmetry demands
that there is no hybridization between the $h$-orbitals at half band-filling,
\begin{equation}
\langle \Psi_0 | 
\hat{h}_{1,\sigma}^+\hat{h}_{2,\sigma}^{\vphantom{+}} 
|\Psi_0\rangle
=0 \; .
\label{eq:h12andh21are zero}
\end{equation}
This relation considerably
simplifies the evaluation of Gutzwiller-correlated wave functions.

\subsection{Limiting cases}
\label{sec:limits}

Before we proceed, we define some parameter limits of interest. 
We compare the Hubbard parameter~$U$ with
the bandwidth of the host electrons~$W$
and the hybridization~$V$. 
The hybridization is always assumed to be small compared to the 
bandwidth, $V\ll W$.

\subsubsection{Atomic limit}
The atomic limit is defined  by $V=0$ so that the $d$-levels are singly occupied 
for all $U>0$. In the presence of a direct electron transfer between
the impurities, $t_{12}\neq 0$, the ground state of the corresponding
two-site Hubbard model is a spin singlet. For $U\gg |t_{12}|$, 
the ground-state energy attains the familiar Heisenberg form,
\begin{equation}
E_{\rm Heis}=-\frac{4|t_{12}|^2}{U} +{\cal O}(1/U^2)\; .
\label{eq:Heisenbergenergy}
\end{equation}
If  for $V\neq 0$ and large interactions
the variational ground-state energy has a contribution 
proportional to $1/U$, eq.~(\ref{eq:Heisenbergenergy}) indicates 
that the impurities are coupled by an effective electron transfer,
see Sect.~\ref{subsec:optimizationofGutzwparameters}.

In the remainder of this section, we 
focus on the case of a vanishingly small direct coupling, $|t_{12}|\to 0$.

\subsubsection{Spin-model, Kondo, and itinerant limits}

When $U$ is the largest energy scale, $V \ll W \ll U$,
the single-impurity Anderson model maps onto the $s$-$d$ (or Kondo) 
model~\cite{Hewson}
\begin{eqnarray}
\hat{H}_{\rm SIAM} &\mapsto& \hat{H}_{\rm K} \; , \label{eq:defSIAM} \\ 
\hat{H}_{\rm SIAM} &=& \hat{T} +  \frac{V}{\sqrt{L}}\sum_{\veck,\sigma}
\left(\hat{c}_{\veck,\sigma}^+\hat{d}_{\sigma}^{\vphantom{+}}
+ \hat{d}_{\sigma}^+\hat{c}_{\veck,\sigma}^{\vphantom{+}}\right) \nonumber \\
&& + U (\hat{n}_{\uparrow}^d-1/2) (\hat{n}_{\downarrow}^d-1/2)\; , \nonumber \\ 
\hat{H}_{\rm K} &=& \hat{T} +  J_{\rm K} \hat{\vecs}\cdot \hat{\vecS} \; ,
\label{eq:defSingleKondo}
\end{eqnarray}
where $\hat{\vecS}$ is the impurity-spin operator and
$\hat{\vecs}$ denotes the host-electron spin at the impurity position at the origin.
Here, the antiferromagnetic Kondo coupling is given by~\cite{Hewson}
\begin{equation}
J_{\rm K}=\frac{4V^2}{U}\; .
\label{eq:Kondocouplingdef}
\end{equation}
A widely used generalization of~(\ref{eq:defSingleKondo}) 
is the two-impurity Kondo model (TIKM), 
\begin{equation}
\hat{H}_{\rm TIKM} = \hat{T} +  
J_{\rm K} \sum_b \hat{\vecs}_{b} \cdot \hat{\vecS}_b
+J_{\rm H} \hat{\vecS}_1\cdot \hat{\vecS}_2\; ,
\label{eq:TIKMdef}
\end{equation}
see, e.g., Refs~\cite{VarmaJonesPRL,VarmaJonesPRB};
it is not clear to us whether or not the TIKM can be derived from the TIAM rigorously.
Therefore, the value of $J_{\rm H}$ as a function of the TIAM parameters is not known.
Consequently, $J_{\rm H}$ is often taken as an independent model 
parameter~\cite{VarmaJonesPRL,VarmaJonesPRB,JonesKotliarMillis,RMFye}.

In the single-impurity Anderson model, the `spin-mo\-del limit' $U\gg W$
is not a prerequisite to find effectively a spin on the impurities, i.e.,
to have the impurity levels almost exactly singly occupied.
As is well known for the SIAM~\cite{Hewson}, the relevant energy scale 
actually is
\begin{equation}
\Gamma=\pi d_{\veczero} V^2 \;,
\end{equation}
where $d_{\veczero}=D_{\sigma,0}(0) \sim 1/W$ 
is the host-electron density of states at the Fermi energy.
Even for $U \ll W$, the `Kondo limit' $\Gamma\ll U\ll W$ guarantees 
that in the ground state there is basically a localized spin at the impurity site.

Lastly, for $U\lesssim \Gamma \ll W$, the system resembles
the features of the non-interacting Anderson model where
the occupation of the impurities is not integer. In this `itinerant limit', 
the impurities experiences an effective RKKY interaction~\cite{nonintTIAM},
\begin{equation}
J_{\rm RKKY}(\vecR)= -2\frac{\Gamma}{\pi}
\left(\frac{d_{\vecR}}{d_{\veczero}}\right)^2 \; ,
\end{equation}
where $d_{\vecR}$ can be expressed as an integral over the Fermi surface, 
see~MBG and Sect.~\ref{sec:tb-simplecubiclattice}.
The RKKY coup\-ling strength vanishes as a function of the impurity separation 
$\vecR=\vecR_1-\vecR_2$.

The key advantage of our variational approach lies in the fact that we can study
the TIAM on equal footing in the whole $(V,U,W)$ parameter space.
In particular, we can treat the spin and Kondo limits analytically to a far extent,
see Sect.~\ref{subsec:optimizationofGutzwparameters}.

\subsubsection{Singlet pairs versus weakly linked impurities}

As pointed out by Jones and Varma~\cite{VarmaJonesPRL,VarmaJonesPRB}
for the two-impurity Kondo model $\hat{H}_{\rm TIKM}$ in eq.~(\ref{eq:TIKMdef}),
there are two competing
mechanisms for singlet formation in the ground state.
For weakly linked impurities where $J_{\rm H}\ll J_{\rm K}$,
the Kondo coupling between the impurities
and the host electrons
leads to individual singlets made from an impurity spin and
its surrounding host electrons (`Kondo effect').
For strongly coupled impurity spins, $J_{\rm H}\gg J_{\rm K}$ 
in $\hat{H}_{\rm TIKM}$,  the two impurity spins form a (Heisenberg) spin singlet.
Correspondingly,
Varma and Jones found a quantum phase transition between these two phases
at $J_{\rm H}\approx J_{\rm K}$.

As we shall show in this work, such a transition generically 
also exists in the two-impurity Anderson
model. Our analytical evaluation in the Kondo limit reveals
a specific dependence of the variational ground-state energy on $U/\Gamma$
in both phases, 
\begin{eqnarray}
E_{\rm var}\sim -\exp\Bigl(-\frac{U\pi}{16\Gamma}\Bigr) &&
\quad \hbox{weakly coupled impurities}\nonumber \\[-3pt]
&& \quad U\ll U_{\rm c}
 \; , \label{eq:wci} \\
E_{\rm var}\sim -\frac{\Gamma}{U} & &\quad \hbox{singlet pair} \nonumber\\[-3pt]
&& \quad U\gg U_{\rm c} \; .
\label{eq:Heisheisheis}
\end{eqnarray}
The numerical minimization of our energy functional shows that 
the critical value for the transition is in the region $(U/\Gamma)_c\approx 12\ldots 16$
for all $2\cdot 10^{-3}W\leq V \leq 2\cdot 10^{-1} W$. 
This implies, however, that the transition cannot be found in the 
spin-model limit $U\gg W$ because it implies 
$V\gg W/\sqrt{(U/\Gamma)_c\pi d_{\veczero}}$
which contradicts our basic assumption $V\ll W$.

Therefore it seems that our findings are in mild conflict with the work of 
Varma and Jones.
Note, however, that $J_{\rm H}$ in $\hat{H}_{\rm TIKM}$ is treated as an 
adjustable parameter so that the competition of the Kondo and Heisenberg singlet
formation can be studied in the two-impurity Kondo model,
independent of the existence and the form 
of a mapping of the two-impurity Anderson  model
to the two-impurity Kondo model.

\section{Gutzwiller variational approach}
\label{sec:GWF}

For the two-impurity Anderson model~(\ref{eq:defH}) 
we propose a Gutzwiller-correlated wave function 
as variational ground state
that we evaluate without approximations.
Therefore, the variational energies obtained in this work provide
upper bounds to the exact ground-state energy.
For comparison, we include results of the Gutzwiller variational approach for
the symmetric single-impurity Anderson model in~\ref{app:d}.

\subsection{Variational state}

In the Gutzwiller approach, we assume that the exact ground state can be approximated
by a normalized single-particle product state $|\varphi_0\rangle$ into which the
so-called Gutzwiller correlator $\hat{P}_{\rm G}$ introduces many-par\-ticle correlations,
\begin{equation}
|\Psi_{\rm G}\rangle 
= \hat{P}_{\rm G} |\varphi_0\rangle \; .
\end{equation}
For the non-interacting case, we recover the exact result by choosing
$|\varphi_0\rangle$ as the exact ground state of $\hat{H}_0$ in eq.~(\ref{eq:defH}),
$|\varphi_0\rangle\equiv | \Phi_0\rangle$, and 
$\hat{P}_{\rm G}(U=0) = \openone$.
In contrast to other variational approaches to the two-impurity Anderson
model~\cite{PhysRevB.48.7322,PhysRevB.44.450,Yanagisawa},
we determine $|\varphi_0\rangle$ fully variationally.

For our two-orbital situation, we employ the most general Hermitian correlator
\begin{equation}
\hat{P}_{\rm G} = \sum_{\Gamma} \lambda_{\Gamma}\hat{m}_{\Gamma}
+ \lambda_m|10\rangle\langle 11| + \lambda_m^*|11\rangle\langle 10| 
\end{equation}
that can be applied to the two-impurity subsystem and does not violate the symmetries.
Here, $\lambda_{\Gamma}$ are real variational parameters 
that control the occupation probabilities of the atomic configuration
$|\Gamma\rangle$ in the single-particle product state 
$|\varphi_0\rangle$. Particle number conservation and spin/particle-hole symmetry 
permit only the states $|10\rangle$ and $|11\rangle$ to be coupled in the correlator,
with the help of a complex parameter $\lambda_m$.

\subsection{Particle-hole symmetry}

We demand that our variational state is invariant under particle-hole symmetry
 at half band-filling, i.e., $\hat{\tau}_{\rm ph}^+
|\Psi_{\rm G}\rangle =|\Psi_{\rm G}\rangle$.
When we work with a particle-hole symmetric single-particle product state,
$\hat{\tau}_{\rm ph}^+|\varphi_0\rangle=|\varphi_0\rangle$,
we must demand that
\begin{equation}
\hat{\tau}_{\rm ph}^+ \hat{P}_{\rm G} \hat{\tau}_{\rm ph}^{\vphantom{+}}
=\hat{P}_{\rm G} \; .
\end{equation}
For $\Gamma=1,\ldots,5$, the projectors $\hat{m}_{\Gamma}$ 
obey
$\hat{\tau}_{\rm ph}^+\hat{m}_{\Gamma}\hat{\tau}_{\rm ph}^{\vphantom{+}}=
\hat{m}_{17-\Gamma}$,
and $\hat{\tau}_{\rm ph}^+\hat{m}_{6}\hat{\tau}_{\rm ph}^{\vphantom{+}}=
\hat{m}_{8}$.
For $\Gamma=7,9,10,11$, the projectors $\hat{m}_{\Gamma}$ 
are invariant under the particle-hole transformation~$\tau_{\rm ph}$,
and $|10\rangle\langle11|$ is equally invariant.
Therefore, to ensure particle-hole symmetry, we must set
$\lambda_{16}=\lambda_{1}$,
$\lambda_{15}=\lambda_{2}$,
$\lambda_{14}=\lambda_{3}$,
$\lambda_{13}=\lambda_{4}$,
$\lambda_{12}=\lambda_{5}$, and
$\lambda_{8}=\lambda_{6}$.
Moreover, due to symmetry under spin-flip $\uparrow\leftrightarrow \downarrow$, 
we set
$\lambda_6=\lambda_7=\lambda_8$,
$\lambda_{4}=\lambda_{5}$,
$\lambda_{3}=\lambda_{2}$.

We are left with eight
variational parameters for the spin and particle-hole symmetric
case whereby
$\lambda_1$, $\lambda_2$, $\lambda_4$,
$\lambda_6$, $\lambda_9$, $\lambda_{10}$, and $\lambda_{11}$ are real and
$\lambda_m$ is complex. We subsume them
in the real vector $\veclambda$,
\begin{equation}
\veclambda=(\lambda_1,
\lambda_2,\lambda_4,\lambda_6,\lambda_9,\lambda_{10},\lambda_{11},
x_m,y_m)
\label{eq:veclambdadef}
\end{equation}
with $x_m={\rm Re}[\lambda_m]$, $y_m={\rm Im}[\lambda_m]$.
At half band-filling, the atomic states $|1\rangle$, $|9\rangle$, and $|16\rangle$
belong to a charge-SU(2) triplet~\cite{Essler}. 
Therefore, we can directly set $\lambda_9=\lambda_1$.
We did not implement this symmetry but verified it numerically
to a high numerical accuracy so that the charge-SU(2)
symmetry is indeed preserved.

\subsection{Constraints}

To facilitate the evaluation of Gutz\-wil\-ler-correlated wave functions,
it is helpful to impose the constraints,
\begin{equation}
\langle \varphi_0|  \hat{P}_{\rm G}^+\hat{P}_{\rm G}^{\vphantom{+}}| \varphi_0\rangle 
=1 \; ,
\label{eq:firstconstraint}
\end{equation}
i.e., for our impurity system we normalize the wave function $|\Psi_{\rm G}\rangle$,
and 
\begin{equation}
\langle \varphi_0|  
\hat{P}_{\rm G}^+\hat{P}_{\rm G}^{\vphantom{+}}
\hat{h}_{b,\sigma}^+\hat{h}_{b,\sigma'}^{\vphantom{+}} | \varphi_0\rangle 
= \delta_{\sigma,\sigma'}
\langle \varphi_0| \hat{h}_{b,\sigma}^+\hat{h}_{b,\sigma}^{\vphantom{+}} 
| \varphi_0\rangle 
\; .
\label{eq:secondconstraint}
\end{equation}
These constraints do not restrict the variational free\-dom. They simply insure
that there are no Hartree bubbles in a diagrammatic
evaluation of Gutzwiller-correlated  wave 
functions~\cite{PhysRevB.41.9452,Buenemann_2012_a,PhysRevB.94.045135}.

Due to particle-hole symmetry, 
Eq.~(\ref{eq:h12andh21are zero})
also holds for the Gutzwiller variational state, i.e.,
\begin{equation}
\langle \varphi_0|  
\hat{P}_{\rm G}^+\hat{P}_{\rm G}^{\vphantom{+}}
\hat{h}_{b,\sigma}^+\hat{h}_{\bar{b},\sigma}^{\vphantom{+}} | \varphi_0\rangle 
= \langle \varphi_0| \hat{h}_{b,\sigma}^+\hat{h}_{\bar{b},\sigma}^{\vphantom{+}} 
| \varphi_0\rangle =0 
\label{eq:thirdconstraint}
\end{equation}
with the notation $\bar{1}\equiv 2$, $\bar{2}\equiv 1$.
The explicit conditions on $\veclambda$
are derived in~\ref{app:a}. 

\subsection{Calculation of expectation values}

\subsubsection{Host electrons}

For the host electrons we need to evaluate
\begin{equation}
\langle \hat{c}_{\veck,\sigma}^+\hat{c}_{\veck,\sigma}^{\vphantom{+}}
\rangle_{\rm G} = \frac{
\langle \Psi_{\rm G} |
\hat{c}_{\veck,\sigma}^+\hat{c}_{\veck,\sigma}^{\vphantom{+}} | \Psi_{\rm G} \rangle
}{\langle \Psi_{\rm G} | \Psi_{\rm G} \rangle} \; .
\end{equation}
By construction, the denominator is unity because
we normalized the Gutzwiller wave function, see eq.~(\ref{eq:firstconstraint}).
The numerator can be cast into the form
\begin{equation}
\langle \Psi_{\rm G} |
\hat{c}_{\veck,\sigma}^+\hat{c}_{\veck,\sigma}^{\vphantom{+}} | \Psi_{\rm G} \rangle
= 
\langle \varphi_0 |
\hat{c}_{\veck,\sigma}^+\hat{c}_{\veck,\sigma}^{\vphantom{+}} 
\hat{P}_{\rm G}^+\hat{P}_{\rm G}^{\vphantom{+}}
| \varphi_0 \rangle \; ,
\label{eq:fourty}
\end{equation}
which can be evaluated with the help of Wick's the\-orem.
All diagrams in eq.~(\ref{eq:fourty}) with lines between $\veck$ and
the impurity system vanish because of the constraint~(\ref{eq:secondconstraint})
and the fact that the constraint~(\ref{eq:thirdconstraint}) is fulfilled due to symmetry.
Consequently,
\begin{equation}
\langle \Psi_{\rm G} |
\hat{c}_{\veck,\sigma}^+\hat{c}_{\veck,\sigma}^{\vphantom{+}} | \Psi_{\rm G} \rangle
= 
\langle \varphi_0 |
\hat{c}_{\veck,\sigma}^+\hat{c}_{\veck,\sigma}^{\vphantom{+}} 
| \varphi_0 \rangle \; .
\end{equation}
This relation is very useful because the correlations are seen to change
only the impurity expectation values but not the host-electron energy,
\begin{eqnarray}
E_{\rm host}&=& \sum_{\veck,\sigma} \epsilon(\veck)
\langle \Psi_{\rm G} |
\hat{c}_{\veck,\sigma}^+\hat{c}_{\veck,\sigma}^{\vphantom{+}} | \Psi_{\rm G} \rangle 
\nonumber \\
&=& \sum_{\veck,\sigma} \epsilon(\veck)
\langle \varphi_0 |
\hat{c}_{\veck,\sigma}^+\hat{c}_{\veck,\sigma}^{\vphantom{+}} 
| \varphi_0 \rangle \; ,
\end{eqnarray}
as for a non-interacting symmetric two-impurity An\-der\-son model
with ground state $|\varphi_0\rangle$.

\subsubsection{Orbital occupancies}

Due to spin-flip symmetry, the orbital occupancies 
do not depend on the spin direction. Moreover, particle-hole
symmetry leads to
\begin{equation}
\langle \Psi_{\rm G} |
\hat{h}_{2,\uparrow}^+\hat{h}_{2,\uparrow}^{\vphantom{+}} 
| \Psi_{\rm G} \rangle
= 1- 
\langle \Psi_{\rm G} |
\hat{h}_{1,\uparrow}^+\hat{h}_{1,\uparrow}^{\vphantom{+}} 
| \Psi_{\rm G} \rangle \; ,
\end{equation}
cf.~eq.~(\ref{eq:btwoandbonedensities}).
We are left with the calculation of
\begin{equation}
\langle \Psi_{\rm G} |
\hat{h}_{1,\uparrow}^+\hat{h}_{1,\uparrow}^{\vphantom{+}} 
| \Psi_{\rm G} \rangle
=
\langle \varphi_0 |
\hat{P}_{\rm G}^+\hat{n}_{1,\uparrow}\hat{P}_{\rm G}^{\vphantom{+}}
| \varphi_0 \rangle \; .
\label{eq:occupancieseval}
\end{equation}
The matrix element is evaluated in~\ref{app:b}.
As a function of $\veclambda$ and $n_{1,\uparrow}^0$ it becomes
\begin{eqnarray}
\langle 
\hat{n}_{1,\uparrow}\rangle_{\rm G}
&=&  \frac{(\lambda_{10}+\lambda_{11}+2x_m)^2}{4} 
(n^0_{1,\uparrow})^4  \nonumber \\
&&
+ 3\lambda_2^2 (n^0_{1,\uparrow})^3\bar{n}^0_{1,\uparrow}
\nonumber \\
&&
+ \frac{(2\lambda_1^2+3\lambda_6^2+\lambda_9^2)}{2} 
(n^0_{1,\uparrow})^2 (\bar{n}^0_{1,\uparrow})^2
\nonumber \\
&&
+ \lambda_4^2 n^0_{1,\uparrow}( \bar{n}^0_{1,\uparrow})^3
\nonumber \\
&&
+ \frac{((\lambda_{11}-\lambda_{10})^2+4y_m^2)}{4} 
(\bar{n}^0_{1,\uparrow})^4
\; ,
\label{eq:correlateddensityfinalsimple}
\end{eqnarray}
where $n^0_{1,\uparrow}
=\langle \varphi_0 | \hat{h}_{1,\uparrow}^+\hat{h}_{1,\uparrow}^{\vphantom{+}}
| \varphi_0\rangle$
and $\bar{n}^0_{1,\uparrow}=1-n^0_{1,\uparrow}$.

\subsubsection{Hybridization}

Due to spin-flip symmetry, the hybridization matrix elements
do not depend on the spin direction. Moreover, particle-hole
symmetry leads to
\begin{equation}
\langle \Psi_{\rm G} |
\hat{c}_{\veck,\uparrow}^+\hat{h}_{2,\uparrow}^{\vphantom{+}} 
| \Psi_{\rm G} \rangle
= 
\langle \Psi_{\rm G} |
\hat{c}_{\vecQ-\veck,\uparrow}^+\hat{h}_{1,\uparrow}^{\vphantom{+}} 
| \Psi_{\rm G} \rangle^* \; .
\end{equation}
Therefore, we are left with the task to calculate
\begin{equation}
\langle \Psi_{\rm G} |
\hat{c}_{\veck,\uparrow}^+\hat{h}_{1,\uparrow}^{\vphantom{+}} 
| \Psi_{\rm G} \rangle
=
\langle \varphi_0 |
\hat{c}_{\veck,\uparrow}^+
\hat{P}_{\rm G}^+\hat{h}_{1,\uparrow}^{\vphantom{+}} \hat{P}_{\rm G}^{\vphantom{+}}
| \varphi_0 \rangle \; .
\label{eq:hybelement}
\end{equation}
This matrix element is evaluated in~\ref{app:b} with the result
\begin{equation}
\langle \Psi_{\rm G} |
\hat{c}_{\veck,\uparrow}^+\hat{h}_{1,\uparrow}^{\vphantom{+}} 
| \Psi_{\rm G} \rangle
=q 
\langle \varphi_0 |
\hat{c}_{\veck,\uparrow}^+\hat{h}_{1,\uparrow}^{\vphantom{+}} 
| \varphi_0 \rangle \; .
\label{eq:hybelementwithq}
\end{equation}
As a function of the Gutzwiller variational parameters $\veclambda$ and
of $n^0_{1,\uparrow}$ we find
 \begin{eqnarray}
q(\veclambda, n_{1,\uparrow}^0)&=& 
 \frac{\lambda_2(\lambda_{10}+\lambda_{11}+2x_m)}{2} 
(n^0_{1,\uparrow})^3\nonumber \\
&& 
+\frac{\lambda_2(2\lambda_1+3\lambda_6+\lambda_9)}{2}
(n^0_{1,\uparrow})^2\bar{n}^0_{1,\uparrow}
\nonumber \\
&& +\frac{\lambda_4(2\lambda_1+3\lambda_6+\lambda_9)}{2} 
n^0_{1,\uparrow}(\bar{n}^0_{1,\uparrow})^2 
\nonumber \\
&& + \frac{\lambda_{4}(\lambda_{10}+\lambda_{11}-2x_m)}{2} 
(\bar{n}^0_{1,\uparrow})^3 \;,
\label{eq:qfactorfinal}
\end{eqnarray}
where $\bar{n}^0_{1,\uparrow}=1-n^0_{1,\uparrow}$.

\subsubsection{Interaction}

For the interaction on the impurity we need to evaluate
\begin{equation}
\langle \hat{H}_{\rm int}\rangle_{\rm G} = 
\sum_{\Gamma} E_{\Gamma} 
\langle \varphi_0 |
\hat{P}_{\rm G}^+\hat{m}_{\Gamma}
\hat{P}_{\rm G}^{\vphantom{+}}
| \varphi_0 \rangle \equiv E_{\rm int}\; .
\label{eq:startHinteval}
\end{equation}
The matrix element is evaluated in~\ref{app:b}.
As a function of $\veclambda$ and $n_{1,\uparrow}^0$ it becomes
\begin{eqnarray}
\frac{2 E_{\rm int}}{U} &=&
(2\lambda_1^2-3\lambda_6^2+\lambda_9^2)
(n^0_{1,\uparrow})^2(\bar{n}^0_{1,\uparrow})^2
\nonumber \\
&& +\frac{\lambda_{11}^2 -\lambda_{10}^2}{2}
( (n^0_{1,\uparrow})^4+ (\bar{n}^0_{1,\uparrow})^4)
\nonumber \\
&& + x_m(\lambda_{11}-\lambda_{10})  ( (n^0_{1,\uparrow})^4- (\bar{n}^0_{1,\uparrow})^4)
\; .
\label{eq:Eintfinalsimple}
\end{eqnarray}

\subsection{Optimization of the single-particle state} 

To determine the variational parameters we must minimize the variational 
ground-state energy,
\begin{eqnarray}
E_{\rm var}(\veclambda,|\varphi_0\rangle) &=&
\langle \Psi_{\rm G} | \hat{H} | \Psi_{\rm G} \rangle \nonumber \\
&=& \sum_{\veck,\sigma} \epsilon(\veck) 
\langle \varphi_0 | \hat{c}_{\veck,\sigma}^+\hat{c}_{\veck,\sigma}^{\vphantom{+}} 
| \varphi_0 \rangle  + E_{\rm int} \nonumber \\
&& + 2 |t_{12}| \left(1-2 \langle \hat{n}_{1,\uparrow}\rangle_{\rm G}\right)\\
&& 
+ \sum_{\veck,b,\sigma}
\left[ qV_{\veck,b}\langle \varphi_0 | 
\hat{c}_{\veck,\sigma}^+\hat{h}_{b,\sigma}^{\vphantom{+}}| \varphi_0 \rangle
+ \hbox{c.c.}\right] \; , \nonumber
\end{eqnarray}
where explicit expressions for~$q(\veclambda,n_{1,\uparrow}^0)$ 
and various other expectation values
can be found in eqs.~(\ref{eq:correlateddensityfinalsimple}), (\ref{eq:qfactorfinal}), 
and~(\ref{eq:Eintfinalsimple}).
In the following we consider the case where there is no direct
coupling between the impurities, $|t_{12}|\to 0$.

To facilitate the optimization with respect to the normalized single-particle
state $|\varphi_0\rangle$, we consider the Lagrange functional
\begin{eqnarray}
{\cal L}&\equiv & {\cal L}\left(|\varphi_0\rangle, 
\tilde{t}_{12}, n_{1,\uparrow}^0\right) \nonumber\; ,\\  
{\cal L}&=&
E_{\rm var}(\veclambda,|\varphi_0\rangle) \nonumber \\
&& -\tilde{t}_{12} \sum_{\sigma}
\left(1-2n_{1,\uparrow}^0+ 
\langle \varphi_0 | \hat{n}_{1,\sigma}- \hat{n}_{2,\sigma}| \varphi_0\rangle\right)\; ,
\end{eqnarray}
where we consider the Gutzwiller parameters $\veclambda$ fixed.
Here, we introduced the Lagrange parameter
$\tilde{t}_{12}$. It guarantees that $n_{1,\uparrow}^0
=\langle \varphi_0 | \hat{n}_{1,\uparrow}| \varphi_0\rangle$ holds for the optimal
$|\varphi_0\rangle$; the other impurity occupancies follow
from particle-hole and spin-flip symmetry. 

The minimization of ${\cal L}$ with respect to the single-particle product state 
$|\varphi_0 \rangle$ shows that $|\varphi_0\rangle$ must be a normalized eigenstate
of an effective, non-interacting two-impurity Anderson model,
see appendix~C of~\cite{Buenemann-2012-b} or appendix~A of~\cite{GutzwillerDFT},
\begin{eqnarray}
\hat{H}_0^{\rm eff} 
|\varphi_0\rangle &=& E_{\rm sp} |\varphi_0\rangle\; , \nonumber \\
\hat{H}_0^{\rm eff}  &=& 
 \sum_{\veck,\sigma} \epsilon(\veck) 
\hat{c}_{\veck,\sigma}^+\hat{c}_{\veck,\sigma}^{\vphantom{+}} +
\tilde{t}_{12} \bigl( 
\hat{h}_{2,\sigma}^+\hat{h}_{2,\sigma}^{\vphantom{+}}
-\hat{h}_{1,\sigma}^+\hat{h}_{1,\sigma}^{\vphantom{+}} 
\bigr) \nonumber \\
&& + \sum_{\veck,b,\sigma}
qV_{\veck,b}\hat{c}_{\veck,\sigma}^+\hat{h}_{b,\sigma}^{\vphantom{+}}
+ qV_{\veck,b}^* 
\hat{h}_{b,\sigma}^+\hat{c}_{\veck,\sigma}^{\vphantom{+}}\; .
\label{eq:defHeffective}
\label{eq:setthree}
\end{eqnarray}
It is natural to choose $|\varphi_0\rangle$ as the normalized ground state of 
$\hat{H}_0^{\rm eff} $. 
Moreover, the derivative of the Lagrange function ${\cal L}$ 
with respect to $\tilde{t}_{12}$ gives back the condition
\begin{equation}
n_{1,\uparrow}^0
=\langle \varphi_0 | \hat{n}_{1,\uparrow}| \varphi_0\rangle 
\label{eq:fix-ttilde-qandn}
\end{equation}
that fixes $\tilde{t}_{12}(q,n_{1,\uparrow}^0)$.

\subsection{Atomic limit}

In the atomic limit, $V\equiv 0$, 
we must project onto the atomic eigenstates with
minimal energy, $E_{\Gamma}=-U/2$. Therefore, 
as seen from table~\ref{tab:Gammas},
we must set
$\lambda_1=\lambda_9=\lambda_{11}=\lambda_m=0$.
Furthermore, $\lambda_2=\lambda_4=0$ guarantees $q=0$.
The two constraints~(\ref{eq:defC1}) and~(\ref{eq:defC2}) 
reduce to ($n_{1,\uparrow}^0\equiv n, \bar{n}=1-n$)
\begin{eqnarray}
1=C_1&=& 3(\lambda_6^{\rm at})^2 n^2\bar{n}^2 +
 (\lambda_{10}^{\rm at})^2 (n^4+\bar{n}^4)/2\nonumber \; , \\[3pt]
2n =2C_2&=&  3(\lambda_6^{\rm at})^2 n^2\bar{n}^2 +
 (\lambda_{10}^{\rm at})^2 n^4 \; .
\end{eqnarray}
This gives
\begin{eqnarray}
(\lambda_6^{\rm at})^2 &=&\frac{1}{3n^2\bar{n}^2}\left(1 - 
\frac{(n-\bar{n})(n^4+\bar{n}^4)}{n^4-\bar{n}^4}\right)
\nonumber \; , \\[3pt]
(\lambda_{10}^{\rm at})^2 &=& \frac{2(n-\bar{n} )}{n^4-\bar{n}^4}
 \; .
\label{eq:atomicvalues}
\end{eqnarray}
{}From eq.~(\ref{eq:Eintfinalsimple}) it follows that
the interaction energy is $E_{\rm int}=-U/2$ for all $n$, as it must.
The occupation probabilities for the spin triplet and the spin singlet are given by
$p_t^{\rm at}=3(\lambda_6^{\rm at})^2 n^2\bar{n}^2$
and
$p_s^{\rm at}= (\lambda_{10}^{\rm at})^2 (n^4+\bar{n}^4)/2$, respectively.
For $n=\bar{n}=1/2$ we find 
$\lambda_6^{\rm at}=\lambda_{10}^{\rm at}=2$ so that
$p_t^{\rm at}=3/4$ and $p_s^{\rm at}=1/4$, 
as it should for two uncoupled spins on the impurity sites.

\section{Tight-binding host electrons}
\label{sec:tb-simplecubiclattice}

To obtain explicit results, we consider host electrons
on a simple cubic lattice with nearest-neighbor hopping 
of band width~$W\equiv 1$,
\begin{equation}
\epsilon(\veck)=-\frac{1}{6}\bigl(\cos(k_x)+\cos(k_y)+\cos(k_z)\bigr) \; .
\label{eq:scdispersion}
\end{equation}
We address the case of a small local hybridization, 
$V_{\veck}\equiv V\ll 1$. In addition, we assume that
the impurities are on different sublattices so that
$\vecR=\vecR_1-\vecR_2\in \hbox{$B$-lattice}$.
In  this case, the hybridization functions between odd and even channels
vanish, $H_{12}(\omega;\vecR)=H_{21}(\omega;\vecR)=0$, see~MBG.
The case where $\vecR_1$ and $\vecR_2$ belong to the same sublattice
is addressed briefly in~\ref{app:RinA}.

In~MBG we derived the single-particle energy, eq.~(\ref{eq:defHeffective}),
and the local particle density, eq.~(\ref{eq:fix-ttilde-qandn}). 
Here, we summarize the results for the non-interacting case with
a purely local hybridization.
For the interacting case, $V$ must be replaced by $qV$.

\subsection{Hybridization functions and density of states}

With the abbreviations
\begin{eqnarray}
R_1(\omega;\vecR)&=& \Lambda_{\veczero}(\omega)- \Lambda_B(\omega;\vecR)\; ,
\nonumber \\
R_2(\omega;\vecR)&=& \Lambda_{\veczero}(\omega)+ \Lambda_B(\omega;\vecR)\; ,
\nonumber \\
I_1(\omega;\vecR)&=&D_{\veczero}(\omega)- D_B(\omega;\vecR)
\; ,\nonumber \\
I_2(\omega;\vecR)&=&D_{\veczero}(\omega)+ D_B(\omega;\vecR)
\label{eq:abbreviations}
\end{eqnarray}
the hybridization functions are given by
\begin{equation}
H_{b,b}(\omega;\vecR)
= V^2R_b(\omega;\vecR)-\rmi \pi V^2 I_b(\omega;\vecR) \; .
\end{equation}
For electrons with nearest-neighbor transfers on a simple-cubic lattice 
at half band-filling the densities $D_{A,B}(\omega;\vecR)$ and their
Hilbert transforms $\Lambda_{A,B}(\omega;\vecR)$
are calculated from
\begin{eqnarray}
\Lambda_A(\omega;\vecR)&=& 
\delta_{\vecR\in A}
(-1)^{(R_x+R_y+R_z)/2} \nonumber \\
&& \times \!\!\int_0^{\infty} \!\rmd t \sin(\omega t)
J_{R_x}\Bigl(\frac{t}{6}\Bigr)J_{R_y}\Bigl(\frac{t}{6}\Bigr)
J_{R_z}\Bigl(\frac{t}{6}\Bigr),
\nonumber \\
D_A(\omega;\vecR)&=&
\delta_{\vecR\in A}
(-1)^{(R_x+R_y+R_z)/2} \nonumber \\
&& \times \!\!\int_0^{\infty}\!\frac{\rmd t}{\pi} \cos(\omega t)
J_{R_x}\Bigl(\frac{t}{6}\Bigr)J_{R_y}\Bigl(\frac{t}{6}\Bigr)
J_{R_z}\Bigl(\frac{t}{6}\Bigr)
\nonumber \\
\label{eq:resultDALamA}
\end{eqnarray}
for $|\omega|\leq 1/2$ and $\vecR\in\hbox{$A$-lattice}$, 
where $J_n(x)$ is the $n$th-order Bessel function.
In particular, $D_{\sigma,0}(\omega)=D_{\veczero}(\omega)=D_A(\omega;\veczero)$ and
$\Lambda_{\veczero}(\omega)=\Lambda_A(\omega;\veczero)$ for the local density
of states and its Hilbert transform.
Moreover,
\begin{eqnarray}
\Lambda_B(\omega;\vecR)&=& 
\delta_{\vecR\in B}
(-1)^{(R_x+R_y+R_z+3)/2}\nonumber \\
&& \times  \!\!\int_{0}^{\infty}\!\!\rmd t \cos(\omega t)
J_{R_x}\Bigl(\frac{t}{6}\Bigr)J_{R_y}\Bigl(\frac{t}{6}\Bigr)J_{R_z}\Bigl(\frac{t}{6}\Bigr),
\nonumber \\
D_B(\omega;\vecR)&=&
\delta_{\vecR\in B}
(-1)^{(R_x+R_y+R_z+1)/2}\nonumber\\
&& \times\!\!\int_{0}^{\infty}\!\frac{\rmd t}{\pi}\sin(\omega t)
J_{R_x}\Bigl(\frac{t}{6}\Bigr)J_{R_y}\Bigl(\frac{t}{6}\Bigr)J_{R_z}\Bigl(\frac{t}{6}\Bigr)
\nonumber \\
\label{eq:resultDBLamB}
\end{eqnarray}
for $|\omega| \leq 1/2$ and $\vecR\in\hbox{$B$-lattice}$.
We note that, for $|\vecR|\gg 1$,  the functions $D_{A,B}(\omega;\vecR)$ 
and $\Lambda_{A,B}(\omega;\vecR)$ oscillate strongly as a function of frequency
whereby their amplitude decays approximately proportional to $1/|\vecR|$.

For $|\omega|>1/2$, the hybridization functions are purely real, 
$I_b(\omega;\vecR)=0^+$, and, due to particle-hole symmetry,
 $R_b(-\omega;\vecR)=-R_b(\omega;\vecR)$ holds. In particular, for $\omega<-1/2$
\begin{eqnarray}
\Lambda_{\veczero}(\omega)&=& - \int_0^{\infty}\rmd \lambda e^{\lambda\omega}
\left[ I_0(\lambda/6)\right] ^3 \; , \nonumber \\
\Lambda_B(\omega;\vecR)&=& -\int_0^{\infty}\rmd \lambda e^{\lambda\omega}
 I_{R_x}(\lambda/6) I_{R_y}(\lambda/6) I_{R_z}(\lambda/6) \, ,\nonumber \\
\label{eq:defLambdaoutside}
\end{eqnarray}
where $I_n(x)$ is the $n$th-order modified Bessel function.

The continuous impurity contributions to the density of states are given by
{\arraycolsep=1pt\begin{eqnarray}
D_{b;\vecR}(\omega) &=&
\frac{V^2I_b(\omega;\vecR)}{N_{b;\vecR}(\omega)} \nonumber \; ,\\ 
N_{b;\vecR}(\omega) &=&
[\omega\pm\tilde{t}_{12}-V^2R_b(\omega;\vecR)]^2
+ 
[\pi V^2I_b(\omega;\vecR)]^2  ,\nonumber \\
\label{eq:D1D2}
\end{eqnarray}}%
where the upper (lower) sign applies to $b=1$ ($b=2$).
In case that the equations
\begin{equation}
\omega_b \pm\tilde{t}_{12}-V^2R_b(\omega_b;\vecR)=0
\label{eq:isthereomegazero}
\end{equation}
have a solution outside the band, i.e., for $\omega_b<-1/2$,
then the impurity density of states has a $\delta$-peak contribution because
$I_b(\omega_0;\vecR)=0^+$. The contribution 
to the impurity density of states is
\begin{eqnarray}
D_{b;\vecR}^{\delta}(\omega)&=&Z_{\vecR,b}\delta(\omega-\omega_b) \; ,
\nonumber \\
Z_{\vecR,b}&=& \frac{1}{1-V^2R_b'(\omega_b;\vecR)}  \; .
\label{eq:defZweight}
\end{eqnarray}
We set $Z_{\vecR,b}\equiv 0$ if eq.~(\ref{eq:isthereomegazero}) has no solution outside
the band. Recall that in eqs.~(\ref{eq:isthereomegazero}) and~(\ref{eq:defZweight})
the functions have to be calculated outside the band, i.e., 
eqs.~(\ref{eq:defLambdaoutside}) must be employed to calculate $R_b(\omega;\vecR)$
and its derivative.

Apart from the bare density of states, the host electron contribute
\begin{equation}
D_{\rm host,b,\vecR}(\omega)=\frac{1}{\pi}{\rm Im}
\left[
\frac{H'_{b,b}(\omega;\vecR)}{\omega\pm\tilde{t}_{12} -H_{b,b}(\omega;\vecR)}
\right] \; ,
\end{equation}
where the prime indicates the partial derivative with respect
to $\omega$. We thus find 
\begin{eqnarray}
D_{\rm host,b,\vecR}(\omega)
&=& -V^2\frac{I_b'(\omega;\vecR)\left(\omega\pm\tilde{t}_{12}
-V^2R_b(\omega;\vecR)\right)}{N_{b;\vecR}(\omega)} \nonumber \\
&& 
-\frac{V^2I_b(\omega;\vecR)R_b'(\omega;\vecR)}{N_{b;\vecR}(\omega)}\; .
\end{eqnarray}

\subsection{Particle density and single-particle energy}

The particle density for given $\vecR$ is obtained from
\begin{equation}
n_{b,\sigma}^0 = Z_{\vecR,b}+\int_{-1/2}^0 \rmd \omega D_{b;\vecR}(\omega)  \; ,
\label{eq:nbsigmafull}
\end{equation}
where we suppressed the lattice index in the particle density to shorten the expressions.
The two levels do not hybridize explicitly for $\vecR\in\hbox{$B$-lattice}$.
Therefore, the ground-state energy
of the non-interacting two-impurity Anderson model
can be cast into the form
\begin{eqnarray}
E_{\rm sp}(V,\tilde{t}_{12})&=&2 (Z_{\vecR,1}\omega_1+Z_{\vecR,2}\omega_2)
\nonumber \\
&& +
\frac{2}{\pi}\sum_b\int_{-1/2}^0\rmd \omega \,
\Cot^{-1} \left[\eta_{b,\vecR}(\omega)\right] 
\label{eq:TIAMRinBenergyfull}
\end{eqnarray}
with the phase-shift function
\begin{equation}
\eta_{b;\vecR}(\omega)=
\frac{\omega\pm \tilde{t}_{12}-V^2R_b(\omega;\vecR)}{\pi V^2I_b(\omega;\vecR)}
 \; .
\end{equation}
Note that we introduced the continuous and continuously differentiable 
function 
$\Cot^{-1}(x) = \cot^{-1}(x) -\pi \Theta(x)$
with the Heaviside step function $\Theta(x)$. 

The expression~(\ref{eq:TIAMRinBenergyfull})
is similar to the ground-state energy of the non-interacting symmetric 
single-impurity Anderson model~\cite{Annalenpaper},
\begin{equation}
E_{\rm SIAM}(V)=
\frac{2}{\pi}\int_{-1/2}^0\rmd \omega 
\cot^{-1}\left(
\frac{\omega-V^2\Lambda_{\veczero}(\omega)}{\pi V^2D_{\veczero}(\omega)}
\right)
\;.
\end{equation}
Since $\Lambda_{\veczero}(0)=0$, we do not have to discriminate be\-tween
${\rm Cot}^{-1}(x) $ and the standard inverse cotangent function $\cot^{-1}(x)$.
Moreover, for the SIAM there is no bound state
outside the band for $V\ll W$.

The energy functional 
is given by
\begin{eqnarray}
\bar{E}_{\rm var}(\veclambda,\tilde{t}_{12},n_{1,\uparrow}^0) 
&=&
E_{\rm sp}(qV,\tilde{t}_{12}) -2\tilde{t}_{12}(1-2n_{1,\uparrow}^0)\nonumber \\
&&+ E_{\rm int}(\veclambda,n_{1,\uparrow}^0)\; ,
\label{eq:Ebardef}
\end{eqnarray}
where $q\equiv q(\veclambda,n_{1,\uparrow}^0)$ from eq.~(\ref{eq:qfactorfinal}).
Moreover, the two constraints~(\ref{eq:firstconstraint}) 
and~(\ref{eq:secondconstraint}) must be obeyed.
They are worked out in~\ref{app:a} 
as eqs.~(\ref{eq:defC1}) and~(\ref{eq:defC2}).
The minimization with respect to $\tilde{t}_{12}$
returns eq.~(\ref{eq:nbsigmafull}). The solution of this implicit equation
determines $\tilde{t}_{12}(n_{1,\uparrow}^0)$.

\subsection{Limit of small hybridization}

For small hy\-bridizations, $V \ll 1$,
we may expand $E_{\rm sp}(qV,\tilde{t}_{12})$ in $V^2$.
To leading order in
$(qV)^2\ln[(qV)^2]$ and $(qV)^2$,
\begin{eqnarray}
E_{\rm sp}(qV, \bar{t}) &=& 
4(qV)^2d_{\veczero}\ln\Bigl[(qV)^2/C\Bigr]\nonumber\\
&& 
\!+2(qV)^2d_{\veczero}\ln\left[(\bar{t}-\pi\tilde{s}_{\vecR}d_{\vecR})^2
+(\pi d_{\veczero})^2\right]\nonumber \\
&& 
\!- 4(qV)^2 \frac{\bar{t}-\pi\tilde{s}_{\vecR}d_{\vecR}}{\pi} 
\arctan\Bigl( \frac{\bar{t}-\pi\tilde{s}_{\vecR}d_{\vecR}}{\pi d_{\veczero}}\Bigr)
\nonumber\\
\end{eqnarray}
with $\bar{t}=\tilde{t}_{12}/(qV)^2$, $C=0.7420 W$, see~\ref{app:cutoff}, 
and $d_{\veczero}=1.712/W$ is the host electron density of states 
at the Fermi energy $E_{\rm F}=0$, see~MBG. Moreover,
\begin{eqnarray}
\tilde{s}_{\vecR}&=& (-1)^{(R_x+R_y+R_z+1)/2}
\nonumber \; , \\
d_{\vecR}&=& \int_0^{\infty}\frac{\rmd t}{\pi} J_{R_x}(t/6)J_{R_y}(t/6)J_{R_z}(t/6) \; ,
\end{eqnarray}
where $d_{\vecR}$ is analyzed in more detail in~MBG.

The minimization with respect to~$\bar{t}$ 
returns $\tilde{t}_{12}=(qV)^2\bar{t}$ explicitly,
\begin{eqnarray}
\tilde{t}_{12}(q,n_{1,\uparrow}^0)
&=& \pi(qV)^2\left(d_{\veczero}\tan(\pi n_{1,\uparrow}^0-\pi/2)
+\tilde{s}_{\vecR}d_{\vecR}\right) \;.\nonumber \\
\label{eq:t12smallVexpression}
\end{eqnarray}
Therefore, we have to address
\begin{eqnarray}
\bar{E}_{\rm var}(\veclambda,n_{1,\uparrow}^0) 
&=&  E_{\rm int}(\veclambda,n_{1,\uparrow}^0)
+E_{\rm sp}(qV,\tilde{t}_{12}(n_{1,\uparrow}^0)\bigr)  \nonumber \\
&& -2\tilde{t}_{12}(q,n_{1,\uparrow}^0)(1-2n_{1,\uparrow}^0))\nonumber \\
&=& 
 E_{\rm int}(\veclambda,n_{1,\uparrow}^0)\nonumber \\
&& + 4d_{\veczero}(qV)^2 \Bigl[
\ln\left(\pi(qV)^2d_{\veczero}/C\right) \nonumber \\
&& \hphantom{+ 4d_{\veczero}(qV)^2 \Bigl[}
-\ln\left[\cos(\pi n_{1,\uparrow}^0-\pi/2)\right]\Bigr]
\nonumber \\
&& + 4(qV)^2(\pi n_{1,\uparrow}^0-\pi/2)\tilde{s}_{\vecR}d_{\vecR}
\label{eq:Ebardefasofn}
\end{eqnarray}
as a function of the Gutzwiller variational parameters~$\veclambda$
and of the level occupancy $n_{1,\uparrow}^0$ in the effective
non-interacting problem. It can be shown analytically that the choice
$y_m=0$ is optimal.

\section{Kondo limit}
\label{subsec:optimizationofGutzwparameters}

In the Kondo limit of large Hubbard interactions, $U\gg \Gamma$.
the Gutzwiller variational energy functional can be minimized analytically
to a far extent. 

\subsection{Simplification of the variational energy func\-tional}
\label{sec:heisenberglimit}

In the Kondo limit $U\gg \Gamma$,
we can safely set $\lambda_1=\lambda_9=0$.
Moreover, to obtain explicit expressions, we set $\lambda_{11}=x_m\equiv 0$.
This is an excellent approximation for all~$n\equiv n_{1,\uparrow}^0$ but finite
$(\lambda_{11},x_m)$ slightly improve the variational energy 
in the limits $n\to 0$ and $n\to 1$.

The two constraints~(\ref{eq:defC1}) and~(\ref{eq:defC2}) become
\begin{eqnarray}
1&=& 3\lambda_6^2 n^2\bar{n}^2 +
4(\lambda_2^2n^3\bar{n}+\lambda_4^2n \bar{n}^3)
+\frac{ \lambda_{10}^2}{2}(n^4+\bar{n}^4)\nonumber \; , \\[3pt]
2n &=&  3\lambda_6^2 n^2\bar{n}^2 +
6\lambda_2^2n^3\bar{n}+2\lambda_4^2n \bar{n}^3 +\lambda_{10}^2 n^4 \; .
\label{eq:givemelam6lam10}
\end{eqnarray}
These two equations can be solved analytically
for $\lambda_6$ and $\lambda_{10}$
as a function of $(\lambda_2,\lambda_4)$.

The interaction energy reads
\begin{equation}
-\frac{2 E_{\rm int}}{U}= 3\lambda_6^2 n^2\bar{n}^2 
+\frac{ \lambda_{10}^2}{2}(n^4+\bar{n}^4) \; .
\end{equation}
Using eq.~(\ref{eq:givemelam6lam10}), it only 
depends on $(\lambda_2,\lambda_4)$,
\begin{equation}
\frac{2}{U}\Bigl(E_{\rm int}+\frac{U}{2}\Bigr) =
4 \lambda_2^2n^3\bar{n}+4 \lambda_4^2n\bar{n}^3 \; .
\label{eq:Elam2lam4}
\end{equation}
Furthermore, to leading order in $1/U$ 
we find for the hybridization renormalization factor 
\begin{eqnarray}
q&=& \frac{\lambda_2\lambda_{10}^{\rm at}}{2}n^3 + 
\frac{3\lambda_2\lambda_6^{\rm at}}{2}n^2\bar{n} + 
\frac{3\lambda_4\lambda_6^{\rm at}}{2}n\bar{n}^2 +
\frac{\lambda_4\lambda_{10}^{\rm at}}{2}\bar{n}^3 \nonumber \\
&\equiv &
\gamma_2(n) \lambda_2+\gamma_4(n) \lambda_4\; .
\label{eq:shortq}
\end{eqnarray}
With the atomic values
for $\lambda_6^{\rm at}$ and $\lambda_{10}^{\rm at}$ from eq.~(\ref{eq:atomicvalues})
we have explicitly
\begin{eqnarray}
\gamma_2\equiv \gamma_2(n) &=& \frac{3}{2}n^2\bar{n}\lambda_6^{\rm at}
+ \frac{1}{2}n^3\lambda_{10}^{\rm at} \nonumber \\
&=& \frac{3}{2}n^2\bar{n}
\sqrt{\frac{1}{3n^2\bar{n}^2}\left(1 - 
\frac{(n-\bar{n})(n^4+\bar{n}^4)}{n^4-\bar{n}^4}\right)}\nonumber \\
&& + \frac{1}{2}n^3\sqrt{\frac{2(n-\bar{n} )}{n^4-\bar{n}^4}}
\; ,\\
\gamma_4\equiv \gamma_4(n) &=& \frac{3}{2}n\bar{n}^2\lambda_6^{\rm at}
+ \frac{1}{2}\bar{n}^3\lambda_{10}^{\rm at} \nonumber \\
&=& \frac{3}{2}n\bar{n}^2
\sqrt{\frac{1}{3n^2\bar{n}^2}\left(1 - 
\frac{(n-\bar{n})(n^4+\bar{n}^4)}{n^4-\bar{n}^4}\right)}\nonumber \\
&& + \frac{1}{2}\bar{n}^3\sqrt{\frac{2(n-\bar{n} )}{n^4-\bar{n}^4}}
\; .
\end{eqnarray}
Eq.~(\ref{eq:shortq}) reveals that $\lambda_2,\lambda_4$ are of the order of~$q$.
We set $\lambda_2=q \mu_2$ and $\lambda_4=q \mu_4$ so that the 
condition~(\ref{eq:shortq}) is fulfilled if the variational parameters  $\mu_2$ 
and $\mu_4$ obey $1=\mu_2\gamma_2+\mu_4\gamma_4$ or
$\mu_4=(1-\mu_2\gamma_2)/\gamma_4$.
Thus, the optimization of~(\ref{eq:Elam2lam4}) with respect to
$\mu_2$ leads to 
\begin{equation}
\lambda_2= 
q \frac{\bar{n}^2\gamma_2}{\gamma_2^2\bar{n}^2+\gamma_4^2 n^2} 
\quad, \quad
\lambda_4= 
q \frac{n^2\gamma_4}{\gamma_2^2\bar{n}^2+\gamma_4^2 n^2} \; ,
\label{eq:lambda2lambda4opt}
\end{equation}
and the remaining variational parameters are $q$ and $n\equiv n_{1,\uparrow}^0$ in
\begin{equation}
\bar{E}^{\rm K}_{\rm int}(q,n)=-\frac{U}{2}+ \frac{Up(n)}{4} q^2 \; , 
\end{equation}
where we introduced the abbreviation
\begin{equation}
p(n)= \frac{8n^3\bar{n}^3}{\bar{n}^2\gamma_2^2+n^2\gamma_4^2}
\end{equation}
with $\gamma_2(1/2)=\gamma_4(1/2)=1/2$ so that $p(1/2)=1$.

Dropping the constant term $-U/2$, in the Kondo limit 
the variational energy as a function of $q$ can be written in the form
\begin{equation}
\bar{E}_{\rm var}^{\rm K}(q,n)=B\left[ q^2 \ln(q^2) + A(n) q^2 \right]
\label{eq:EbarKondoAB}
\end{equation}
with
\begin{eqnarray}
A(n)&=& \frac{U p(n)}{16 d_{\veczero}V^2 } +
(n\pi-\pi/2)\frac{\tilde{s}_{\vecR}d_{\vecR}}{d_{\veczero}}
\nonumber\\
&& 
+ \ln\left(\pi V^2 d_{\veczero}/C\right) -\ln\left[\cos(\pi n -\pi/2)\right]\; ,\nonumber \\
B&=& 4d_{\veczero}V^2 \; .
\end{eqnarray}
The minimization of $\bar{E}_{\rm var}^{\rm K}(q,n)$ 
in eq.~(\ref{eq:EbarKondoAB}) with respect to~$q$ thus gives
$q=0$ (atomic limit) or
\begin{equation}
[q(n)]^2=\exp\left[-\bigr(1+A(n)\bigl)\right]
\label{eq:qoptimized}
\end{equation}
for the optimal $q$ as a function of~$n\equiv n_{1,\uparrow}^0$.

\subsection{Optimization of the density parameter}
\label{subsec:notransitioninKondolimit}

The optimal value for the level occupancy remains to be determined.
We insert the optimal value for~$q(n)$ from eq.~(\ref{eq:qoptimized})
in eq.~(\ref{eq:EbarKondoAB}) to find ($\ln(e)=1$)
\begin{eqnarray}
\bar{E}_{\rm var}^{\rm K}(n)
&=&\bar{E}_{\rm var}^{\rm K}(q(n),n)=-B[q(n)]^2=
-\frac{B}{e}e^{-A(n)} \nonumber \\
&=& -\frac{4C}{\pi e} \cos(\pi n -\pi/2)
\nonumber \\
&& \times \exp\Bigl (-\frac{\pi p(n) U}{16 \Gamma}
-(n\pi-\pi/2) \frac{\tilde{s}_{\vecR}d_{\vecR}}{d_{\veczero}}\Bigr)
\label{eq:EK-RKKYofn}
\end{eqnarray}
with $\Gamma=\pi d_{\veczero}V^2$.
The result suggests a Kondo-type variational energy 
as in the single-impurity case, see eq.~(\ref{eq:wci}) and~\ref{app:d},
eq.~(\ref{eq:GutzenergySIAM}). 
This, however, is actually not the case in the present Kondo limit, as we show now.

The density-dependent factor $p(n)$ monotonically increases (decreases)
for $n<1/2$ ($n>1/2$) and reaches its minimum at $n=0$ ($n=1$),
$p(0)=p(1)=0$. Therefore, we expand $A(n)$ around $n=0$ ($n=1$). 
For small Kondo couplings, we expand
$\bar{E}_{\rm var}^{\rm K}(n)$ 
to linear order in~$n$ for $\tilde{s}_{\vecR}d_{\vecR}>0$
and to linear order in~$(1-n)$ for $\tilde{s}_{\vecR}d_{\vecR}<0$,
respectively. With $\widetilde{n}\equiv n$ for $n\to 0$ and $\widetilde{n}=1-n$ 
for $n\to 1$ we find
\begin{eqnarray}
\bar{E}_{\rm var}^{\rm K}(\widetilde{n}\to 0)
&\approx &
-\frac{4C}{e}  \exp\Bigl[\frac{|\pi d_{\vecR}|}{2 d_{\veczero}}\Bigr] 
\nonumber \\
&& \times  \widetilde{n} \exp\Bigl[-\Bigl(\frac{\pi  U}{\Gamma}+
\frac{|\pi d_{\vecR}|}{d_{\veczero}}\Bigr)\widetilde{n}\Bigr]\; .
\end{eqnarray}
The minimum is found at
\begin{equation}
\widetilde{n}_{\rm opt}=\Bigl(\frac{\pi U}{\Gamma}+
\frac{|\pi d_{\vecR}|}{d_{\veczero}}\Bigr)^{-1}
\approx \frac{\Gamma}{\pi U}
\label{eq:noptTIAMnonint-final}
\end{equation}
with 
\begin{equation}
\bar{E}_{\rm var}^{\rm K, opt}
=-  \frac{4 C \Gamma }{\pi U} 
\exp\Bigl[\frac{|\pi d_{\vecR}|}{2 d_{\veczero}}-2\Bigr] +{\cal O}(1/U^2) \; .
\label{eq:Evarfinal}
\end{equation}
Eq.~(\ref{eq:Evarfinal}) shows the absence of a Kondo screening.
The two magnetic impurities sense each other 
via the host-electron bath. They form a spin singlet with a
Heisenberg-type energy gain proportional to $\Gamma/U$.
It is spatially modulated by a distance-depend\-ent factor 
of order unity, $\exp|\pi d_{\vecR}/(2d_{\veczero})|={\cal O}(1)$.

Note that $n\equiv n_{1,\uparrow}^0$ which is close to zero or unity, has no
physical significance. In fact, the physical level occupancy 
$\langle \hat{n}_{1,\uparrow}\rangle_{\rm G}=
\langle \hat{h}_{1,\uparrow}^+\hat{h}_{1,\uparrow}^{\vphantom{+}}
\rangle_{\rm G}$ 
in eq.~(\ref{eq:correlateddensityfinalsimple})
is close to half filling. For strong coupling where
$\lambda_1=0$, $\lambda_9=0$, $\lambda_m=0$, $\lambda_{11}=0$,
and $\lambda_2$, $\lambda_4$ are given by eq.~(\ref{eq:lambda2lambda4opt}),
we find for $n_{\rm opt}\to 1$
\begin{equation}
\langle \hat{n}_{1,\uparrow}\rangle_{\rm G}=
\langle \hat{h}_{1,\uparrow}^+\hat{h}_{1,\uparrow}^{\vphantom{+}}
\rangle_{\rm G}=
\frac{1}{2} +2(q^{\rm opt})^2 \widetilde{n}_{\rm opt}
\label{eq:nGopt}
\end{equation}
with the optimal value for the hybridization reduction factor
\begin{equation}
(q^{\rm opt})^2= \frac{C}{U} \exp\Bigl[\frac{|\pi d_{\vecR}|}{2 d_{\veczero}}-2\Bigr] 
+{\cal O}(1/U^2) \; .
\label{eq:qoptquadrat}
\end{equation}
Note that we derived these results under the condition $q\ll 1$
which required $U\gg \Gamma$, i.e., we had to address the Kondo limit.
In eq.~(\ref{eq:qoptquadrat}) only the ratio $W/U$ enters
and $q\ll 1$ also requires $U\gtrsim W$, i.e., we should not be far from the 
spin-model limit. Under these conditions, 
the impurity levels are almost exactly half filled, 
with corrections of the order $1/U^2$.

The energy gain~(\ref{eq:Evarfinal}) can be interpreted in terms
of an effective direct electron transfer between the impurities,
see our discussion in Sect.~\ref{sec:limits} and eq.~(\ref{eq:Heisenbergenergy}),
\begin{equation}
|t_{12}^{\rm direct}(\vecR)|=
\sqrt{C \Gamma/\pi}
\exp\Bigl[\frac{|\pi d_{\vecR}|}{4 d_{\veczero}}-1\Bigr]  \sim V\; .
\label{eq:t12directheislimit}
\end{equation}
Recall that, for $V\ll W$,  $E_{\rm var}^{\rm K, opt}$ is an exact variational
bound to the ground-state energy in the Kondo limit.
The result~(\ref{eq:t12directheislimit}) suggests a direct electron transfer 
proportional to~$V$ also in the exact solution.
This is indeed the case for a few-orbital toy model, 
as we show in~\ref{app:foursitemodel}, and has also been found in 
the original antiferromagnetic Hartree-Fock study~\cite{AlexanderAnderson}.

Moreover, the induced direct coupling does not vanish for large impurity
separations, $|\vecR|\to \infty$.
The origin of this somewhat 
surprising result lies in the fact that, for $U\approx W\gg \Gamma$,
the impurity electrons couple to all electrons in the system so that 
the physical distance of the impurity levels is of minor importance.
This is also reflected in the energetic position of the impurity levels 
of the effective non-interacting two-impurity Hamiltonian.
They are located  deep in the bare band 
at $|\tilde{t}_{12}^{\rm opt}(|\vecR|\gg1)|\approx (C/e^2)W ={\cal O}(W/2)$
so that not only energy levels close to the Fermi energy are involved
in the exchange interaction.

In summary, the results in this section show that, in the Kondo limit $U\gg \Gamma$
and for $U\approx W$, there is only a phase with singlet pairs and 
no transition to a phase with weakly coupled impurities can be realized.
Therefore, the transition is only conceivable for $U/\Gamma$-values 
that require a numerical minimization of the variational energy functional.
This will be discussed in Sect.~\ref{numminHeislimit}.
Nevertheless, as we discuss next, signatures of the transition are already discernible
in the Kondo-limit energy functional~(\ref{eq:EK-RKKYofn}).

\subsection{Quantum phase transition from the Kondo energy functional
for small hybridizations}

When the hybridization is (unrealistically) strong, we can obtain 
a nontrivial Kondo-type solution of the variational 
energy functional~(\ref{eq:EK-RKKYofn})
that permits an analytic discussion of the Varma-Jones scenario on the basis
of our Gutzwiller variational approach.

To obtain an exponential energy dependence, the minimum 
in eq.~(\ref{eq:EK-RKKYofn}) has to show up for values of~$n$ that are {\em not\/} 
close to zero or unity. This is possible if 
\begin{equation}
\frac{U}{\Gamma}< 8\pi[1-(d_{\vecR}/d_{\veczero})^2]
\label{eq:Jlargeenough}
\end{equation}
for some distance~$\vecR$. 
Since the right-hand-side of this equation is largest
for $|\vecR| \to \infty $,
we must demand $U/\Gamma<8\pi\approx 25$.
However, this contradicts our assumption 
$U\gg \Gamma$ in the Kondo limit.

For the sake of argument let us assume that the energy functional
is still valid in the critical regime, $U/\Gamma \approx 12\ldots 16$, 
see Sect.~\ref{numminHeislimit}.
Then, we find 
\begin{equation}
n_{\rm K}^{\rm opt} \approx \frac{1}{2} -
\frac{8(\Gamma/U)\tilde{s}_{\vecR}d_{\vecR}/d_{\veczero}}{
8 \pi (\Gamma/U) (1- (d_{\vecR}/d_{\veczero})^2)-1}
\label{eq:nkoptRKKY}
\end{equation}
remains in the vicinity of one half,
and the variational ground-state energy has a Kondo form,
\begin{equation}
E_{\rm var}^{\rm K}
=\bar{E}_{\rm var}^{\rm K}(n_{\rm K}^{\rm opt})=
2 E_{\rm opt}^{\rm SIAM} \left(1+\varepsilon_{\rm RKKY}^{B}(\vecR)\right)\; , 
\label{eq:Kondo-RKKYsinglet}
\end{equation}
as in eq.~(\ref{eq:wci})
with the single-impurity Gutzwiller variation\-al Kondo energy, see~\ref{app:d},
eq.~(\ref{eq:GutzenergySIAM}).
The RKKY energy enhancement is given by
\begin{equation}
\varepsilon_{\rm RKKY}^B(\vecR) = 
\frac{4\pi (\Gamma/U)(d_{\vecR}/d_{\veczero})^2}{
8 \pi (\Gamma/U)(1- (d_{\vecR}/d_{\veczero})^2)-1} \ll 1 \; ,
\label{eq:epsfactorB}
\end{equation}
where the upper index `$B$' indicates that the impurities are on different sublattices.
The enhancement vanishes for infinite 
impurity distances because \hbox{$d_{|\vecR|\to \infty}\to 0$}.
Equation~(\ref{eq:Kondo-RKKYsinglet}) permits 
a simple interpretation of the ground-state energy in terms of 
the Kondo and RKKY physics. Apparently, 
the two impurities are (partially) Kondo-screened and weakly
correlated by the RKKY interaction.

For a fixed impurity separation~$\vecR$, the scenario of Varma and Jones
can be realized as a function of $U/\Gamma$.
For a small enough $U/\Gamma$, we have $n\approx 1/2$ in the effective
single-particle Hamiltonian, and
the impurities represent weakly interacting
Kondo-screened spins, as expressed by eq.~(\ref{eq:Kondo-RKKYsinglet}).
Upon in\-creas\-ing $U/\Gamma$, the impurity spins (discontinuously) 
bind into a Heisenberg-type singlet, we find $|n-1/2|\approx 0.4$,
and the Kondo screening is absent.
The same scenario can be obtained for a suitable fixed $U/\Gamma$
as a function of~$\vecR$. For short distances, the impurity spins
are bound into Heisenberg singlets. Beyond a critical separation, 
$|\vecR|>R_c$, weakly interacting Kondo-screened spins appear.

\section{Numerical minimization}
\label{numminHeislimit}

We  minimize numerical\-ly
the full energy functional 
in eq.~(\ref{eq:Ebardef})
using a conjugate gradient method in combination
with the augmented penalty 
method~\cite{num-mini-book,PhysRevB.94.035116,pssgutzi}.
On a modern CPU, 
the optimization for fixed model parameters is a matter of seconds
if we use the small-$V$ approximation 
for the single-particle energy~(\ref{eq:Ebardefasofn}).

For the case of a general~$V$, we evaluate and store
$10^6$ values in the interval $[-1/2,0]$
for the densities $D_{A,B}(\omega_j;\vecR)$
and their Hilbert transforms $\Lambda_{A,B}(\omega_j;\vecR)$
for each $\vecR$. These discrete values provide
sampling points for the frequency integrations.
The relative accuracy of all data is better than $10^{-6}$.
We do not encounter bound states, $Z_{\vecR,b}=0$ in 
eqs.~(\ref{eq:nbsigmafull}), (\ref{eq:TIAMRinBenergyfull}).

\subsection{Ground-state energy and phase transition}

\begin{figure}[b]
\includegraphics[width=\columnwidth]{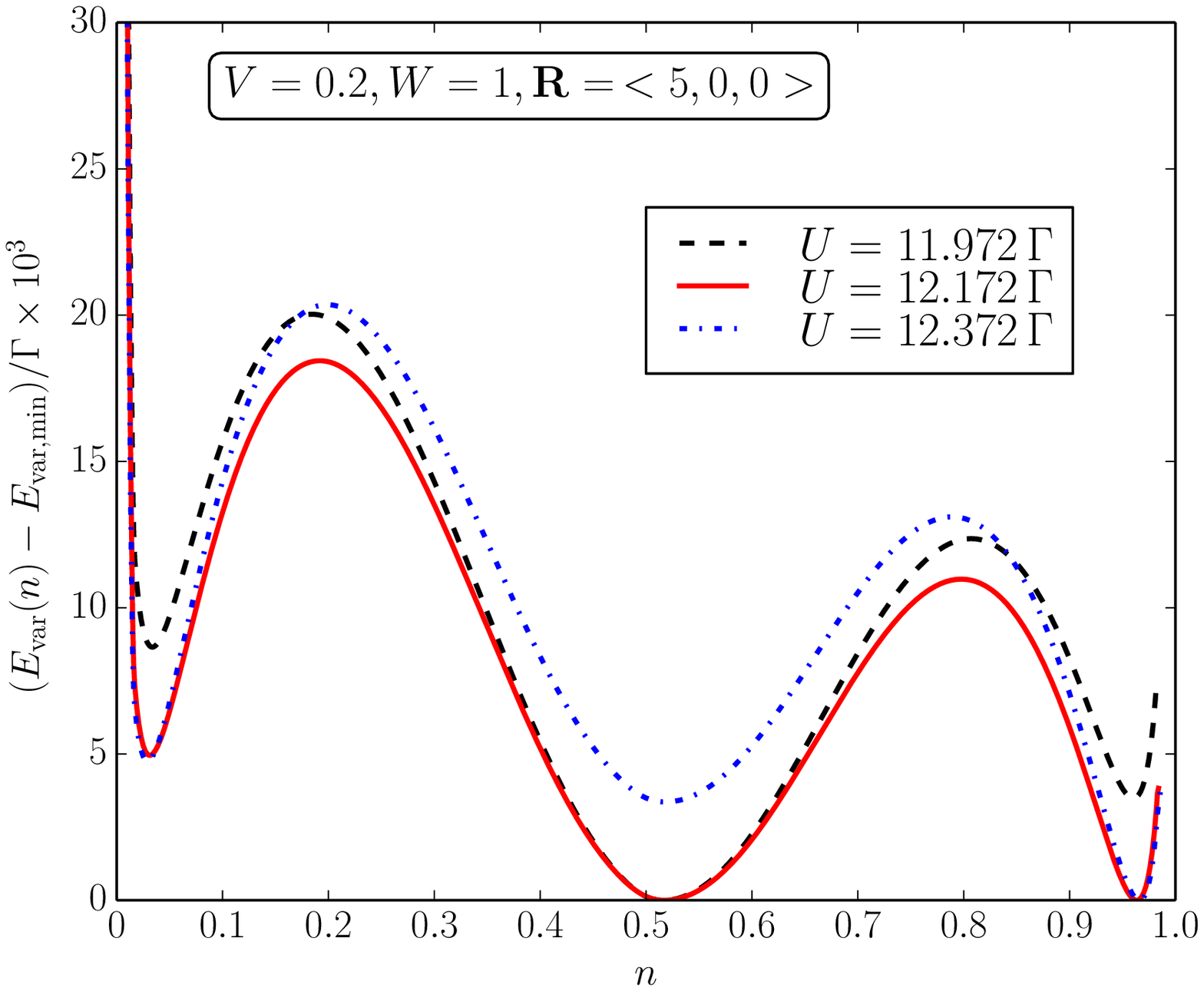}
\caption{Variational ground-state energy $E_{\rm var}(n)$ 
as a function of the density $n=n_{1,\uparrow}^0$
in the effective non-interacting two-impurity model~(\protect\ref{eq:defHeffective})
at $\vecR=<\!\!5,0,0\!\!>$
for $V=0.2$, $W=1$ and three values of~$U$ at and in the vicinity
of the critical value $U_{\rm c}(\vecR)=12.172\Gamma $.
The energies are shifted by their value at the minimum, 
$E_{\rm var,min}/\Gamma=-6.157732, -6.250614,-6.347167$ for
$U/\Gamma=11.972,12.172,12.372$, respectively. 
For $U\lesssim U_{\rm c}(\vecR)$, we observe weakly RKKY-interacting
Kondo-screened spins, $n^{\rm opt}\approx 0.517$,
for $U\gtrsim U_{\rm c}(\vecR)$ the impurity spins are bound 
into Heisenberg-type singlets, 
$n^{\rm opt}\approx 0.966$.\label{fig:EvarHeisenberglimit}}
\end{figure}

In Fig.~\ref{fig:EvarHeisenberglimit} we show the 
variational ground-state energy as a function of the density $n=n_{1,\uparrow}^0$
to illustrate the variational transition 
at $\vecR=<\!5,0,0\!>$ for $V=0.2$ and $W=1$.
Below the transition, $U\lesssim U_{\rm c}(\vecR)=12.172\Gamma$, 
the optimal variational wave function describes
weakly RKKY-interacting Kondo spins, compare eq.~(\ref{eq:Kondo-RKKYsinglet}), with
$n^{\rm opt}\approx 0.517$, see eq.~(\ref{eq:nkoptRKKY}).
Above the transition, $U\gtrsim U_{\rm c}(\vecR)$, the system prefers to form 
a Heisenberg-type singlet at $n^{\rm opt}\approx 0.966$.

As seen from the figure, the energy functional resembles that of a
fourth-order Landau functional 
for phase transitions~\cite{Landaubook} with even and odd powers,
where $n$ is the order parameter, $\Gamma/U$ acts as temperature
and $d_{\vecR}$ plays the role of an external field.
Therefore, we find a discontinuity in $n$ and a tricritical point~\cite{Lawrie}.

In general, at the transition the value of $n$ jumps from $n_{\rm K}$ for $U<U_{\rm c}$
to $n_{\rm H}$ for $U>U_{\rm c}$. 
As we discuss in more detail in Sect.~\ref{subsec:ptpstau},
this jump can also be seen in 
the multiplet occupations and in the expectation value
for the inter-impurity electron transfer,
\begin{equation}
\tau = - \frac{\langle \hat{T}_d\rangle}{2|t_{12}|}= 
\langle \hat{n}_{1,\uparrow}\rangle_{\rm G}-\frac{1}{2} \; .
\label{eq:taudef}
\end{equation}

\begin{figure}[b]
\includegraphics[width=\columnwidth]{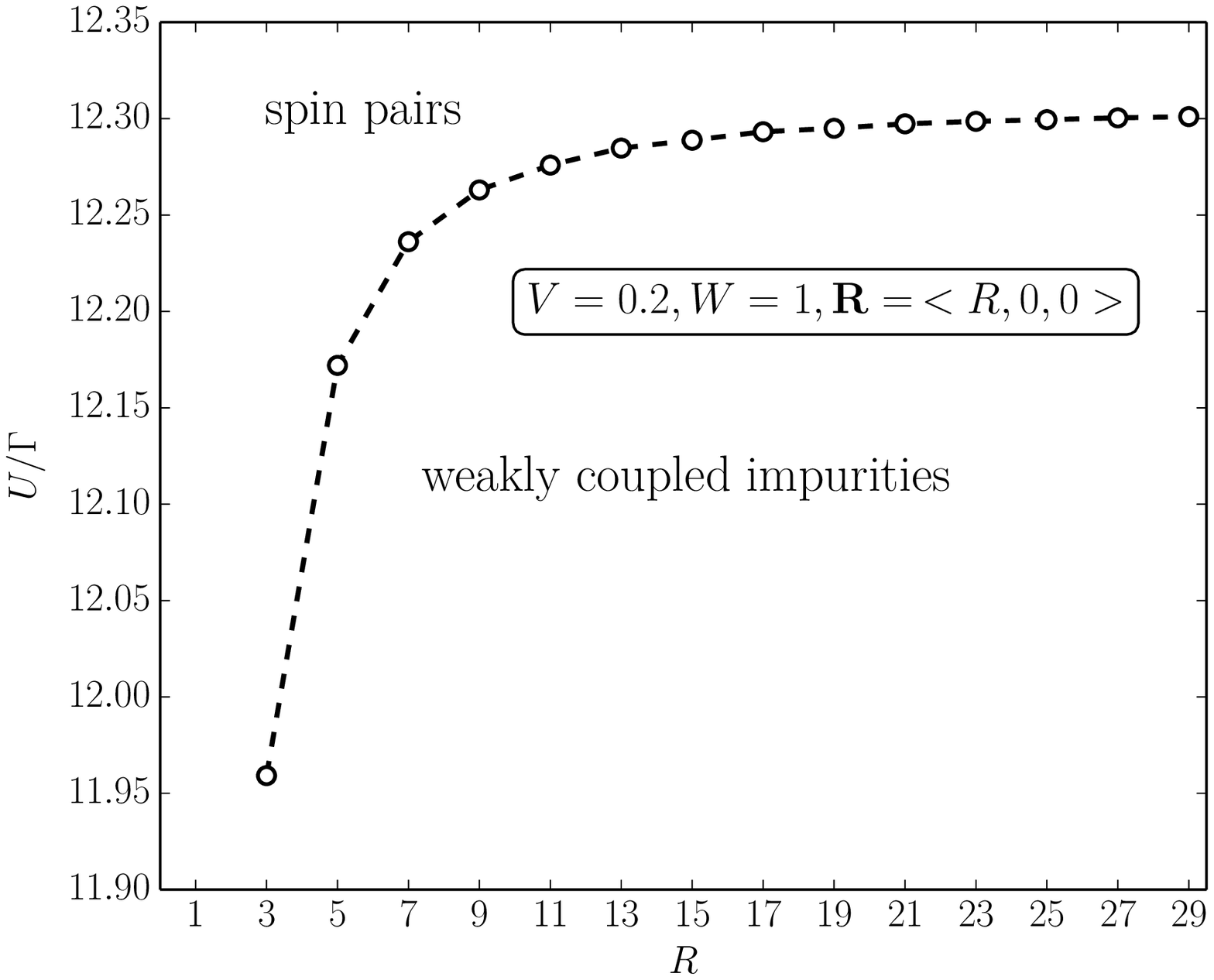}
\includegraphics[width=\columnwidth]{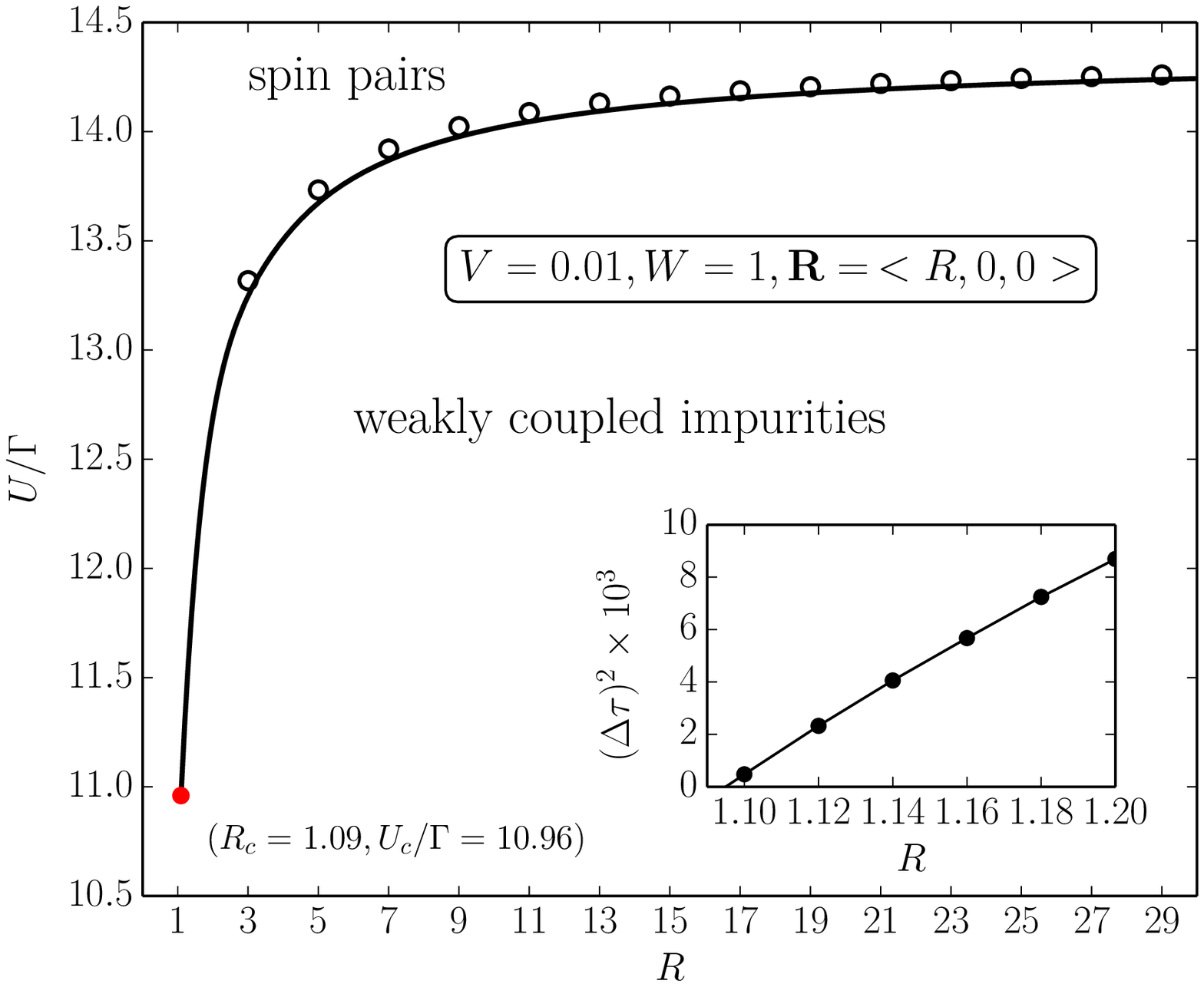}
\caption{Ground-state phase diagram for $\vecR=<\! R,0,0\!>$
and odd~$R$ for $V=0.2$ (upper part) and $V=0.01$ (lower part).
Weakly interacting impurities are found below the critical curve
$U_{\rm c}(\vecR)$, the phase with spin pairs is found above.
The line marks discontinuous transitions and ends at a tricritical point,
$(R_{\rm c},U_{\rm c}/\Gamma)\approx (1.09,10.96)$ for $V=0.01$.
The transition is continuous below $R_{\rm c}$,
i.e., for $\vecR=<\!1,0,0\!>$.
At $|\vecR|\to\infty$ we observe $U_{\rm c}^{\infty}/\Gamma=12.305$ 
for $V=0.2$ and $U_{\rm c}^{\infty}/\Gamma=14.33$ for $V=0.01$,
respectively. The continuous line gives the values for continuous~$R$ 
in the evaluation of $d_{\vecR}$ using the small-$V$ 
expression~(\protect\ref{eq:Ebardefasofn}), the dashed line is a guide to the eyes.
Inset: jump discontinuity $\Delta \tau=\tau(U_{\rm c}^{+})-
\tau(U_{\rm c}^{-})$ at the transition for $V=0.01$ from the small-$V$ 
expression.\label{fig:qpt}}
\end{figure}

\subsection{Phase diagram}

In Fig.~\ref{fig:qpt} we show the ground-state phase diagram
for $V=0.2$ and $V=0.01$ ($W=1$), respectively.
Below the transition line~$U_{\rm c}(\vecR)$,
weakly interacting impurities are observed. The critical line
marks discontinuous changes in the variational parameter $n\equiv n_{1,\uparrow}^0$
that show up, e.g., in the inter-impurity transfer matrix element~$\tau$, 
eq.~(\ref{eq:taudef}). As seen in the inset, the jump $\Delta \tau$ 
goes to zero at the critical endpoint $(R_{\rm c},U_{\rm c})$.
The mean-field exponent of one half is seen from the linear behavior
of $(\Delta \tau)^2$.

The transition is continuous below $R_{\rm c}$ so that 
for $\vecR=<\!\!1,0,0\!\!>$ 
the singlet state continuously forms from weakly-coupled impurities.
Therefore, in numerical simulations of the two-impurity Anderson model, 
the impurity distance must not be chosen too small to observe a conceivable 
transition, as suggested by our variational approach.

\begin{figure}[htb]
\includegraphics[width=\columnwidth]{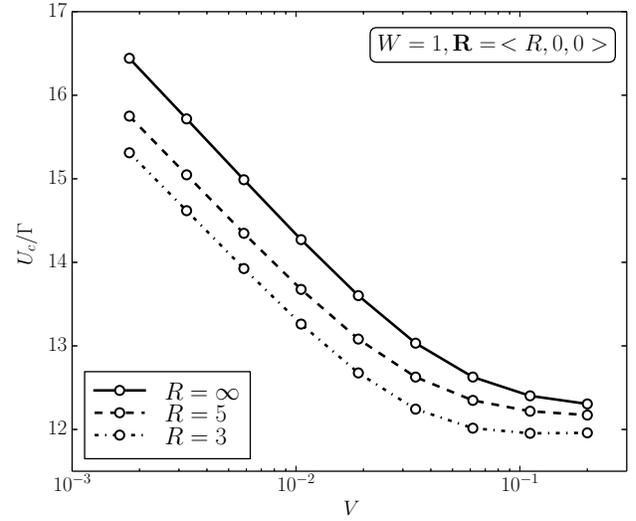}
\caption{Critical interaction strength $U_{\rm c}/\Gamma$ 
as a function of $2\cdot 10^{-3}\leq V \leq 2\cdot 10^{-1}$
for various values of $\vecR=<\!R,0,0\!>$ ($R=3,5,\infty)$; 
note the logarithmic scale on the abscissa. 
In the numerically accessible region,
the critical parameter lies in the range $U_{\rm c}/\Gamma\approx 12 \ldots 16$.
\label{fig:criticalUc}}
\end{figure}

In Fig.~\ref{fig:criticalUc} we show the 
critical interaction strength $U_{\rm c}/\Gamma$ as a function of~$V$
for various values of $\vecR=\hbox{$<\!R,0,0\!>$}$ ($R=3,5,\infty)$.
In the numerically accessible region, $2\cdot 10^{-3}\leq V\leq 2\cdot 10^{-1}$
the critical parameter lies in the range $U_{\rm c}/\Gamma=12 \ldots 16$.
This demonstrates that $\Gamma$ is indeed the relevant energy scale
with which~$U$ must be compared.
Furthermore, this confirms out previous claim 
in Sect.~\ref{subsec:notransitioninKondolimit} that
the transition never occurs in the Kondo or spin-model limits, $U\gg \Gamma$
or $U\gg W$, respectively. Therefore, the transition is difficult to access 
using effective spin models that approximate the two-impurity Anderson model,
or by any perturbative method~\cite{PhysRevB.49.6746,PhysRevB.62.12577,Grewe}.

\subsection{Multiplet occupations and effective 
inter-im\-purity transfer matrix element}
\label{subsec:ptpstau}

To gain further insight into the properties of the two different 
Gutzwiller variational states,
we discuss the probability to find the two impurities in a spin triplet 
and a spin singlet configuration. From spin symmetry we find that 
$\langle \hat{m}_6\rangle_{\rm G}=\langle \hat{m}_7\rangle_{\rm G}
=\langle \hat{m}_8\rangle_{\rm G}$ so that
\begin{equation}
p_t= 3\langle \hat{m}_6\rangle_{\rm G}=3 \lambda_6^2 n^2\bar{n}^2
\label{eq:deftripletprob}
\end{equation}
is the probability to find a triplet state.
The prob\-abil\-ity for a spin-singlet configuration is given by
\begin{eqnarray}
p_s&=&\langle \hat{m}_{10}\rangle_{\rm G}=
\frac{\lambda_{10}^2+x_m^2}{2}(n^4+\bar{n}^4) 
+ \lambda_{10}x_m(n^4-\bar{n}^4) \, .\nonumber \\
\label{eq:defsingletprob}
\end{eqnarray}

\begin{figure}[b]
\includegraphics[width=\columnwidth]{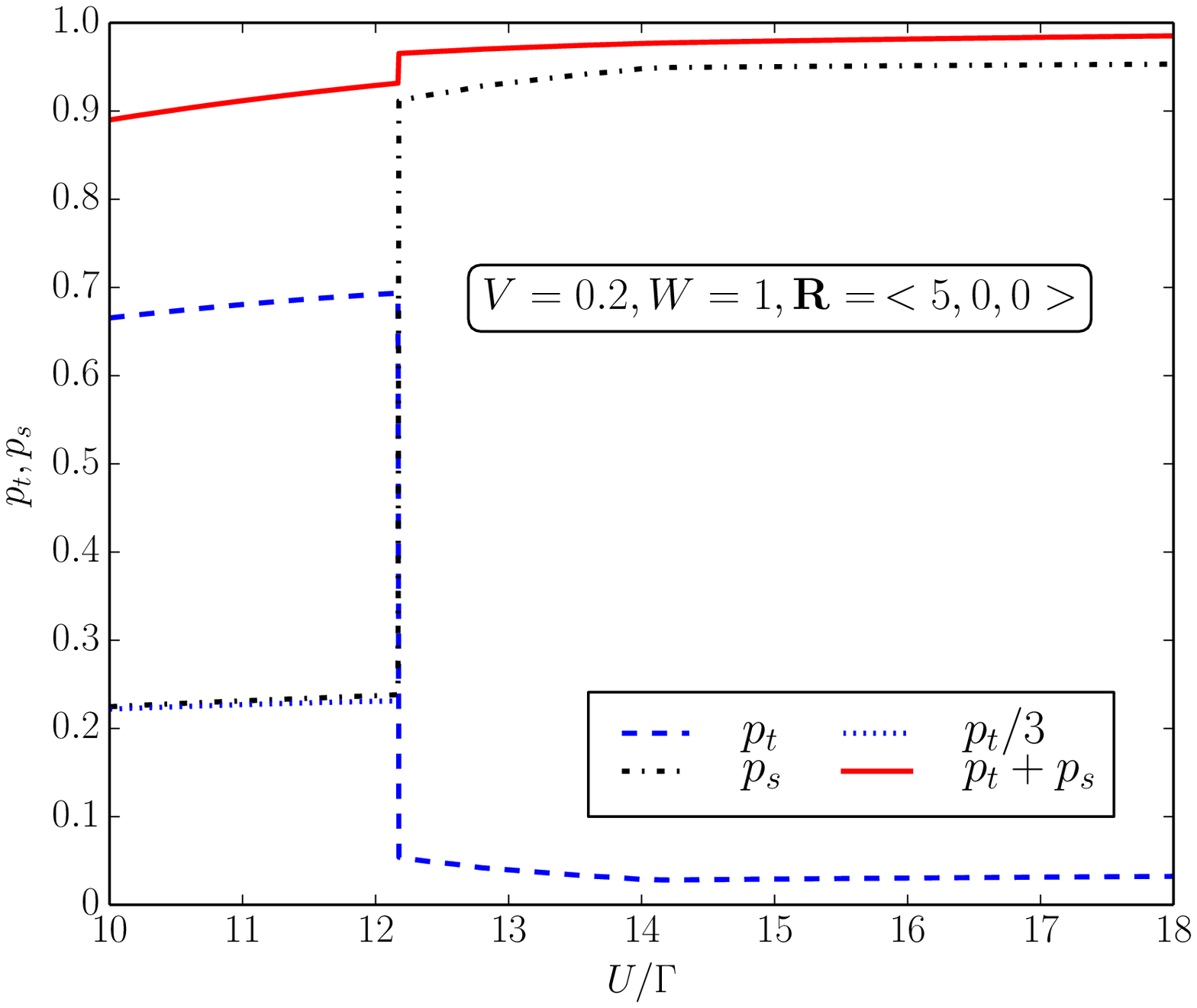}
\includegraphics[width=\columnwidth]{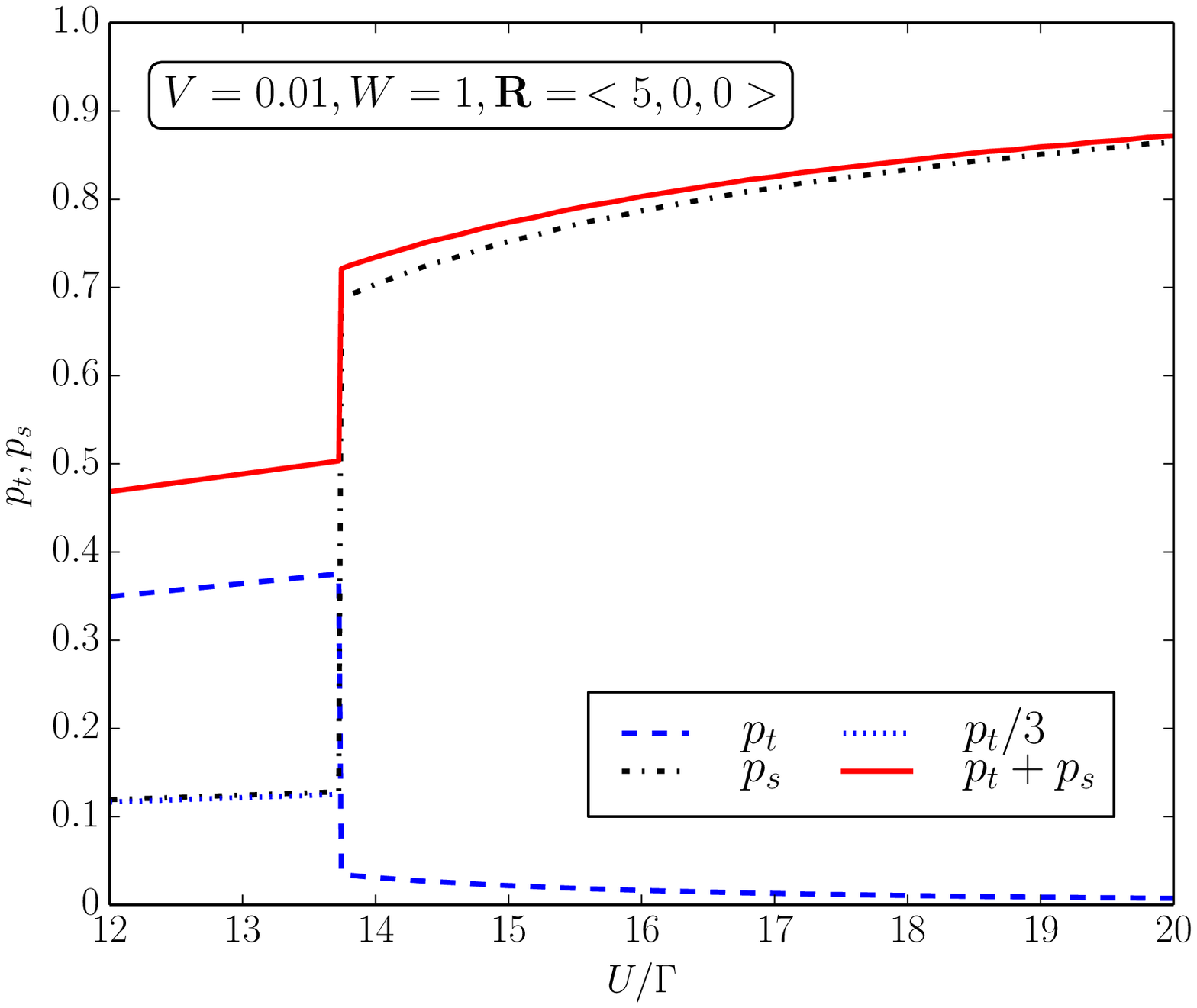}
\caption{Probabilities $p_t$ and $p_s$ to find the impurity spins
in the triplet and the singlet configuration 
at $\vecR=<\!5,0,0\!>$ as a function of the interaction strength~$U/\Gamma$
for $V=0.2$, upper part ($V=0.01$, lower part) and $W=1$.
At the critical value $U_{\rm c}(\vecR)=12.172\Gamma$ 
($U_{\rm c}(\vecR)=13.73\Gamma$) the system changes from
weakly coupled impurities to the spin-singlet phase.\label{fig:ptps}}
\end{figure}

In Fig.~\ref{fig:ptps} we show $p_t$ and $p_s$ 
for $\vecR=<\!\!5,0,0\!\!>$ as a function of~$U/\Gamma$ 
across the transition. For $V=0.2$,
$p_t+p_s\approx 1$ for all~$U\gtrsim 12\Gamma$ so that we are close to
the Kondo limit where the impurities are singly occupied.
For $U>U_{\rm c}(\vecR)$, the probability $p_t$ 
to find one of the three triplet states is small compared to $p_s$ so that
we can safely argue that 
the two impurity spins form a Heisenberg-type singlet state.
In contrast, for $U<U_{\rm c}(\vecR)$, the probability $p_s$ 
for a singlet configuration
is only marginally enhanced over the probability $p_t/3$ for one of the 
triplet states. This shows that we have two almost independent (Kondo-screened)
spins that display only a small RKKY-interaction induced 
tendency towards forming a singlet.

For $V=0.01$, the transition occurs far from the Kondo limit.
As seen from~\ref{app:d},  
the single-impurity Anderson model for $V=0.01$
quantitatively enters the Kondo regime for $U/\Gamma \gtrsim 50$.
Therefore, in contrast to the case $V=0.2$, the impurity electrons remain fairly
itinerant across the transition. This can be seen in Fig.~\ref{fig:ptps}
as the sum of the probabilities for spin singlet and triplet configurations 
is noticeably below unity.

\begin{figure}[htb]
\includegraphics[width=\columnwidth]{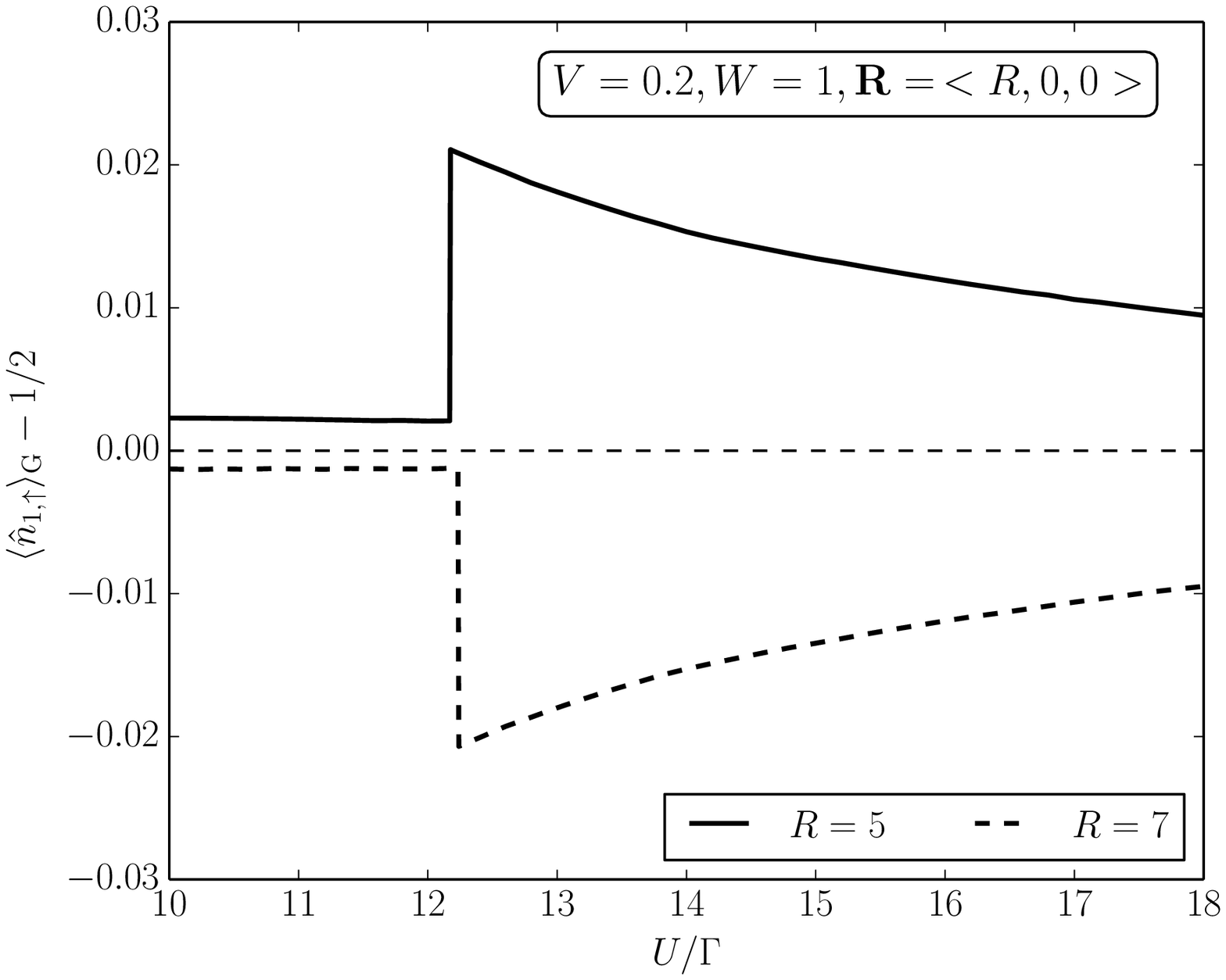}
\includegraphics[width=\columnwidth]{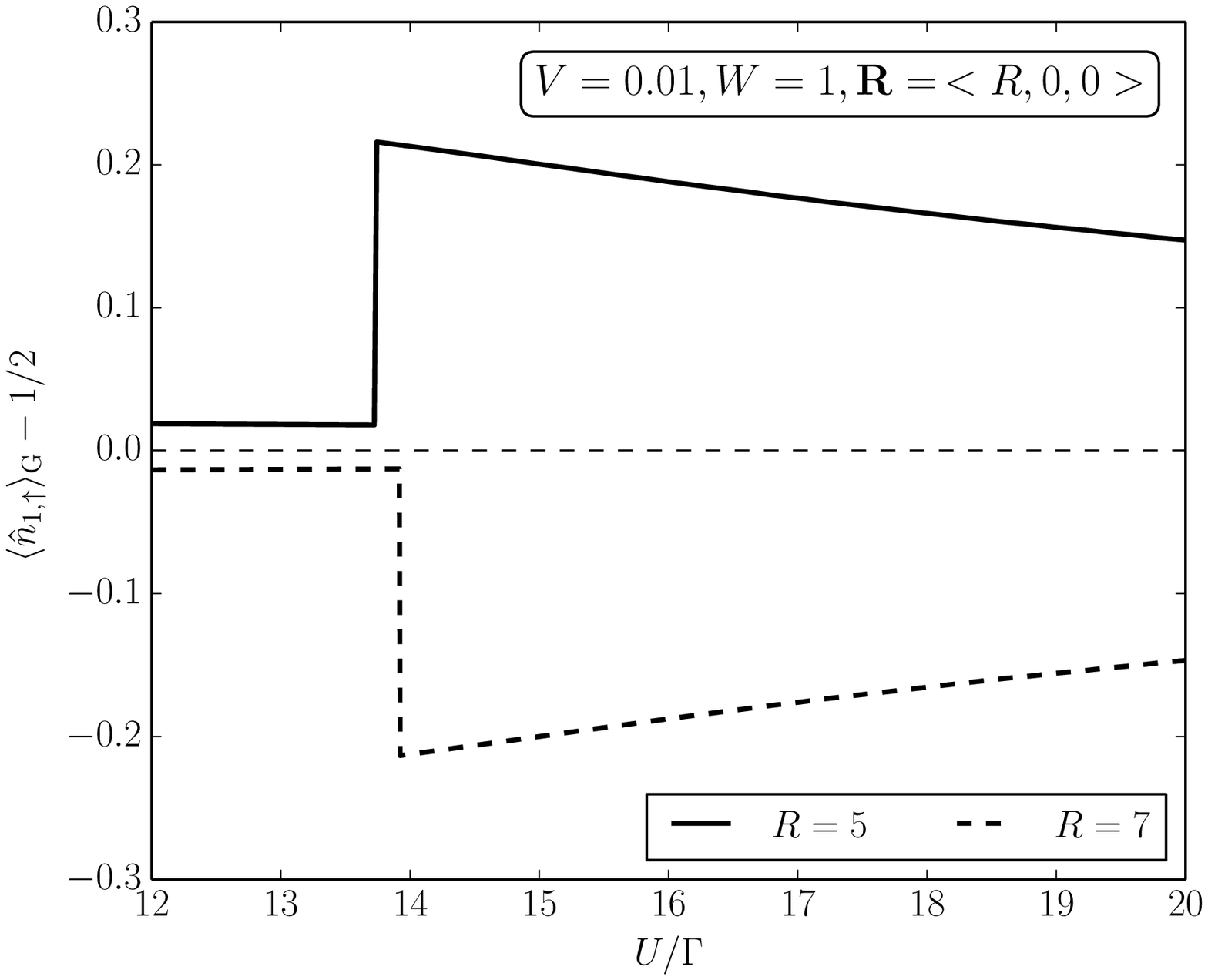}
\caption{Inter-impurity transfer matrix 
element~$\tau=\langle \hat{n}_{1,\uparrow}\rangle_{\rm G}-1/2$, 
eq.~(\protect\ref{eq:taudef}),
as a function of the interaction strength~$U/\Gamma$ 
for $V=0.2$, upper part ($V=0.01$, lower part)
at $\vecR=<\!(5,7),0,0\!>$ 
and $W=1$.
At the critical value $U_{\rm c}(\vecR)$ the system changes from
weakly coupled impurities to the spin-pair phase
with sizable inter-impurity electron transfer.\label{fig:nGsigma}}
\end{figure}

In Fig.~\ref{fig:nGsigma} we show the inter-impurity transfer matrix element~$\tau$
at $\vecR=<\!(5,7),0,0\!>$ as a function of the interaction strength~$U/\Gamma$
for $V=0.2$ ($V=0.01$), $W=1$.
The transfer matrix element~$\tau$ differs slightly from zero 
for all~$U/\Gamma$ because
the $h$-orbital symmetry is broken already at $U=0$, see~MBG.
At the transition to the spin-pair phase, 
the absolute value of the inter-impurity transfer matrix element~$\tau$ increases
discontinuously, in general.
In the Kondo limit, $U\gg \Gamma$, the effective transfer matrix element
decays proportional to $1/U^2$, see eq.~(\ref{eq:nGopt}).

\begin{figure}[htb]
\includegraphics[width=\columnwidth]{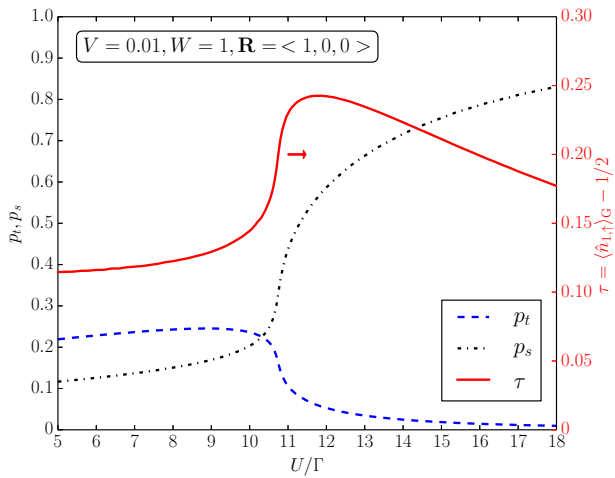}
\caption{Probabilities for singlet and triplet configurations, $p_s$ and $p_t$,
and inter-impurity transfer matrix 
element, $\tau=\langle \hat{n}_{1,\uparrow}\rangle_{\rm G}-1/2$, 
eq.~(\protect\ref{eq:taudef}),
as a function of the interaction strength~$U/\Gamma$ 
for $V=0.01$ at $\vecR=<\!1,0,0\!>$ 
and $W=1$. At $U/\Gamma\approx 11$, the system continuously changes
from weakly coupled impurities to spin pairs.\label{fig:NNtau}}
\end{figure}

It is only for $\vecR=\langle 1,0,0\rangle$ and for small hy\-brid\-izations~$V$
that the Gutzwiller approach for
the two-impurity Anderson model describes a crossover from
weakly correlated impurities to spin pairs. In Fig.~\ref{fig:NNtau} we show
the inter-impurity transfer matrix element~$\tau$ as a function
of $U/\Gamma$ for $V=0.01$. Around $U/\Gamma=11$, the symmetry
breaking parameter increases strongly before it decreases again 
for large interaction strengths. Fig.~\ref{fig:NNtau}
also shows the corresponding crossovers of the impurities' singlet and triplet
occupancies. 

We remind the reader that variational approaches
have a tendency to predict discontinuous 
quantum phase transitions. For example,
the Mott metal-insulator transition 
in the $1/r$-Hubbard model~\cite{GebhardRuckenstein}
is continuous as a function of the Hubbard interaction
but even elaborate variational wave functions 
predict it to be discontinuous~\cite{Florianmodelvariational,Dionysredo}.
Another example is provided by 
the two-impurity Kondo model with a Heisenberg exchange
between the impurities. When the impurities are on different sublattices 
there is no quantum 
phase transition~\cite{PhysRevLett.68.1046,AffleckLudwigJones}.
It might also be difficult to reproduce this scenario using
variational wave functions.

The reason for this shortcoming is fairly obvious.
When we compare the energies of two variational states 
that describe different
physical situations, we observe a level-crossing as a function of a
control parameter, typically some interaction strength.
The variational energy remains continuous but, in general, its derivatives
are discontinuous. In our case, we see that 
the probabilities 
for spin singlet and spin triplet display jump discontinuities 
at some critical Hubbard interaction.
Given the conceptual problems of variational approaches,
variational predictions for discontinuous quantum phase transitions
should not be overrated.

\section{Conclusions}
\label{sec:conclusions}

In this work, we analyzed Gutzwiller-correlated vari\-ational wave functions
as possible ground states for the particle-hole and spin
symmetric two-impurity Anderson model.
The single-particle product state
permits orbital-symmetry breaking 
in the two-level description that corresponds
to a finite single-electron transfer matrix element
between the two impurities. As known from the two-site Hubbard model,
the two impurities thus have a strong tendency to build a singlet state.
As a consequence, we find quantum phase transitions between
a regime with weakly coupled, partly Kondo-screened impurities to
a spin-pair regime  where the impurities form a spin singlet.

It is an advantage of our variational method that it covers 
effortless the whole parameter regime, i.e., we can readily
optimize the variational energy function for all
$(V,U,W)$ and all impurity separations~$\vecR$.
For host electrons that move between nearest neighbors
on a simple cubic lattice, 
we generically find a discontinuous 
quantum phase transition in the range $U_{\rm c}/\Gamma=12 \ldots 16$ 
for $2\cdot 10^{-3}W\leq V\leq 2\cdot 10^{-1} W$ where $\Gamma=\pi d_{\veczero}V^2$
and $d_{\veczero}\sim 1/W$ is the density of states at
of the host electrons at the Fermi energy. Since $U>3 \Gamma$,
the transition cannot be reached using weak-coupling perturbation
theory. For small $V\ll W$, the transition is also far from the
Kondo and spin limits where the impurities are singly occupied.
Therefore, the transition in the two-impurity Anderson model
cannot be described in terms of the two-impurity Kondo limit, in general.
It is only in the (unrealistic) case of fairly large hybridizations,
$V=0.2W$, that we approach the Kondo limit where 
the impurities are (almost) only singly occupied.

The main difference between our present study and previous approaches to
the two-impurity Anderson model lies in the fact that our variational state
permits an effective electron transfer between
the impurity sites ($h$-orbital symmetry breaking).
Note that the $h$-orbital symmetry is broken already for $U=0$ 
when the impurities are on different sublattices.
Therefore, this feature is generic for the two-impurity Anderson model.
For large interactions, $U\gg \Gamma$, the ground-state energy
is bound from above by an the energy gain proportional to $V^2/U$.
Since this is an exact bound we argue that a direct coupling of
the impurities via an electron transfer proportional to~$V$ 
also is a feature of the exact ground state, 
up to possible non-analytic corrections.
This picture was seen in Alexander and Anderson's
antiferromagnetic Hartree-Fock analysis~\cite{AlexanderAnderson} and
is supported by our few-orbital toy-model study.

For a qualitative understanding of our results 
we refer to our central finding in~MBG,
namely, that even at $U=0$ we must consider two hybridized impurities.
The RKKY approximation starts from bare impurities and thus 
does not give the correct size and distance-dependence of the
interaction between the impurities. 
These effects are generally not included in analytic approaches, e.g.,
the real part of the hybridization functions is often ignored.
For this reason, the $h$-orbital symmetry breaking is frequently excluded
from the beginning.

\ack
Z.M.M.\ Mahmoud thanks the Fachbereich Physik 
at the Philipps Universit\"at Marburg
for its hospitality.


\appendix

\section{Explicit form of the constraints}
\label{app:a}

\subsection{Square of the Gutzwiller correlator}
In general, the product $\hat{P}_{\rm G}^+\hat{P}_{\rm G}^{\vphantom{+}}$ has 
the form
\begin{eqnarray}
\hat{P}_{\rm G}^+\hat{P}_{\rm G}^{\vphantom{+}}
&=& \sum_{\Gamma} \lambda_{\Gamma}^2 \hat{m}_{\Gamma}
+ |\lambda_m|^2 \left(\hat{m}_{10}+\hat{m}_{11}\right)\nonumber \\
&&
+\left[\lambda_m(\lambda_{10}+\lambda_{11}) |10\rangle \langle 11|+\hbox{h.c.}\right]
\nonumber \\
&=& \openone + \left.\hat{P}_{\rm G}^+\hat{P}_{\rm G}^{\vphantom{+}}\right|_{\rm s}
+ \left.\hat{P}_{\rm G}^+\hat{P}_{\rm G}^{\vphantom{+}}\right|_{\rm d}
\label{eq:Gutzcorrsquare}
\end{eqnarray}
because $\hat{m}_{\Gamma}$ are projection operators, and
we separated the spin-flip and orbital-flip terms from the density
terms. We have 
\begin{eqnarray}
\left.\hat{P}_{\rm G}^+\hat{P}_{\rm G}^{\vphantom{+}}\right|_{\rm s}
&=& \Bigl[\frac{(\lambda_7^2-\lambda_9^2)}{2} 
\hat{h}_{1,\uparrow}^+\hat{h}_{1,\downarrow}^{\vphantom{+}}
\hat{h}_{2,\downarrow}^+\hat{h}_{2,\uparrow}^{\vphantom{+}}
\nonumber \\
&& +\frac{1}{2} 
\hat{h}_{1,\uparrow}^+\hat{h}_{1,\downarrow}^+
\hat{h}_{2,\downarrow}^{\vphantom{+}}\hat{h}_{2,\uparrow}^{\vphantom{+}}
\nonumber \\
&&\hphantom{\frac{\alpha^2}{2}\!}
\times
[\lambda_{11}^2-\lambda_{10}^2+(\lambda_m-\lambda_m^*)
(\lambda_{10}+\lambda_{11})]\Bigr]\nonumber \\
&& +\hbox{h.c.} \; , 
\end{eqnarray}
and 
\begin{eqnarray}
\left.\hat{P}_{\rm G}^+\hat{P}_{\rm G}^{\vphantom{+}}\right|_{\rm d}
&=& -\openone +
\lambda_1^2 \bar{n}_{1,\uparrow}\bar{n}_{1,\downarrow}
\bar{n}_{2,\uparrow}\bar{n}_{2,\downarrow} \nonumber \\
&& 
+ \lambda_2^2 \hat{n}_{1,\uparrow}\bar{n}_{1,\downarrow}
\bar{n}_{2,\uparrow}\bar{n}_{2,\downarrow} \nonumber \\
&& 
+ \lambda_3^2 \bar{n}_{1,\uparrow}\hat{n}_{1,\downarrow}
\bar{n}_{2,\uparrow}\bar{n}_{2,\downarrow} \nonumber \\
&&
+\lambda_4^2 \bar{n}_{1,\uparrow}\bar{n}_{1,\downarrow}
\hat{n}_{2,\uparrow}\bar{n}_{2,\downarrow} \nonumber \\
&&
+\lambda_5^2 \bar{n}_{1,\uparrow}\bar{n}_{1,\downarrow}
\bar{n}_{2,\uparrow}\hat{n}_{2,\downarrow} \nonumber \\
&& 
+ \lambda_6^2 \hat{n}_{1,\uparrow}\bar{n}_{1,\downarrow}
\hat{n}_{2,\uparrow}\bar{n}_{2,\downarrow} \nonumber \\
&&
+ \frac{(\lambda_7^2+\lambda_9^2)}{2}\hat{n}_{1,\uparrow}\bar{n}_{1,\downarrow}
\bar{n}_{2,\uparrow}\hat{n}_{2,\downarrow} 
\nonumber \\
&&
+ \frac{(\lambda_7^2+\lambda_9^2)}{2} 
\bar{n}_{1,\uparrow}\hat{n}_{1,\downarrow}
\hat{n}_{2,\uparrow}\bar{n}_{2,\downarrow}
 \nonumber \\
&& 
+ \lambda_8^2 \bar{n}_{1,\uparrow}\hat{n}_{1,\downarrow}
\bar{n}_{2,\uparrow}\hat{n}_{2,\downarrow} \nonumber \\
&&
+ \frac{(\lambda_{10}^2+\lambda_{11}^2+2|\lambda_m|^2)}{2}
\hat{n}_{1,\uparrow}\hat{n}_{1,\downarrow}
\bar{n}_{2,\uparrow}\bar{n}_{2,\downarrow}
\nonumber \\
&& 
+ \frac{(\lambda_{10}^2+\lambda_{11}^2+2|\lambda_m|^2)}{2}
\bar{n}_{1,\uparrow}\bar{n}_{1,\downarrow}\hat{n}_{2,\uparrow}\hat{n}_{2,\downarrow}
 \nonumber \\
&&
+ \frac{1}{2}(\lambda_m+\lambda_m^*)(\lambda_{10}+\lambda_{11})
\hat{n}_{1,\uparrow}\hat{n}_{1,\downarrow}
\bar{n}_{2,\uparrow}\bar{n}_{2,\downarrow}
 \nonumber \\
&&
- \frac{1}{2}(\lambda_m+\lambda_m^*)(\lambda_{10}+\lambda_{11})
\bar{n}_{1,\uparrow}\bar{n}_{1,\downarrow}\hat{n}_{2,\uparrow}\hat{n}_{2,\downarrow} 
\nonumber \\
&&
+\lambda_{12}^2 \hat{n}_{1,\uparrow}\bar{n}_{1,\downarrow}
\hat{n}_{2,\uparrow}\hat{n}_{2,\downarrow} \nonumber \\
&& 
+ \lambda_{13}^2 \bar{n}_{1,\uparrow}\hat{n}_{1,\downarrow}
\hat{n}_{2,\uparrow}\hat{n}_{2,\downarrow} \nonumber \\
&& 
+ \lambda_{14}^2 \hat{n}_{1,\uparrow}\hat{n}_{1,\downarrow}
\hat{n}_{2,\uparrow}\bar{n}_{2,\downarrow} \nonumber \\
&& 
+ \lambda_{15}^2 \hat{n}_{1,\uparrow}\hat{n}_{1,\downarrow}
\bar{n}_{2,\uparrow}\hat{n}_{2,\downarrow} \nonumber \\
&& 
+ \lambda_{16}^2 \hat{n}_{1,\uparrow}\hat{n}_{1,\downarrow}
\hat{n}_{2,\uparrow}\hat{n}_{2,\downarrow}\; ,
\label{eq:fullPGdens}
\end{eqnarray}
where we used the abbreviation $\bar{n}_{b,\sigma}=1-\hat{n}_{b,\sigma}$.
Due to the constraints~(\ref{eq:firstconstraint}),
(\ref{eq:secondconstraint}), and (\ref{eq:thirdconstraint}),
we can cast
$\hat{P}_{\rm G}^+\hat{P}_{\rm G}^{\vphantom{+}}\bigr|_{\rm d}$
into a form where local Hartree bubbles are absent,
{\arraycolsep=0pt\begin{eqnarray}
\left.\hat{P}_{\rm G}^+\hat{P}_{\rm G}^{\vphantom{+}}\right|_{\rm d}
&=& \sum_{(b,\sigma)<(b',\sigma')} 
X_{b\sigma;b'\sigma'} \delta\hat{n}_{b,\sigma}\delta\hat{n}_{b',\sigma'}
\nonumber \\
&& + \!\sum_{{{(b,\sigma)}\atop{<(b',\sigma')}}\atop {<(b'',\sigma'')}} 
\!Y_{b\sigma;b'\sigma';b'',\sigma''}
\delta\hat{n}_{b,\sigma}\delta\hat{n}_{b',\sigma'}
\delta\hat{n}_{b'',\sigma''}
\nonumber \\
&&+Z 
\delta\hat{n}_{1,\uparrow}
\delta\hat{n}_{1,\downarrow}
\delta\hat{n}_{2,\uparrow}
\delta\hat{n}_{2,\downarrow} \; ,
\label{eq:PG2withXYZ}
\end{eqnarray}}%
where we introduced the abbreviation
$\delta\hat{n}_{b,\sigma}
=\hat{n}_{b,\sigma}
-n_{b,\sigma}^0$,
$n_{b,\sigma}^0=\langle \varphi_0 | \hat{n}_{b,\sigma} | \varphi_0\rangle$,
and the orbital level order
$\hbox{$(1,\uparrow)$}<\hbox{$(1,\downarrow)$}<(2,\uparrow)<(2,\downarrow)$.
The spin/orbital-flip contribution
$\hat{P}_{\rm G}^+\hat{P}_{\rm G}^{\vphantom{+}}\bigr|_{\rm s}$
is free of Hartree bubbles due to the 
constraint~(\ref{eq:thirdconstraint}).
For the calculation of the variational ground-state energy, 
we do not have to know the coefficients $X$, $Y$, and $Z$
explicitly. 

\subsection{Constraints}

The representation~(\ref{eq:PG2withXYZ}) 
requires the constraints~(\ref{eq:firstconstraint})
and~(\ref{eq:secondconstraint}) to be fulfilled.
Using eq.~(\ref{eq:fullPGdens}) we find ($\bar{n}_{b,\sigma}^0=1-n_{b,\sigma}^0$)
\begin{eqnarray}
1&=& 
\lambda_1^2 (\bar{n}^0_{1,\uparrow}\bar{n}^0_{1,\downarrow}
\bar{n}^0_{2,\uparrow}\bar{n}^0_{2,\downarrow} 
+ n^0_{1,\uparrow}n^0_{1,\downarrow}n^0_{2,\uparrow}n^0_{2,\downarrow})
\nonumber \\
&& 
+ \lambda_2^2 (n^0_{1,\uparrow}\bar{n}^0_{1,\downarrow}
\bar{n}^0_{2,\uparrow}\bar{n}^0_{2,\downarrow} 
+ n^0_{1,\uparrow}n^0_{1,\downarrow}
\bar{n}^0_{2,\uparrow}n^0_{2,\downarrow})\nonumber \\
&& 
+ \lambda_2^2 (\bar{n}^0_{1,\uparrow}n^0_{1,\downarrow}
\bar{n}^0_{2,\uparrow}\bar{n}^0_{2,\downarrow} 
+ n^0_{1,\uparrow}n^0_{1,\downarrow}n^0_{2,\uparrow}\bar{n}^0_{2,\downarrow})
\nonumber \\
&&
+\lambda_4^2 (\bar{n}^0_{1,\uparrow}\bar{n}^0_{1,\downarrow}
n^0_{2,\uparrow}\bar{n}^0_{2,\downarrow} 
+ \bar{n}^0_{1,\uparrow}n^0_{1,\downarrow}n^0_{2,\uparrow}n^0_{2,\downarrow})
\nonumber \\
&&
+\lambda_4^2 (\bar{n}^0_{1,\uparrow}\bar{n}^0_{1,\downarrow}
\bar{n}^0_{2,\uparrow}n^0_{2,\downarrow} 
+n^0_{1,\uparrow}\bar{n}^0_{1,\downarrow} n^0_{2,\uparrow}n^0_{2,\downarrow} )
\nonumber \\
&& 
+ \lambda_6^2 (n^0_{1,\uparrow}\bar{n}^0_{1,\downarrow}
n^0_{2,\uparrow}\bar{n}^0_{2,\downarrow}
+\bar{n}^0_{1,\uparrow}n^0_{1,\downarrow}\bar{n}^0_{2,\uparrow}n^0_{2,\downarrow} )
 \nonumber \\
&&
+ \frac{(\lambda_6^2+\lambda_9^2)}{2}(n^0_{1,\uparrow}\bar{n}^0_{1,\downarrow}
\bar{n}^0_{2,\uparrow}n^0_{2,\downarrow} 
+\bar{n}^0_{1,\uparrow}n^0_{1,\downarrow}
n^0_{2,\uparrow}\bar{n}^0_{2,\downarrow})
 \nonumber \\
&&
+ \frac{(\lambda_{10}^2+\lambda_{11}^2+2|\lambda_m|^2)}{2}
n^0_{1,\uparrow}n^0_{1,\downarrow}
\bar{n}^0_{2,\uparrow}\bar{n}^0_{2,\downarrow}
\nonumber \\
&& 
+ \frac{(\lambda_{10}^2+\lambda_{11}^2+2|\lambda_m|^2)}{2}
\bar{n}^0_{1,\uparrow}\bar{n}^0_{1,\downarrow}n^0_{2,\uparrow}n^0_{2,\downarrow}
 \nonumber \\
&&
+ \frac{1}{2}(\lambda_m+\lambda_m^*)(\lambda_{10}+\lambda_{11})
n^0_{1,\uparrow}n^0_{1,\downarrow}
\bar{n}^0_{2,\uparrow}\bar{n}^0_{2,\downarrow}
 \nonumber \\
&&
- \frac{1}{2}(\lambda_m+\lambda_m^*)(\lambda_{10}+\lambda_{11})
\bar{n}^0_{1,\uparrow}\bar{n}^0_{1,\downarrow}n^0_{2,\uparrow}n^0_{2,\downarrow}\; .
\label{eq:normfulfil}
\end{eqnarray}
Particle-hole symmetry and spin symmetry permit to express the constraint 
solely as a function of the Gutzwiller parameters $\veclambda$
and of
$n_{1,\uparrow}^0=n_{1,\downarrow}^0=\bar{n}_{2,\uparrow}^0=
\bar{n}_{2,\downarrow}^0$,
 \begin{eqnarray}
1 &=& 
(2\lambda_1^2+3\lambda_6^2+\lambda_9^2)
(n_{1,\uparrow}^0)^2(\bar{n}_{1,\uparrow}^0)^2\nonumber \\
&& +4\lambda_2^2(n_{1,\uparrow}^0)^3\bar{n}_{1,\uparrow}^0
+ 4\lambda_4^2n_{1,\uparrow}^0(\bar{n}_{1,\uparrow}^0)^3
\nonumber \\
&& + \frac{(\lambda_{10}^2+\lambda_{11}^2)}{2}
\bigl((n_{1,\uparrow}^0)^4+(\bar{n}_{1,\uparrow}^0)^4\bigr)\nonumber\\
&& + |\lambda_m|^2
\bigl((n_{1,\uparrow}^0)^4+(\bar{n}_{1,\uparrow}^0)^4\bigr)\nonumber\\
&& + x_m(\lambda_{10}+\lambda_{11})\bigl( (n_{1,\uparrow}^0)^4
- (\bar{n}_{1,\uparrow}^0)^4\bigr) \; . 
\label{eq:defC1}
\end{eqnarray}
Using eq.~(\ref{eq:secondconstraint})  with $b=1,\sigma=\uparrow$ we find 
\begin{eqnarray}
n_{1,\uparrow}^0&=& 
\lambda_1^2 n^0_{1,\uparrow}n^0_{1,\downarrow}
n^0_{2,\uparrow}n^0_{2,\downarrow}\nonumber \\
&& +
\lambda_2^2 n^0_{1,\uparrow}\bar{n}^0_{1,\downarrow}
\bar{n}^0_{2,\uparrow}\bar{n}^0_{2,\downarrow} \nonumber \\
&& 
+ \lambda_2^2 n^0_{1,\uparrow}n^0_{1,\downarrow}
n^0_{2,\uparrow}\bar{n}^0_{2,\downarrow} \nonumber \\
&& 
+ \lambda_2^2 n^0_{1,\uparrow}n^0_{1,\downarrow}
\bar{n}^0_{2,\uparrow}n^0_{2,\downarrow} \nonumber \\
&& 
+ \lambda_{4}^2 n^0_{1,\uparrow}\bar{n}^0_{1,\downarrow}
n^0_{2,\uparrow}n^0_{2,\downarrow} \nonumber \\
&& 
+ \lambda_6^2 n^0_{1,\uparrow}\bar{n}^0_{1,\downarrow}
n^0_{2,\uparrow}\bar{n}^0_{2,\downarrow} \nonumber \\
&&
+ \frac{(\lambda_6^2+\lambda_9^2)}{2}n^0_{1,\uparrow}\bar{n}^0_{1,\downarrow}
\bar{n}^0_{2,\uparrow}n^0_{2,\downarrow} 
\nonumber \\
&&
+ \frac{(\lambda_{10}^2+\lambda_{11}^2+2|\lambda_m|^2)}{2}
n^0_{1,\uparrow}n^0_{1,\downarrow}
\bar{n}^0_{2,\uparrow}\bar{n}^0_{2,\downarrow}
\nonumber \\
&&
+ \frac{1}{2}(\lambda_m+\lambda_m^*)(\lambda_{10}+\lambda_{11})
n^0_{1,\uparrow}n^0_{1,\downarrow}
\bar{n}^0_{2,\uparrow}\bar{n}^0_{2,\downarrow}\; .
\nonumber \\
\label{eq:fullfillsecondbisone}
\end{eqnarray}
As a function of the Gutzwiller variational parameters $\veclambda$ and
of $n^0_{1,\uparrow}$ we find
\pagebreak[3]
 \begin{eqnarray}
n^0_{1,\uparrow} &=& 
(\lambda_1^2+3\lambda_6^2/2+\lambda_9^2/2)
(n_{1,\uparrow}^0)^2(\bar{n}_{1,\uparrow}^0)^2\nonumber \\
&&+ 3\lambda_2^2(n_{1,\uparrow}^0)^3\bar{n}_{1,\uparrow}^0
+ \lambda_4^2n_{1,\uparrow}^0(\bar{n}_{1,\uparrow}^0)^3\nonumber \\
&& + \frac{\lambda_{10}^2+\lambda_{11}^2+2|\lambda_m|^2}{2}(n_{1,\uparrow}^0)^4
\nonumber \\
&&+ x_m(\lambda_{10}+\lambda_{11})(n_{1,\uparrow}^0)^4\; .
\label{eq:defC2}
\end{eqnarray}
Eq.~(\ref{eq:secondconstraint})  with $b=2,\sigma=\uparrow$ 
is fulfilled due to particle-hole symmetry that leads to $n_{1,\sigma}^0+n_{2,\sigma}^0=1$.
The two equations~(\ref{eq:normfulfil}) and (\ref{eq:fullfillsecondbisone})
fix $\lambda_6$ and $\lambda_{10}$
as a function of the remaining variational parameters 
$\lambda_1$, $\lambda_2$, $\lambda_4$, $\lambda_9$, $\lambda_{11}$ and
$\lambda_m$. 
The equations for $(b,\downarrow)$ do not provide new information because
we impose spin-flip symmetry.

\section{Calculation of matrix elements}
\label{app:b}

In this appendix we calculate the matrix elements for the orbital occupancies,
the hybridization, and the interaction energy.

\subsection{Orbital occupancies}

For the evaluation of the matrix element~(\ref{eq:occupancieseval}) we first calculate
{\arraycolsep=0pt\begin{eqnarray}
\hat{n}_{1,\uparrow}\hat{P}_{\rm G}^{\vphantom{+}}
&=& \lambda_2 
\hat{n}_{1,\uparrow} \bar{n}_{1,\downarrow}
\bar{n}_{2,\uparrow} \bar{n}_{2,\downarrow}  
+ \lambda_6 
\hat{n}_{1,\uparrow} \bar{n}_{1,\downarrow}
\hat{n}_{2,\uparrow} \bar{n}_{2,\downarrow}  \nonumber \\
&&
\!+ \frac{(\lambda_7+\lambda_9)}{2} \hat{n}_{1,\uparrow} \bar{n}_{1,\downarrow}
\bar{n}_{2,\uparrow} \hat{n}_{2,\downarrow}  \nonumber \\
&&
\!+ \frac{(\lambda_7-\lambda_9)}{2} 
\hat{h}_{1,\uparrow}^+ \hat{h}_{1,\downarrow}^{\vphantom{+}}
\hat{h}_{2,\downarrow}^+ \hat{h}_{2,\uparrow}^{\vphantom{+}}
\nonumber \\
&&
\!+ \frac{(\lambda_{10}+\lambda_{11}+\lambda_m+\lambda_m^*)}{2} 
\hat{n}_{1,\uparrow} \hat{n}_{1,\downarrow}
\bar{n}_{2,\uparrow} \bar{n}_{2,\downarrow}  \nonumber \\
&&
\!+ \frac{\alpha^2(\lambda_{11}-\lambda_{10}+\lambda_m-\lambda_m^*)}{2} 
\hat{h}_{1,\uparrow}^+ \hat{h}_{1,\downarrow}^+
\hat{h}_{2,\downarrow}^{\vphantom{+}} \hat{h}_{2,\uparrow}^{\vphantom{+}}
\nonumber \\
&&
\!+ \lambda_{12} \hat{n}_{1,\uparrow} \bar{n}_{1,\downarrow}
\hat{n}_{2,\uparrow} \hat{n}_{2,\downarrow}
+ \lambda_{14} \hat{n}_{1,\uparrow} \hat{n}_{1,\downarrow}
\hat{n}_{2,\uparrow} \bar{n}_{2,\downarrow}  \nonumber \\
&&
\!+ \lambda_{15} \hat{n}_{1,\uparrow} \hat{n}_{1,\downarrow}
\bar{n}_{2,\uparrow} \hat{n}_{2,\downarrow}
+ \lambda_{16} \hat{n}_{1,\uparrow} \hat{n}_{1,\downarrow}
\hat{n}_{2,\uparrow} \hat{n}_{2,\downarrow}  \; .\nonumber \\
\end{eqnarray}}%
Then, we use $\hat{P}_{\rm G}^+
\hat{n}_{1,\uparrow}\hat{P}_{\rm G}^{\vphantom{+}}
= (\hat{n}_{1,\uparrow}\hat{P}_{\rm G})^+
(\hat{n}_{1,\uparrow}\hat{P}_{\rm G}^{\vphantom{+}})$
to find
{\arraycolsep=0pt\begin{eqnarray}
\hat{P}_{\rm G}^+\hat{n}_{1,\uparrow}\hat{P}_{\rm G}^{\vphantom{+}}
&=& \lambda_2^2 
\hat{n}_{1,\uparrow} \bar{n}_{1,\downarrow}
\bar{n}_{2,\uparrow} \bar{n}_{2,\downarrow}  
+ \lambda_6^2
\hat{n}_{1,\uparrow} \bar{n}_{1,\downarrow}
\hat{n}_{2,\uparrow} \bar{n}_{2,\downarrow}  \nonumber \\
&&
+ \frac{(\lambda_7+\lambda_9)^2}{4} \hat{n}_{1,\uparrow} \bar{n}_{1,\downarrow}
\bar{n}_{2,\uparrow} \hat{n}_{2,\downarrow}  \nonumber \\
&&
+ \frac{(\lambda_7-\lambda_9)^2}{4} 
\bar{n}_{1,\uparrow} \hat{n}_{1,\downarrow}
\hat{n}_{2,\uparrow} \bar{n}_{2,\uparrow}
\nonumber \\
&&
+ \frac{(\lambda_{10}+\lambda_{11}+2{\rm Re}\lambda_m)^2}{4} 
\hat{n}_{1,\uparrow} \hat{n}_{1,\downarrow}
\bar{n}_{2,\uparrow} \bar{n}_{2,\downarrow}  \nonumber \\
&&
+ \frac{|\lambda_{11}-\lambda_{10}+2{\rm i}{\rm Im}\lambda_m|^2}{4} 
\bar{n}_{1,\uparrow} \bar{n}_{1,\downarrow}
\hat{n}_{2,\uparrow} \hat{n}_{2,\downarrow}
\nonumber \\
&&
+ \lambda_{12}^2 \hat{n}_{1,\uparrow} \bar{n}_{1,\downarrow}
\hat{n}_{2,\uparrow} \hat{n}_{2,\downarrow}\nonumber \\
&&
+ \lambda_{14}^2 \hat{n}_{1,\uparrow} \hat{n}_{1,\downarrow}
\hat{n}_{2,\uparrow} \bar{n}_{2,\downarrow}  \nonumber \\
&&
+ \lambda_{15}^2 \hat{n}_{1,\uparrow} \hat{n}_{1,\downarrow}
\bar{n}_{2,\uparrow} \hat{n}_{2,\downarrow}\nonumber \\
&& 
+ \lambda_{16}^2 \hat{n}_{1,\uparrow} \hat{n}_{1,\downarrow}
\hat{n}_{2,\uparrow} \hat{n}_{2,\downarrow}  \;.
\end{eqnarray}}%
For the correlated impurity occupancy we thus find
\begin{eqnarray}
\langle 
\hat{n}_{1,\uparrow}\rangle_{\rm G}
&=& \lambda_1^2 n^0_{1,\uparrow} n^0_{1,\downarrow}
n^0_{2,\uparrow} n^0_{2,\downarrow}  
+ \lambda_2^2 n^0_{1,\uparrow} n^0_{1,\downarrow}
\bar{n}^0_{2,\uparrow} n^0_{2,\downarrow}
\nonumber \\
&& +\lambda_2^2 
n^0_{1,\uparrow} \bar{n}^0_{1,\downarrow}
\bar{n}^0_{2,\uparrow} \bar{n}^0_{2,\downarrow}  
+ \lambda_6^2
n^0_{1,\uparrow} \bar{n}^0_{1,\downarrow}
n^0_{2,\uparrow} \bar{n}^0_{2,\downarrow}  \nonumber \\
&&
+ \lambda_2^2 n^0_{1,\uparrow} n^0_{1,\downarrow}
n^0_{2,\uparrow} \bar{n}^0_{2,\downarrow}  
+ \lambda_4^2 n^0_{1,\uparrow} \bar{n}^0_{1,\downarrow}
n^0_{2,\uparrow} n^0_{2,\downarrow}
\nonumber \\
&&
+ \frac{(\lambda_6+\lambda_9)^2}{4} n^0_{1,\uparrow} \bar{n}^0_{1,\downarrow}
\bar{n}^0_{2,\uparrow} n^0_{2,\downarrow}  \nonumber \\
&&
+ \frac{(\lambda_6-\lambda_9)^2}{4} 
\bar{n}^0_{1,\uparrow} n^0_{1,\downarrow}
n^0_{2,\uparrow} \bar{n}^0_{2,\uparrow}
\nonumber \\
&&
+ \frac{(\lambda_{10}+\lambda_{11}+\lambda_m+\lambda_m^*)^2}{4} 
n^0_{1,\uparrow} n^0_{1,\downarrow}
\bar{n}^0_{2,\uparrow} \bar{n}^0_{2,\downarrow}  \nonumber \\
&&
+ \frac{|\lambda_{11}-\lambda_{10}+\lambda_m-\lambda_m^*|^2}{4} 
\bar{n}^0_{1,\uparrow} \bar{n}^0_{1,\downarrow}
n^0_{2,\uparrow} n^0_{2,\downarrow}
\nonumber  \; .\\
\label{eq:correlateddensityfinal}
\end{eqnarray}

\subsection{Interaction}

Using the atomic spectrum $E_{\Gamma}$ we find from eq.~(\ref{eq:startHinteval})
\begin{eqnarray}
\frac{2 E_{\rm int}}{U} &=&
\lambda_1^2\langle \varphi_0 |\hat{m}_1+\hat{m}_{16}| \varphi_0 \rangle 
+ \lambda_9^2 
\langle \varphi_0 |\hat{m}_9 | \varphi_0 \rangle 
\nonumber \\
&& -  \lambda_{6}^2 
\langle \varphi_0 |\hat{m}_6+ \hat{m}_7 +\hat{m}_8
| \varphi_0 \rangle 
\nonumber \\
&&
-(\lambda_{10}^2 -|\lambda_m|^2 )
\langle \varphi_0 |\hat{m}_{10} | \varphi_0 \rangle  \nonumber \\
&&+(\lambda_{11}^2 -|\lambda_m|^2 )
\langle \varphi_0 |\hat{m}_{11} | \varphi_0 \rangle \nonumber
\\
&&+ \left[\lambda_m^*(\lambda_{11}-\lambda_{10})
\langle \varphi_0 | \bigl[ |11\rangle \langle 10| \bigr] | \varphi_0 \rangle  +\hbox{c.c.} \right]
\; .\nonumber\\
\end{eqnarray}
After evaluation of the expectation values we find
\begin{eqnarray}
\frac{2 E_{\rm int}}{U} &=&
\lambda_1^2(
\bar{n}^0_{1,\uparrow}\bar{n}^0_{1,\downarrow}\bar{n}^0_{2,\uparrow}
\bar{n}^0_{2,\downarrow}
+
n^0_{1,\uparrow}n^0_{1,\downarrow}n^0_{2,\uparrow}n^0_{2,\downarrow}
)
\nonumber \\
&& -  \lambda_{6}^2 
(
n^0_{1,\uparrow}\bar{n}^0_{1,\downarrow}n^0_{2,\uparrow}\bar{n}^0_{2,\downarrow}
+
\bar{n}^0_{1,\uparrow}n^0_{1,\downarrow}\bar{n}^0_{2,\uparrow}n^0_{2,\downarrow}
)
\nonumber \\
&& + \frac{\lambda_9^2-\lambda_6^2}{2}\nonumber \\
&&\hphantom{+\frac{1}{2}}\times
(n^0_{1,\uparrow}\bar{n}^0_{1,\downarrow}\bar{n}^0_{2,\uparrow}n^0_{2,\downarrow}
+ \bar{n}^0_{1,\uparrow}n^0_{1,\downarrow}n^0_{2,\uparrow}\bar{n}^0_{2,\downarrow})
\nonumber \\
&&
+\frac{\lambda_{11}^2 -\lambda_{10}^2}{2}\nonumber \\
&&\hphantom{+\frac{1}{2}}\times
(
n^0_{1,\uparrow}n^0_{1,\downarrow}\bar{n}^0_{2,\uparrow}\bar{n}^0_{2,\downarrow}
+
\bar{n}^0_{1,\uparrow}\bar{n}^0_{1,\downarrow}n^0_{2,\uparrow}n^0_{2,\downarrow}
)
\nonumber \\
&&+ \frac{(\lambda_m^*+\lambda_m)(\lambda_{11}-\lambda_{10})}{2}\nonumber \\
&& \hphantom{+\frac{1}{2}}\times
(
n^0_{1,\uparrow}n^0_{1,\downarrow}\bar{n}^0_{2,\uparrow}\bar{n}^0_{2,\downarrow}
-
\bar{n}^0_{1,\uparrow}\bar{n}^0_{1,\downarrow}n^0_{2,\uparrow}n^0_{2,\downarrow}
)
\; .\nonumber\\
\label{eq:Eintfinal}
\end{eqnarray}

\subsection{Hybridization}

For the evaluation of the matrix element~(\ref{eq:hybelement}) we must express
$| \Gamma\rangle \langle \Gamma | 
\hat{h}_{1,\uparrow}^{\vphantom{+}}
| \Gamma' \rangle \langle \Gamma' |$ 
in second quantization. Due to the action of the annihilation operator,
the number of impurity electrons
in $|\Gamma\rangle$ ($|\Gamma'\rangle$)
is $n_{\Gamma}=0,1,2,3$ ($n_{\Gamma'}=n_{\Gamma}+1$).
The non-vanishing matrix elements are 
\begin{eqnarray}
n=0 &:& | 1\rangle \langle 1 | \hat{h}_{1,\uparrow}^{\vphantom{+}}| 2 \rangle \langle 2 |
= \hat{h}_{1,\uparrow}^{\vphantom{+}} \bar{n}_{1,\downarrow} 
\bar{n}_{2,\uparrow} \bar{n}_{2,\downarrow}  \; ; 
\end{eqnarray}
\begin{eqnarray}
n=1&:&
| 3\rangle \langle 3 | \hat{h}_{1,\uparrow}^{\vphantom{+}}| 10 \rangle \langle 10 |
=\frac{1}{2} \bigl[
\hat{h}_{1,\uparrow}^{\vphantom{+}} n_{1,\downarrow} 
\bar{n}_{2,\uparrow} \bar{n}_{2,\downarrow}  \nonumber \\
&& \hphantom{| 3\rangle \langle 3 | \hat{h}_{1,\uparrow}^{\vphantom{+}}
| 10 \rangle \langle 10 |= \frac{1}{2} \bigl[}
- \hat{h}_{1,\downarrow}^+ \bar{n}_{1,\uparrow} 
\hat{h}_{2,\downarrow}^{\vphantom{+}} \hat{h}_{2,\uparrow}^{\vphantom{+}} \bigr] 
\nonumber \; ,\\
&&
| 3\rangle \langle 3 | \hat{h}_{1,\uparrow}^{\vphantom{+}}| 11 \rangle \langle 11 |
= \frac{1}{2} \bigl[
\hat{h}_{1,\uparrow}^{\vphantom{+}} n_{1,\downarrow} 
\bar{n}_{2,\uparrow} \bar{n}_{2,\downarrow}  \nonumber \\
&& \hphantom{| 3\rangle \langle 3 | \hat{h}_{1,\uparrow}^{\vphantom{+}}
| 10 \rangle \langle 10 |= \frac{1}{2} \bigl[}
+ \hat{h}_{1,\downarrow}^+ \bar{n}_{1,\uparrow} 
\hat{h}_{2,\downarrow}^{\vphantom{+}} \hat{h}_{2,\uparrow}^{\vphantom{+}} \bigr] 
\nonumber \; ,\\
&&
| 3\rangle \langle 3 | \hat{h}_{1,\uparrow}^{\vphantom{+}}| 10 \rangle \langle 11 |
= 
| 3\rangle \langle 3 | \hat{h}_{1,\uparrow}^{\vphantom{+}}| 11 \rangle \langle 11 |
\nonumber \; ,\\
&&
| 3\rangle \langle 3 | \hat{h}_{1,\uparrow}^{\vphantom{+}}| 11 \rangle \langle 10 |
= 
| 3\rangle \langle 3 | \hat{h}_{1,\uparrow}^{\vphantom{+}}| 10 \rangle \langle 10 |
\nonumber \; ,\\
&&
| 4\rangle \langle 4 | \hat{h}_{1,\uparrow}^{\vphantom{+}}|6 \rangle \langle 6 |
= \hat{h}_{1,\uparrow}^{\vphantom{+}} \bar{n}_{1,\downarrow} 
n_{2,\uparrow} \bar{n}_{2,\downarrow}  \nonumber \; , \\
&&
| 5\rangle \langle 5 | \hat{h}_{1,\uparrow}^{\vphantom{+}}|7 \rangle \langle 7 |
= \frac{1}{2} \bigl[
\hat{h}_{1,\uparrow}^{\vphantom{+}} \bar{n}_{1,\downarrow} 
\bar{n}_{2,\uparrow} n_{2,\downarrow}  \nonumber \\
&& \hphantom{| 5\rangle \langle 5 | \hat{h}_{1,\uparrow}^{\vphantom{+}}
|7 \rangle \langle 7 |= \frac{1}{2} \bigl[}
+ \hat{h}_{1,\downarrow}^{\vphantom{+}} \bar{n}_{1,\uparrow} 
\hat{h}_{2,\downarrow}^+ \hat{h}_{2,\uparrow}^{\vphantom{+}} \bigr] 
\nonumber \; ,\\
&&
| 5\rangle \langle 5 | \hat{h}_{1,\uparrow}^{\vphantom{+}}|9 \rangle \langle 9 |
= \frac{1}{2} \bigl[
\hat{h}_{1,\uparrow}^{\vphantom{+}} \bar{n}_{1,\downarrow} 
\bar{n}_{2,\uparrow} n_{2,\downarrow}  \nonumber \\
&& \hphantom{| 5\rangle \langle 5 | \hat{h}_{1,\uparrow}^{\vphantom{+}}
|9 \rangle \langle 9 |= \frac{1}{2} \bigl[}
- \hat{h}_{1,\downarrow}^{\vphantom{+}} \bar{n}_{1,\uparrow} 
\hat{h}_{2,\downarrow}^+ \hat{h}_{2,\uparrow}^{\vphantom{+}} \bigr]  \; ;
\end{eqnarray}
\begin{eqnarray}
n=2 &:&
|10\rangle \langle 10 | \hat{h}_{1,\uparrow}^{\vphantom{+}}|12 \rangle \langle 12 |
= \frac{1}{2} \bigl[
\hat{h}_{1,\uparrow}^{\vphantom{+}} \bar{n}_{1,\downarrow} 
n_{2,\uparrow} n_{2,\downarrow}  \nonumber \\
&& \hphantom{|10\rangle \langle 10 | \hat{h}_{1,\uparrow}^{\vphantom{+}}|12 
\rangle \langle 12 |= }
+ \hat{h}_{1,\downarrow}^+ n_{1,\uparrow} 
\hat{h}_{2,\downarrow}^{\vphantom{+}} \hat{h}_{2,\uparrow}^{\vphantom{+}} \bigr] 
\; , \nonumber \\
&&
|11\rangle \langle 11 | \hat{h}_{1,\uparrow}^{\vphantom{+}}|12 \rangle \langle 12 |
= \frac{1}{2} \bigl[
\hat{h}_{1,\uparrow}^{\vphantom{+}} \bar{n}_{1,\downarrow} 
n_{2,\uparrow} n_{2,\downarrow}  \nonumber \\
&& \hphantom{|11\rangle \langle 11 | \hat{h}_{1,\uparrow}^{\vphantom{+}}|12 
\rangle \langle 12 |= }
- \hat{h}_{1,\downarrow}^+ n_{1,\uparrow} 
\hat{h}_{2,\downarrow}^{\vphantom{+}} \hat{h}_{2,\uparrow}^{\vphantom{+}} \bigr] 
\; , \nonumber \\
&&
|10\rangle \langle 11 | \hat{h}_{1,\uparrow}^{\vphantom{+}}|12 \rangle \langle 12 |
= \frac{1}{2} \bigl[
-\hat{h}_{1,\uparrow}^{\vphantom{+}} \bar{n}_{1,\downarrow} 
n_{2,\uparrow} n_{2,\downarrow}  \nonumber \\
&& \hphantom{|10\rangle \langle 11 | \hat{h}_{1,\uparrow}^{\vphantom{+}}|12 
\rangle \langle 12 |= }
- \hat{h}_{1,\downarrow}^+ n_{1,\uparrow} 
\hat{h}_{2,\downarrow}^{\vphantom{+}} \hat{h}_{2,\uparrow}^{\vphantom{+}} \bigr] 
\; , \nonumber \\
&&
|11\rangle \langle 10 | \hat{h}_{1,\uparrow}^{\vphantom{+}}|12 \rangle \langle 12 |
= \frac{1}{2} \bigl[
-\hat{h}_{1,\uparrow}^{\vphantom{+}} \bar{n}_{1,\downarrow} 
n_{2,\uparrow} n_{2,\downarrow}  \nonumber \\
&& \hphantom{|11\rangle \langle 10 | \hat{h}_{1,\uparrow}^{\vphantom{+}}|12 
\rangle \langle 12 |= }
+ \hat{h}_{1,\downarrow}^+ n_{1,\uparrow} 
\hat{h}_{2,\downarrow}^{\vphantom{+}} \hat{h}_{2,\uparrow}^{\vphantom{+}} \bigr] 
\; , \nonumber\\
&&|7\rangle \langle 7 | \hat{h}_{1,\uparrow}^{\vphantom{+}}|14 \rangle \langle 14 |
= \frac{1}{2} \bigl[
\hat{h}_{1,\uparrow}^{\vphantom{+}} 
n_{1,\downarrow} n_{2,\uparrow} \bar{n}_{2,\downarrow}  \nonumber \\
&& \hphantom{|7\rangle \langle 7 | \hat{h}_{1,\uparrow}^{\vphantom{+}}
|14 \rangle \langle 14 |= \frac{1}{2} \bigl[}
-\hat{h}_{1,\downarrow}^{\vphantom{+}} n_{1,\uparrow} 
\hat{h}_{2,\downarrow}^+ \hat{h}_{2,\uparrow}^{\vphantom{+}} \bigr] 
\; , \nonumber \\
&&
|9\rangle \langle 9 | \hat{h}_{1,\uparrow}^{\vphantom{+}}|14 \rangle \langle 14 |
= \frac{1}{2} \bigl[
\hat{h}_{1,\uparrow}^{\vphantom{+}} n_{1,\downarrow} 
n_{2,\uparrow} \bar{n}_{2,\downarrow}  \nonumber \\
&& \hphantom{|9\rangle \langle 9 | \hat{h}_{1,\uparrow}^{\vphantom{+}}
|14 \rangle \langle 14 |= \frac{1}{2} \bigl[}
+\hat{h}_{1,\downarrow}^{\vphantom{+}} n_{1,\uparrow} 
\hat{h}_{2,\downarrow}^+ \hat{h}_{2,\uparrow}^{\vphantom{+}} \bigr] 
\; , \nonumber \\
&&
|8\rangle \langle 8 | \hat{h}_{1,\uparrow}^{\vphantom{+}}|15 \rangle \langle 15 |
= \hat{h}_{1,\uparrow}^{\vphantom{+}} n_{1,\downarrow} 
\bar{n}_{2,\uparrow} n_{2,\downarrow}  \; ;
\end{eqnarray}
\begin{eqnarray}
n=3 &:&
|13\rangle \langle 13 | \hat{h}_{1,\uparrow}^{\vphantom{+}}| 16 \rangle \langle 16 |
= \hat{h}_{1,\uparrow}^{\vphantom{+}} n_{1,\downarrow} 
n_{2,\uparrow} n_{2,\downarrow} \; .
\end{eqnarray}
When we calculate the expectation value in eq.~(\ref{eq:hybelement}),
we realize that we must contract $\hat{c}_{\veck,\uparrow}^+$ with
$\hat{h}_{1,\uparrow}^{\vphantom{+}}$ because we otherwise
generate at least one vanishing contraction among the impurity operators,
see eqs.~(\ref{eq:secondconstraint}) and~(\ref{eq:thirdconstraint}).
Therefore, eq.~(\ref{eq:hybelement}) reduces to eq.~(\ref{eq:hybelementwithq})
with~$q$ given by
\begin{eqnarray}
q&=& \lambda_1\lambda_2 \bar{n}^0_{1,\downarrow}\bar{n}^0_{2,\uparrow}
\bar{n}^0_{2,\downarrow}
+\lambda_4\lambda_6 \bar{n}^0_{1,\downarrow}n^0_{2,\uparrow}
\bar{n}^0_{2,\downarrow}
\nonumber \\
&& +\frac{\lambda_2(\lambda_{10}+\lambda_{11}+\lambda_m+\lambda_m^*)}{2} 
n^0_{1,\downarrow}\bar{n}^0_{2,\uparrow}\bar{n}^0_{2,\downarrow} \nonumber \\
&& +\frac{\lambda_4(\lambda_6+\lambda_9)}{2} 
\bar{n}^0_{1,\downarrow}\bar{n}^0_{2,\uparrow}n^0_{2,\downarrow}
\nonumber \\
&& +\frac{\lambda_{4}(\lambda_{10}+\lambda_{11}-\lambda_m-\lambda_m^*)}{2} 
\bar{n}^0_{1,\downarrow}n^0_{2,\uparrow}n^0_{2,\downarrow}
\nonumber\\
&&+ \frac{\lambda_{2}(\lambda_6+\lambda_9)}{2} 
n^0_{1,\downarrow}n^0_{2,\uparrow}\bar{n}^0_{2,\downarrow}
\nonumber \\
&& + \lambda_6\lambda_{2} n^0_{1,\downarrow}\bar{n}^0_{2,\uparrow}
n^0_{2,\downarrow}
+ \lambda_{4}\lambda_{1} n^0_{1,\downarrow}n^0_{2,\uparrow}
n^0_{2,\downarrow} \; .
\label{eq:qfactor}
\end{eqnarray}

\section{Results for $\vecR\in A$-lattice}
\label{app:RinA}
In this appendix we collect some results for the case that the two impurities 
lie on the same sublattice.
We restrict ourselves to the case of small hybridizations, $V\ll W$,
and use approximate analytic formulae for the single-particle energy contribution.

\subsection{Single-particle quantities}

To leading order in $(qV)^2\ln[(qV)^2]$ and $(qV)^2$
we have from~MBG
\begin{eqnarray}
E_{\rm sp}(q, x) &=& 
4(qV)^2d_{\veczero}\Bigl( 
\ln\Bigl[(qV)^22\pi d_{\veczero}/(C x)\Bigr]\nonumber\\
&& \hphantom{4(qV)^2d_{\veczero}\Bigl( }
+\frac{x^2-4}{4x} G(x) \Bigr) \; ,
\label{eqapp:Espofx}
\end{eqnarray}
where $G(0<x<2)\equiv G_1(x)$ and
$G(x\geq 2)\equiv G_2(x)$ with
\begin{eqnarray}
G_1(x)&=&\frac{\pi}{\sqrt{4-x^2}} - \frac{2}{\sqrt{4-x^2}} 
\arctan\left(\frac{x^2-2}{x\sqrt{4-x^2}}\right) \;, \nonumber \\
G_2(x)&=& 
-\frac{1}{\sqrt{x^2-4}}\ln \left( 
\frac{x^2-2-x\sqrt{x^2-4}}{
x^2-2+x\sqrt{x^2-4}}
\right)\; ,
\end{eqnarray}
and
\begin{equation}
 x\equiv x_{\vecR}(\bar{t})
= \frac{2\pi d_{\veczero}}{\sqrt{\bar{t}^2+\pi^2(d_{\veczero}^2-d_{\vecR}^2)}}
\end{equation}
or 
\begin{equation}
\bar{t}(x) = \pm \frac{\pi\sqrt{d_{\veczero}^2(4-x^2)+d_{\vecR}^2x^2}}{x} \; .
\end{equation}
For a real $\bar{t}$ we must restrict $x$ to the region
$0\leq x\leq 2d_{\veczero}/\sqrt{d_{\veczero}^2-d_{\vecR}^2}$.

The minimization of $\bar{E}_{\rm var}(\veclambda,\bar{t},n_{1,\uparrow}^0)$
with respect to $\bar{t}$ links $n_{1,\uparrow}^0$ to~$x$,
\begin{equation}
n_{1,\uparrow}^0(x)-\frac{1}{2}
=\pm \frac{\sqrt{(4-x^2)d_{\veczero}^2+x^2d_{\vecR}^2}G(x)}{4\pi d_{\veczero}}\; ,
\label{eqapp:xofn}
\end{equation}
see~MBG. We may use $x$ instead of $n_{1,\uparrow}^0$ as our variational parameter.
Eq.~(\ref{eqapp:xofn}) shows that for every solution 
of the minimization equations with 
$0\leq x< 2d_{\veczero}/\sqrt{d_{\veczero}^2-d_{\vecR}^2}$
we obtain two equivalent solutions for the density, $n_{1,\uparrow,a}^0>1/2$
and $n_{1,\uparrow,b}^0=1-n_{1,\uparrow,a}^0<1/2$.
In the following we shall investigate solutions with $n_{1,\uparrow,a}^0\geq 1/2$.

The variational energy functional to be minimized is given by
\begin{eqnarray}
\bar{E}_{\rm var}(\veclambda,x) 
&=&  E_{\rm int}(\veclambda,n_{1,\uparrow}^0(x))
+E_{\rm sp}(q,x)  \nonumber \\
&& -2(qV)^2\bar{t}(x)[1-2n_{1,\uparrow}^0(x)] 
\nonumber \\
&=& E_{\rm int}(\veclambda,n_{1,\uparrow}^0(x)) \nonumber \\
&& + 4 (qV)^2 d_{\veczero}\biggl[ 
\ln \frac{2\pi d_{\veczero}(qV)^2}{Cx} 
\nonumber \\
&& \hphantom{+ 4 (qV)^2 d_{\veczero}\biggl[ }
+ \frac{x G(x)}{4} \left( \frac{d_{\vecR}}{d_{\veczero}}\right)^2\biggr]
\label{eq:Ebardefasofn-RinA}
\end{eqnarray}
as a function of the Gutzwiller variational parameters~$\veclambda$
and of the parameter~$x$. As in the main text, we choose $y_m=0$ in the following.

\subsection{Analytical expressions in the Kondo limit for small hybridizations}

When we repeat the steps in Sect.~\ref{sec:heisenberglimit} we find
\begin{equation}
\bar{E}_{\rm var}^{\rm K}(x)
= -\frac{2Cx}{\pi e} \exp\Bigl(-\frac{\pi p[n(x)] U}{16 \Gamma}
-\frac{xG(x)}{4}\frac{d_{\vecR}^2}{d_{\veczero}^2}\Bigr) 
\label{eq:EK-RKKYofnRinA}
\end{equation}
for the variational energy in the Kondo limit with $\Gamma=\pi d_{\veczero}V^2$.
When $x=2d_{\veczero}/\sqrt{d_{\veczero}^2-d_{\vecR}^2})$ ($n=1/2$)
is the minimum of $\bar{E}_{\rm var}^{\rm K}(x)$, the variational state
describes two weakly interacting Kondo-screened impurity spins,
\begin{equation}
E_{\rm var}^{\rm K}=2E_{\rm opt}^{\rm SIAM}(1-\varepsilon_{\rm RKKY}^A(\vecR)) \; ,
\end{equation}
compare eq.~(\ref{eq:Kondo-RKKYsinglet}), with the RKKY energy reduction 
\begin{equation}
\varepsilon_{\rm RKKY}^A(\vecR)=\frac{1}{2} \frac{d_{\vecR}^2}{d_{\veczero}^2}
\ll 1\; ,
\label{eq:epsfactorA}
\end{equation}
compare eq.~(\ref{eq:epsfactorB}),
where the upper index `$A$' indicates that the impurities are on the same sublattice.

\begin{figure}[b]
\includegraphics[width=\columnwidth]{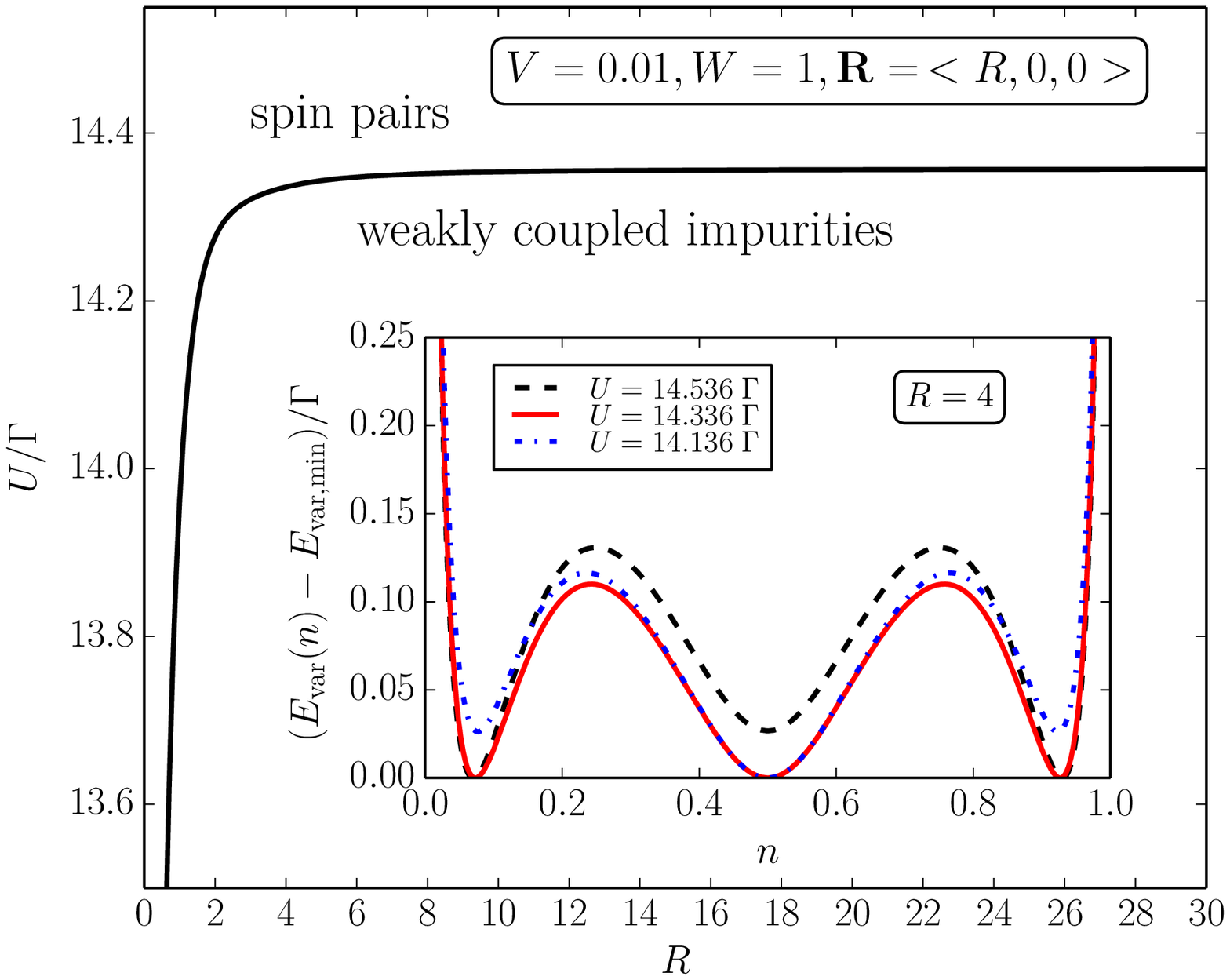}
\caption{Ground-state phase diagram for $\vecR=<\! R,0,0\!>$
and even~$R$ for $V=0.01$.
Weakly interacting impurities are found below the critical curve
$U_{\rm c}(\vecR)$, the phase with spin pairs is found above.
The line gives the values for continuous~$R$ 
in the evaluation of $d_{\vecR}$ using the small-$V$ 
expression~(\protect\ref{eq:Ebardefasofn}).
The phase transition is discontinuous for all even~$R$.
Inset:
Variational ground-state energy $E_{\rm var}(n)$ 
as a function of the density $n=n_{1,\uparrow}^0$
in the effective non-interacting two-impurity model~(\protect\ref{eq:defHeffective})
at $\vecR=<\!4,0,0\!>$
for $V=0.01$, $W=1$ and three values of~$U$ at and in the vicinity
of the critical value $U_{\rm c}(\vecR)\approx 14.336\Gamma$
using the small-$V$ expression~(\protect\ref{eq:Ebardefasofn}).\label{fig:qptRinA}}
\end{figure}

For large~$U$, the two impurity spins are coupled into
a singlet. With $n(x)\approx 1-x/(2\pi)$ and $p[n(x)]\approx 8x/\pi$
and neglecting high-order corrections in $1/U$ we find
$n_{\rm opt}=1-\Gamma/(\pi U)$ as in eq.~(\ref{eq:noptTIAMnonint-final})
and 
\begin{equation}
\bar{E}_{\rm var}^{\rm K, opt}
=-  \frac{4C}{e^2} \frac{\Gamma}{\pi U}+{\cal O}(1/U^2) \; ,
\label{eq:EvarfinalRinA}
\end{equation}
as in eq.~(\ref{eq:Evarfinal}). Since there is no level splitting 
for $V=0$, $\vecR$-dependent corrections to first order in $1/U$
are absent for $\vecR\in\hbox{$A$-lattice}$.

\subsection{Ground-state phase diagram}

Fig.~\ref{fig:qptRinA} shows the ground-state phase diagram.
The critical line does not terminate at a tricritical point
because a level splitting is absent at the RKKY level.
For $\vecR\in\hbox{$A$-lattice}$,
the quantum phase transition is discontinuous for all~$R$ 
because the $h$-orbital symmetry is not broken for $0\leq U\leq U_{\rm c}(\vecR)$,
and there is no direct electron transfer 
between the impurities, $\tau(U<U_{\rm c}(\vecR))=0$.
This is seen in the inset of Fig.~\ref{fig:qptRinA} where we show
the variational energy close to the transition for $\vecR=<\!4,0,0\!>$.

Since the RKKY interaction does not
split the $h$-orbital energies, we always find an extremum at $n=1/2$.
Correspondingly, we have $p_s=p_t/3$ for the singlet and triplet
occupation probabilities below the transition. 
Above the transition, $n\neq 1/2$ holds for the optimal variational energy
and $\tau$ jumps to a finite value. 
Likewise, $p_s$ and $p_t$ are discontinuous.
Above the transition, $\tau(U>U_{\rm c}(\vecR))$ 
decays proportional to $1/U^2$ for large $U/\Gamma$.

Apart from the behavior below the transition and apart from the case
of neighboring impurities, the differences between
odd an even impurity separations are small.

\section{Gutzwiller approach to the 
single-impurity Anderson model (SIAM)}
\label{app:d}

For comparison and future reference,  in this appendix
we collect the results for the symmetric SIAM~(\ref{eq:defSIAM}).
The results 
were derived earlier from Gutzwiller vari\-ation\-al wave functions, see, e.g.,
Ref.~\cite{GebhardPRB1991}, and using
Kotliar-Ruckenstein slave bosons, see, e.g., Ref.~\cite{Schoenhammer1990}.

\subsection{Gutzwiller variational ground state}

As our variational ground state we use the Gutzwiller Ansatz
\begin{equation}
| \Psi_{\rm G}\rangle = \hat{P}_{\rm G} | \varphi_0\rangle \quad , \quad
\hat{P}_{\rm G}=
\sum_I\lambda_I\hat{m}_{\Gamma}\; .
\label{eq:Gutzdef}
\end{equation}
Here, 
\begin{equation}
| \varphi_0\rangle = \prod_{k,\sigma}{}^{{}^\prime} \hat{a}_{k,\sigma}^+|{\rm vac}\rangle
\end{equation}
is the ground state of some effective single-particle Hamiltonian,
\begin{equation}
\hat{H}_0^{\rm eff} | \varphi_0\rangle = 
E_{\rm sp}| \varphi_0\rangle \; .
\label{eq:H0effintro}
\end{equation}
The effective single-particle Hamiltonian can be cast into the form
\begin{equation}
\hat{H}_0^{\rm eff}=\hat{T}+q\hat{V}\; . 
\label{eq:H0effdef}
\end{equation}
In equation~(\ref{eq:Gutzdef}) we employ the projection operators onto the four possible
impurity configurations, $I\in\left\{ \emptyset,
\uparrow,\downarrow,d\right\}$,
\begin{eqnarray}
\hat{m}_{\emptyset} =
\Bigl(1-\hat{d}_{\uparrow}^+\hat{d}_{\uparrow}^{\vphantom{+}}\Bigr) 
\Bigl(1-\hat{d}_{\downarrow}^+\hat{d}_{\downarrow}^{\vphantom{+}}\Bigr) 
& ,& \hat{m}_{d} = 
\hat{d}_{\uparrow}^+\hat{d}_{\uparrow}^{\vphantom{+}}
\hat{d}_{\downarrow}^+\hat{d}_{\downarrow}^{\vphantom{+}}
 \; ,
\nonumber \\
\hat{m}_{\uparrow} = 
\hat{d}_{\uparrow}^+\hat{d}_{\uparrow}^{\vphantom{+}}
\Bigl(1-\hat{d}_{\downarrow}^+\hat{d}_{\downarrow}^{\vphantom{+}}\Bigr) 
 & ,& 
\hat{m}_{\downarrow} = 
\hat{d}_{\downarrow}^+\hat{d}_{\downarrow}^{\vphantom{+}}
\Bigl(1-\hat{d}_{\uparrow}^+\hat{d}_{\uparrow}^{\vphantom{+}}\Bigr) , \nonumber \\
\label{eq:mdefs}
\end{eqnarray}
and $\lambda_I$ are real-valued variational parameters.
We demand that
\begin{equation}
\hat{P}_{\rm G}^2 = 1+ x
\left(\hat{d}_{\uparrow}^+\hat{d}_{\uparrow}^{\vphantom{+}}-1/2\right) 
\left(\hat{d}_{\downarrow}^+\hat{d}_{\downarrow}^{\vphantom{+}}-1/2\right)  
\label{eq:defPGsquered}
\end{equation}
for the paramagnetic half-filled system.
This leads to the conditions
\begin{equation}
\lambda_{\emptyset}=\lambda_d \quad ,\quad
\lambda_{\sigma}=\sqrt{2-\lambda_d^2}\quad ,\quad
x=4(\lambda_d^2-1) \; ,
\label{eq:onlylambdad}
\end{equation}
so that $\lambda_d$ is the only remaining variational parameter.

Our choice for the variational parameters $\lambda_{\emptyset}$ and
$\lambda_{\sigma}$ guarantees that the Gutzwiller variational state is normalized,
\begin{equation}
\langle \Psi_{\rm G} | \Psi_{\rm G}\rangle = 
\langle \varphi_0 |\hat{P}_{\rm G}^2| \varphi_0\rangle = 1
\end{equation}
because 
\begin{equation}
\langle \varphi_0 |\hat{d}_{\sigma}^+\hat{d}_{\sigma}^{\vphantom{+}}| \varphi_0\rangle 
= 1/2
\label{eq:ddisonehalf}
\end{equation}
at particle-hole symmetry.
Likewise,
\begin{eqnarray}
\langle \Psi_{\rm G} |
\hat{d}_{\sigma}^+\hat{d}_{\sigma}^{\vphantom{+}}
| \Psi_{\rm G}\rangle &=& 
\langle \varphi_0| 
\hat{d}_{\sigma}^+\hat{d}_{\sigma}^{\vphantom{+}}
| \varphi_0\rangle 
=1/2\label{eq:halfremainshalf}
\end{eqnarray}
so that the Gutzwiller variational ground state~(\ref{eq:Gutzdef}) 
respects particle-hole symmetry.

\subsection{Calculation of the variational energy}


For the operator of the kinetic energy we find
\begin{eqnarray}
\langle \Psi_{\rm G} |\hat{T} | \Psi_{\rm G}\rangle &=& 
\sum_{\veck,\sigma} \epsilon(\veck) 
\langle \varphi_0 | \hat{c}_{\veck,\sigma}^+\hat{c}_{\veck,\sigma}^{\vphantom{+}}
\hat{P}_{\rm G}^2| \varphi_0\rangle \nonumber \\
&=& 
\sum_{\veck,\sigma} \epsilon(\veck) 
\langle \varphi_0 | \hat{c}_{\veck,\sigma}^+\hat{c}_{\veck,\sigma}^{\vphantom{+}}
| \varphi_0\rangle \; ,
\end{eqnarray}
where we used eqs.~(\ref{eq:defPGsquered}) and~(\ref{eq:ddisonehalf}).


For the hybridization operator we find
\begin{eqnarray}
\langle \Psi_{\rm G} |\hat{V} | \Psi_{\rm G}\rangle &= &
\frac{1}{\sqrt{L}}
\sum_{\veck,\sigma}
V_{\veck} \langle \varphi_0 | \hat{P}_{\rm G} \hat{d}_{\sigma}^+\hat{P}_{\rm G}
\hat{c}_{\veck,\sigma}^{\vphantom{+}}| \varphi_0\rangle
+ {\rm c.c.}\nonumber \\
&=& 
q \frac{1}{\sqrt{L}}
\sum_{\veck,\sigma} \left(V_{\veck} 
\langle \varphi_0 | \hat{d}_{\sigma}^+\hat{c}_{\veck,\sigma}^{\vphantom{+}}
| \varphi_0\rangle + {\rm c.c.}\right)
\; ,\nonumber \\
\label{eq:sqrtqV}
\end{eqnarray}
where we used eqs.~(\ref{eq:Gutzdef}), (\ref{eq:mdefs}), and~(\ref{eq:onlylambdad})
to arrive at
\begin{eqnarray}
 \hat{P}_{\rm G} \hat{d}_{\sigma}^+\hat{P}_{\rm G}
&=& \left[\lambda_{\sigma}(1-\hat{d}_{-\sigma}^+\hat{d}_{-\sigma}^{\vphantom{+}})
+\lambda_d\hat{d}_{-\sigma}^+\hat{d}_{-\sigma}^{\vphantom{+}})\right]
\hat{d}_{\sigma}^+ \nonumber \\
&& \times
\left[\lambda_{\emptyset}(1-\hat{d}_{-\sigma}^+\hat{d}_{-\sigma}^{\vphantom{+}})
+\lambda_{-\sigma}\hat{d}_{-\sigma}^+\hat{d}_{-\sigma}^{\vphantom{+}})\right]
\nonumber \\
&=& \lambda_{\sigma}\lambda_d \hat{d}_{\sigma}^+ \; .
\end{eqnarray}
Moreover, we introduced the abbreviation
\begin{equation}
q^2=\lambda_d^2\left(2-\lambda_d^2\right) \quad , \quad
\lambda_d^2=1-\sqrt{1-q^2}\; .
\label{eq:introduceq}
\end{equation}
Equation~(\ref{eq:sqrtqV}) shows that 
the Gutzwiller correlator rescales the hybridization by the factor~$q$.


For the operator of the interaction energy we find from 
eqs.~(\ref{eq:defHint}) and~(\ref{eq:halfremainshalf})
\begin{eqnarray}
\langle \Psi_{\rm G} |\hat{H}_{\rm int} | \Psi_{\rm G}\rangle &= &
-\frac{U}{4} + U
\langle \varphi_0 | \hat{m}_d \hat{P}_{\rm G}^2| \varphi_0\rangle\nonumber \\
&=& -\frac{U}{4} + \frac{U}{4}\lambda_d^2 \; ,
\end{eqnarray}
where we also used eq.~(\ref{eq:Gutzdef}).

\subsection{Minimization of the variational energy}
\label{app:C3}

Using the results of the previous subsection, we can express
the variational energy as a function of the single variational parameter~$q$.

\subsubsection{Optimization of the single-particle state}

The variational energy can be cast into the form
\begin{eqnarray}
E_{\rm var}\left(q,|\varphi_0\rangle\right)
&=& \frac{\langle \Psi_{\rm G} | \hat{H}|\Psi_{\rm G}\rangle
}{
\langle \Psi_{\rm G} |\Psi_{\rm G}\rangle}\nonumber \\
&=& \langle \varphi_0 |\hat{T} +q \hat{V} |\varphi_0\rangle
+ \frac{U(1-\sqrt{1-q^2})}{4}
\; ,\nonumber \\
\end{eqnarray}
where we dropped the constant $-U/4$.

The optimization of the variational energy with respect to
the single-particle state~$| \varphi_0\rangle$ returns the
single-particle Schr\"odinger equation~(\ref{eq:H0effintro})
with $H_0^{\rm eff}$ from eq.~(\ref{eq:H0effdef}).

\begin{figure}[hb]
\includegraphics[width=\columnwidth]{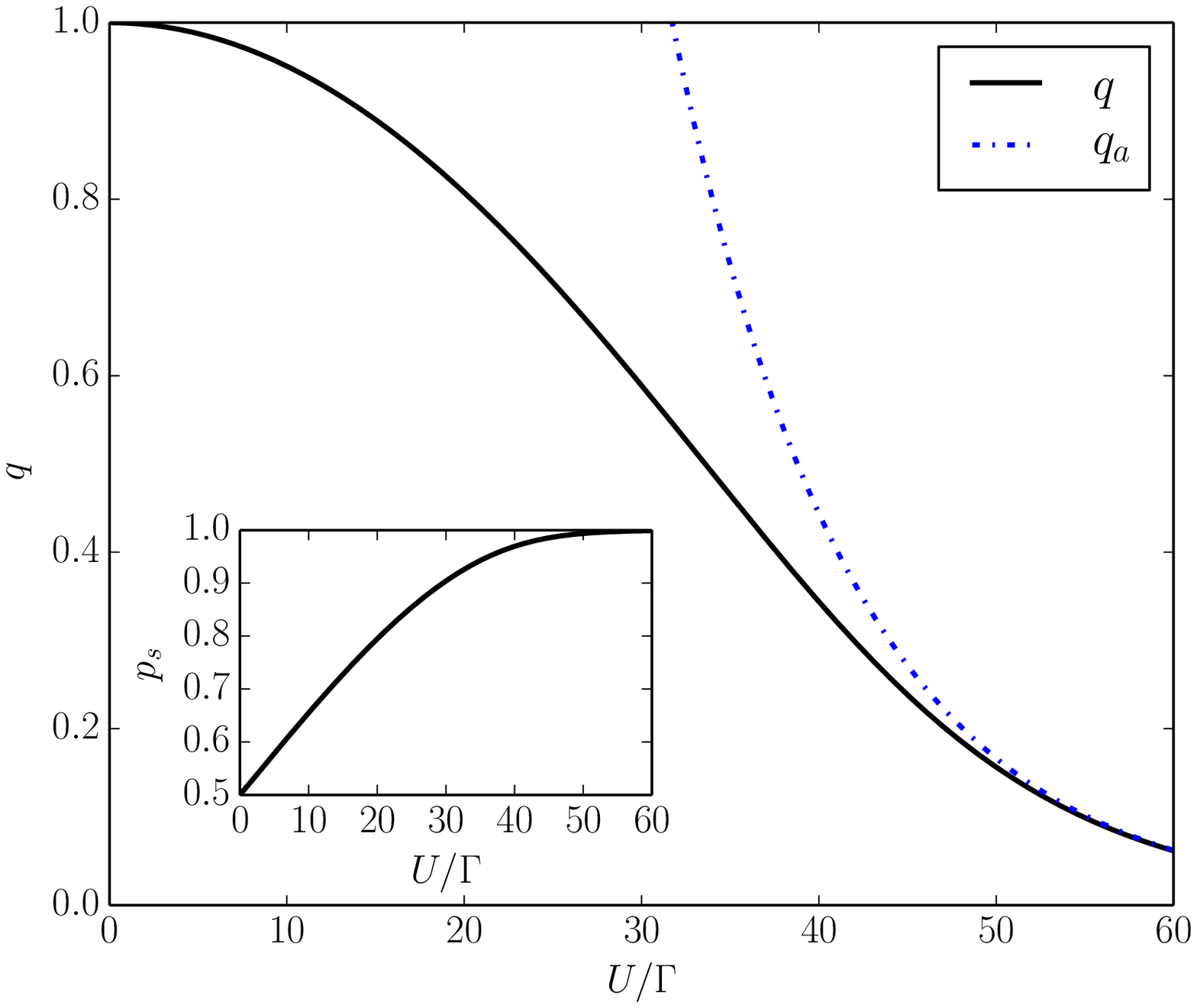}
\caption{Hybridization reduction factor~$q$ for the 
symmetric single-impurity Anderson model as a function of~$U/\Gamma$
($\Gamma=\pi d_{\veczero}V^2$, $V=0.01$, $d_{\veczero}=1.712$), 
in comparison with the analytical
strong-coupling expansion, eq.~(\ref{eq:qa}).
The inset shows the probability for a single occupancy as a function 
of~$U/\Gamma$.\label{fig:SIAM}}
\end{figure}

\subsubsection{Gutzwiller variational ground-state energy in the strong-coupling
limit}

For the optimal single-particle state~$|\varphi_0\rangle$,
the variational ground-state energy in presence of the impurity can be written as
\begin{equation}
E_{\rm var}(q)=E_0(qV) + \frac{U(1-\sqrt{1-q^2})}{4}
\; ,
\end{equation}
where $E_0(V)$ is the energy for the non-interacting SIAM.
For a small, constant hybridization $V_{\veck}\equiv V\ll 1$ we have
\begin{eqnarray}
E_{\rm var}(q)&=&
2d_{\veczero}(qV)^2\ln\left(\frac{\pi d_{\veczero}(qV)^2}{C}\right)\nonumber \\
&& +\frac{U(1-\sqrt{1-q^2})}{4} \; ,
\label{eqapp:EzerosmallV}
\end{eqnarray}
where $d_{\veczero}$ is the density of states per spin direction at the Fermi energy 
and $C$ is a constant that depends on the form of the density of states.
The minimization of $E_{\rm var}(q)$ with respect to~$q$
leads to an implicit equation for $q(U)$.
Its solution for $V=0.01$ is shown in Fig.~\ref{fig:SIAM}.

For large interactions, $U\gg \Gamma=\pi d_{\veczero}V^2$, 
we have $q\to 0$ and the minimization equation 
with respect to $q^2$ becomes
\begin{equation}
0 = \ln(q^2) +1 + \ln(\Gamma/C) + \frac{\pi U}{16\Gamma}
\end{equation}
with the solution ($\ln(e)=1$)
\begin{equation}
q_a^2=
\frac{C}{e \Gamma}\exp\left(-\frac{\pi U}{16\Gamma}\right)  \; .
\label{eq:qa}
\end{equation}
With $J_{\rm K}=4V^2/U$ 
the optimized variational ground-state energy in the Kondo limit reads
\begin{equation}
E_{\rm opt}^{\rm SIAM}(J_{\rm K}\to 0)= -\frac{2C}{\pi e}
\exp\left(-\frac{1}{4d_{\veczero} J_{\rm K}}\right) \; .
\label{eq:GutzenergySIAM}
\end{equation}
The Gutzwiller variational energy reproduces the ex\-ponentially small
binding energy but lacks a factor of two in the exponent, i.e., 
the exact Kondo temperature 
obeys $T_{\rm K}\sim \exp[-1/(2 J_{\rm K}d_{\veczero})]$~\cite{Hewson}.

\subsubsection{Kondo limit for the two-impurity Anderson model 
for a half-filled effective single-particle Hamiltonian} 

When we restrict ourselves to the case $n_{b,\sigma}=n=\bar{n}=1/2$
for the half-filled effective single-particle Hamiltonian,
the analysis in Sect.~\ref{sec:heisenberglimit}
carries over to the Kondo limit because $p(1/2)=1$, and all 
density-dependent asymmetric terms vanish. Therefore, in the Kondo limit
this variational state
describes two isolated impurities with energy
\begin{eqnarray}
E_{\rm opt}^{\rm TIAM}(J_{\rm K}\to 0)&=& 
2 E_{\rm opt}^{\rm SIAM}(J_{\rm K}\to 0)\nonumber \\
&=& -\frac{4C}{\pi e}
\exp\left(-\frac{1}{4d_{\veczero}J_{\rm K}}\right) \; .
\end{eqnarray}

\section{Cut-off energies for small hybridizations}
\label{app:cutoff}
For completeness, we derive an expression for the constant~$C$ 
in eq.~(\ref{eqapp:EzerosmallV}) for all density of states.

\subsection{Single-impurity Anderson model}

For a general density of states and all~$V\ll 1$ 
in the non-interacting single-impurity Anderson model,
the ground-state energy correction due to the hybridization of the
impurity with the host electrons is given by~\cite{Annalenpaper}
\begin{eqnarray}
E_{\rm SIAM}&=&\frac{2}{\pi} \int_{-1/2}^0 \rmd \epsilon \cot^{-1}\left[
\frac{\epsilon- v \Lambda_{\veczero}(\epsilon)}{\pi v D_{\veczero}(\epsilon)}\right] 
\nonumber \\
&=& 2v d_{\veczero}\ln\left[\frac{\pi v d_{\veczero} }{C}\right] +{\cal O}\left(v^2\ln v\right) \; ,
\label{eqapp:findC-1}
\end{eqnarray}
where $v=V^2$, $D_{\veczero}(\epsilon)$ is the density of states,
and $\Lambda_{\veczero}(\epsilon)$ is its Hilbert transform.
In the second step, we used the approximation for small~$V$ employed in 
eq.~(\ref{eqapp:EzerosmallV}).
Therefore, for $v\to 0$ we find by differentiating both sides 
of eq.~(\ref{eqapp:findC-1}) with respect to~$v$ that $L(v)= R(v)$ with
\begin{eqnarray}
L(v)&=& \int_{-1/2}^0 \rmd \epsilon 
\frac{D_{\veczero}(\epsilon)\epsilon}{(\epsilon-v\Lambda_{\veczero}(\epsilon))^2+
(\pi v D_{\veczero}(\epsilon))^2}  \; , \nonumber \\
R(v) &=& d_{\veczero}\left(\ln(v)+1+\ln(\pi d_{\veczero}/C)\right)
\label{eqapp:findC-2}
\end{eqnarray}
to order $\ln(v)$ and order unity. We add and subtract an integral that can be
evaluated analytically and write
\begin{eqnarray}
L(v)&=& L(v) - \int_{-1/2}^0 \rmd \epsilon 
\frac{d_{\veczero} \epsilon}{\epsilon^2+
(\pi v d_{\veczero})^2}  \nonumber \\
&&+d_{\veczero}\left(\ln(v) + \ln(2) + \ln(\pi d_{\veczero})\right) \; .
\end{eqnarray}
We thus find from $L(v)=R(v)$ in the limit $v\to 0$ that
\begin{eqnarray}
\ln[e/(2C)] &=&\int_{-1/2}^0 \rmd \epsilon 
\biggl[  
\frac{D_{\veczero}(\epsilon)\epsilon/d_{\veczero}}{(\epsilon-v\Lambda_0(\epsilon))^2+
(\pi v D_{\veczero}(\epsilon))^2}  \nonumber \\
&& \hphantom{\int_{-1/2}^0 \rmd \epsilon \biggl[  }
-\frac{\epsilon}{\epsilon^2+(\pi v d_{\veczero})^2} 
\biggr] \; .
\end{eqnarray}
Letting $v\to 0$ in this expression gives
\begin{equation}
C=\frac{e}{2} \exp\left[
-\int_{-1/2}^0\rmd \epsilon \frac{D_{\veczero}(\epsilon)-d_{\veczero}}{d_{\veczero}\epsilon}
\right] \; .
\label{eq:DosgamesforC}
\end{equation}
For a constant density of states 
with $D_{\rm cons}(\epsilon)=1$ for $|\epsilon|\leq 1/2$
we thus obtain $C^{\rm cons}=e/2\approx 1.36$, and
for a semi-elliptic density of states 
with $D_{\rm se}(\epsilon)=(4/\pi)\sqrt{1-4\epsilon^2}$ ($|\epsilon|\leq1/2$), 
we get $C^{\rm se}=1$~\cite{Annalenpaper}.
For the simple-cubic lattice with electron dispersion~(\ref{eq:scdispersion})
we may rewrite the integral in eq.~(\ref{eq:DosgamesforC})
to find
\begin{eqnarray}
\ln C^{\rm sc}&=&1-\ln(2) \label{eq:DosgamesforCsc}
\\
&& -\frac{6}{\pi d_{\veczero}^{\rm sc}}
\int_0^{\infty}\rmd x [J_0(x)]^3[\gamma+\ln(3x)-\Ci (3x)]\; ,\nonumber 
\end{eqnarray}
where $J_0(x)$ is the Bessel function to order $n=0$,
$\gamma$ is Euler's constant, $\Ci(x)$ is the cosine integral, and
$d_{\veczero}^{\rm sc}=1.712$. 
The integral is readily evaluated numerically~\cite{Mathematica}
to $C^{\rm sc}\approx 0.7420$.

\subsection{Two-impurity Anderson model}

The ground-state energy can be calculated using the density of states,
\begin{eqnarray}
E_{\rm TIAM}(V)&=& E_{\rm TIAM}^{\rm d}(V)
+ E_{\rm TIAM}^{\rm host}(V) \nonumber \; ,\\
E_{\rm TIAM}^{\rm d}(V)
&=& 2 \sum_b \int_{-1/2}^{0} \rmd \omega \omega D_b(\omega) \; ,\\
E_{\rm TIAM}^{\rm host}(V)
&=& 2 \int_{-1/2}^{0} \sum_b \rmd \omega \omega D_{\rm host,b}(\omega) \; ,
\label{eq:hostenergy}
\end{eqnarray}
where we suppress the $\vecR$-dependence for convenience.

\subsubsection{Impurity contribution}

Using $d_{\veczero}=D_{\veczero}(0)$ and the abbreviations
$\alpha\equiv V^2(\bar{t}-\pi \tilde{s}_{\vecR}d_{\vecR})$,
$\beta\equiv \pi V^2d_{\veczero}$, and $\tilde{t}_{12}=V^2\bar{t}$
we eliminate the logarithmically divergent terms in the integrand, 
\begin{eqnarray}
\frac{E_{\rm TIAM}^{\rm d}(V)}{2V^2 d_{\veczero}}
 &=& 
\int_{-1/2}^0 \rmd \omega \omega 
\biggl(\frac{D_1(\omega)+D_2(\omega)}{V^2d_{\veczero}}
\nonumber \\
&& 
-\frac{1}{(\omega+\alpha)^2+\beta^2}-\frac{1}{(\omega-\alpha)^2+\beta^2}
\biggr)\nonumber \\
&& + \ln\left(\alpha^2+\beta^2\right)
-\frac{1}{2}\ln[(\alpha+1/2)^2+\beta^2] \nonumber \\
&& -2 \frac{\alpha}{\beta} \arctan\left(\frac{\alpha}{\beta}\right)
\nonumber \\
&& +\frac{\alpha}{\beta} \arctan\left(\frac{\alpha\pm1/2}{\beta}\right) \; .
\end{eqnarray}
Now, we are in the position to let $V\to 0$ both in the integrand as well 
as in all other terms ($\alpha\to 0$, $\beta\to 0$, and $\alpha/\beta$ remains finite),
\begin{eqnarray}
\frac{\Delta E_{\rm TIAM}^{\rm imp}(V)}{2V^2 d_{\veczero}}
 &\approx& 2 
\int_{-1/2}^0 \rmd \omega \frac{D_{\veczero}(\omega)-d_{\veczero}}{d_{\veczero}\omega} 
+ 2\ln(V^2)
\nonumber \\[6pt]
&& 
\!+\ln (4) 
+\ln\left[ (\bar{t}-\pi \tilde{s}_{\vecR}d_{\vecR})^2+(\pi d_{\veczero})^2
\right]
\nonumber \\[6pt]
&& \!-2 \Bigl(\frac{\bar{t}-\pi \tilde{s}_{\vecR}d_{\vecR}}{\pi d_{\veczero}}\Bigr)
\tan^{-1}\Bigl( \frac{\bar{t}-\pi \tilde{s}_{\vecR}d_{\vecR}}{\pi d_{\veczero}}\Bigr).
\nonumber \\
\end{eqnarray}

\subsubsection{Host contribution}

For small~$V$ we have 
\begin{eqnarray}
D_{\rm host}(\omega)&\approx&
\frac{V^2}{\pi} 
{\rm Im} \left[ 
\frac{R_1'(\omega;\vecR)-\rmi \pi I_1'(\omega;\vecR))}{\omega-\alpha+\rmi \beta}
\right]\nonumber \\
&& + 
\frac{V^2}{\pi} {\rm Im} \left[ 
\frac{R_2'(\omega;\vecR)
-\rmi \pi I_2'(\omega;\vecR)}{\omega+\alpha+\rmi \beta}
\right] \; .
\end{eqnarray}
In the energy integral~(\ref{eq:hostenergy})
 we may safely let $V\to 0$ to obtain the second-order term,
\begin{eqnarray}
E_{\rm TIAM}^{\rm host}(V) &\approx &
-4 V^2 \int_{-1/2}^{0} \rmd \omega \omega 
\frac{D'_{\veczero}(\omega)}{\omega}= -4V^2d_{\veczero} \; ,\nonumber \\
\end{eqnarray}
where we used $D_{\veczero}(0)=d_{\veczero}$ and $D_{\veczero}(-1/2)=0$.

\subsubsection{Calculation of $C$}

Altogether, we find up to and including all terms to order~$V^2$
\begin{eqnarray}
\frac{E_{\rm TIAM}(V)}{4V^2d_{\veczero}}  &\approx& \ln\Bigl(\frac{V^2}{C}\Bigr) 
+ \!\frac{1}{2}
\ln\bigl[ (\bar{t}-\pi \tilde{s}_{\vecR}d_{\vecR})^2+(\pi d_{\veczero})^2\bigr]
\nonumber \\[3pt]
&& -
\Bigl(\frac{\bar{t}-\pi \tilde{s}_{\vecR}d_{\vecR}}{\pi d_{\veczero}}\Bigr)
\tan^{-1}\Bigl( \frac{\bar{t}-\pi \tilde{s}_{\vecR}d_{\vecR}}{\pi d_{\veczero}}\Bigr)
\nonumber \\
\end{eqnarray}
with
\begin{equation}
\ln\Bigl(\frac{1}{C}\Bigr)= \ln(2)-1 
+ \int_{-1/2}^0 \rmd \omega \frac{D_{\veczero}(\omega)-d_{\veczero}}{d_{\veczero}\omega} 
\end{equation}
\pagebreak[3]
or 
\begin{equation}
C=\frac{e}{2} \exp\left[-
\int_{-1/2}^0 \rmd \omega \frac{D_{\veczero}(\omega)-d_{\veczero}}{d_{\veczero}\omega} 
\right]
\; ,
\end{equation}
in agreement with the result for the single-impurity Anderson model,
eq.~(\ref{eq:DosgamesforC}).

\section{Four-orbital toy model}
\label{app:foursitemodel}

For a simple illustration of a central result of this work, we address
an exactly solvable four-orbital model at half band-filling, $L=2$ in eq.~(\ref{eq:defH}).
It consists of only two sites with one host-electron orbital and one impurity orbital
on each site. The electron transfer amplitude between the host-electron orbitals 
is~$t=-1/2$ ($W=1$), 
the local hybridization between host-electron orbital and impurity orbital is~$V$,
and the electrons in the impurity orbitals interact via the Hubbard interaction 
given by eq.~(\ref{eq:defHint}). We study the case of half band-filling
with $N=2L+2=4$ electrons in the system.

\begin{figure}[htb]
\includegraphics[width=\columnwidth]{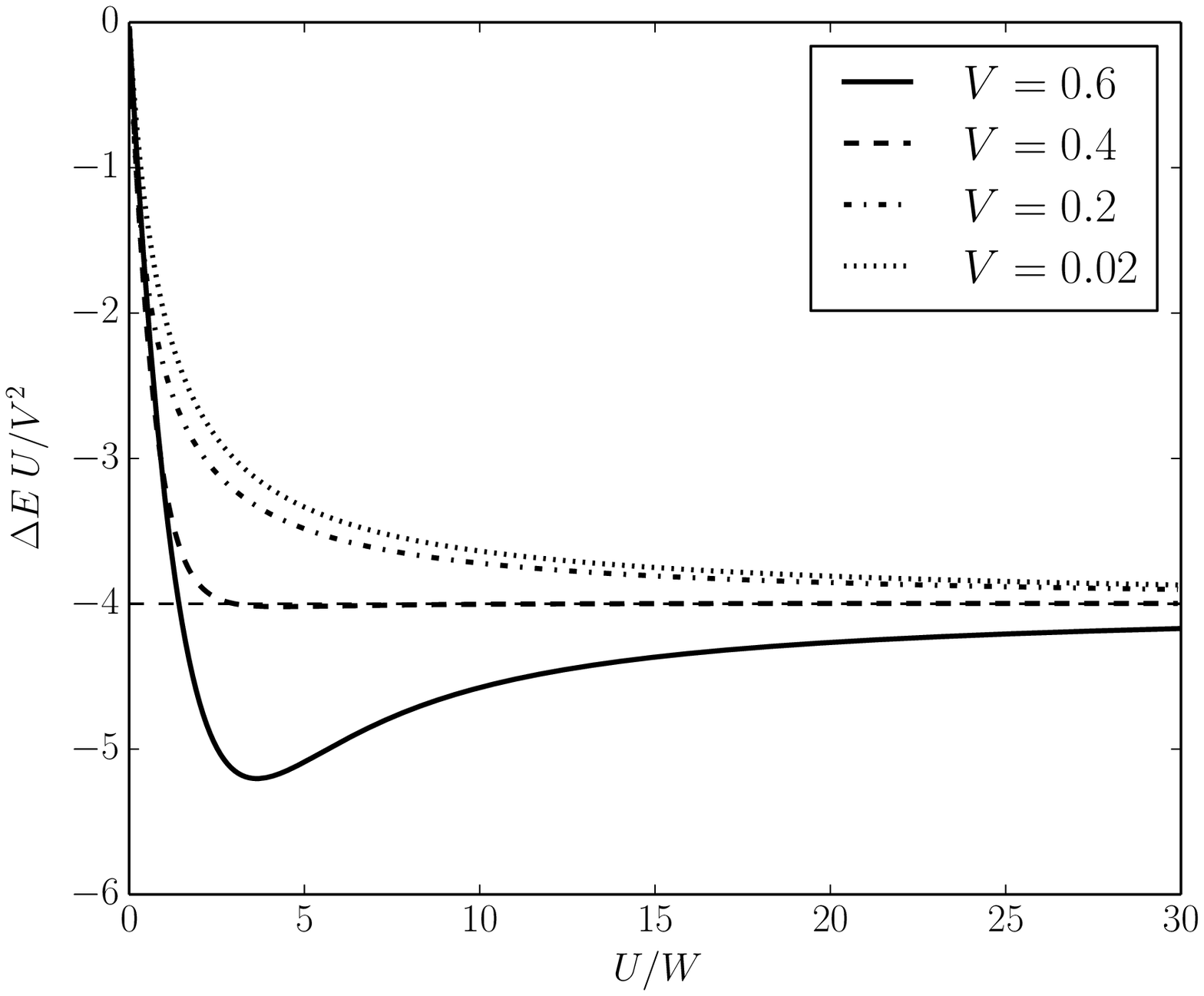}
\caption{Ground-state energy $\Delta E(U,V)=E_0(U,V)+W+U/2$ scaled by its
limiting large-$U$ behavior as a function of $U$ for $V=0.6,0.4,0.2,0.02$.
Apparently, $\Delta E(U,V)\sim - 4V^2/U$.\label{fig:gsenegrytoy}}
\end{figure}

In Fig.~\ref{fig:gsenegrytoy} we show the ground-state energy 
as a function of~$U$ for $V=0.6,0.4,0.2,0.02$. The $V$-de\-pen\-dent energy correction
is given by
\begin{equation}
\Delta E(U,V)=E_0(U,V)+W+U/2 \;, 
\end{equation}
where $E_0(U,V)$ is the ground-state energy.
For large values~$U/V$ we see that
\begin{equation}
\Delta E(U,V)\sim -4 \frac{V^2}{U}\; ,
\end{equation}
as in our variational description, eq.~(\ref{eq:Evarfinal}). This in\-di\-cates
that there is an effective direct electron trans\-fer between
the impurity orbitals of the order~$V$
in the spin limit, see eq.~(\ref{eq:t12directheislimit}).

\begin{figure}[htb]
\includegraphics[width=\columnwidth]{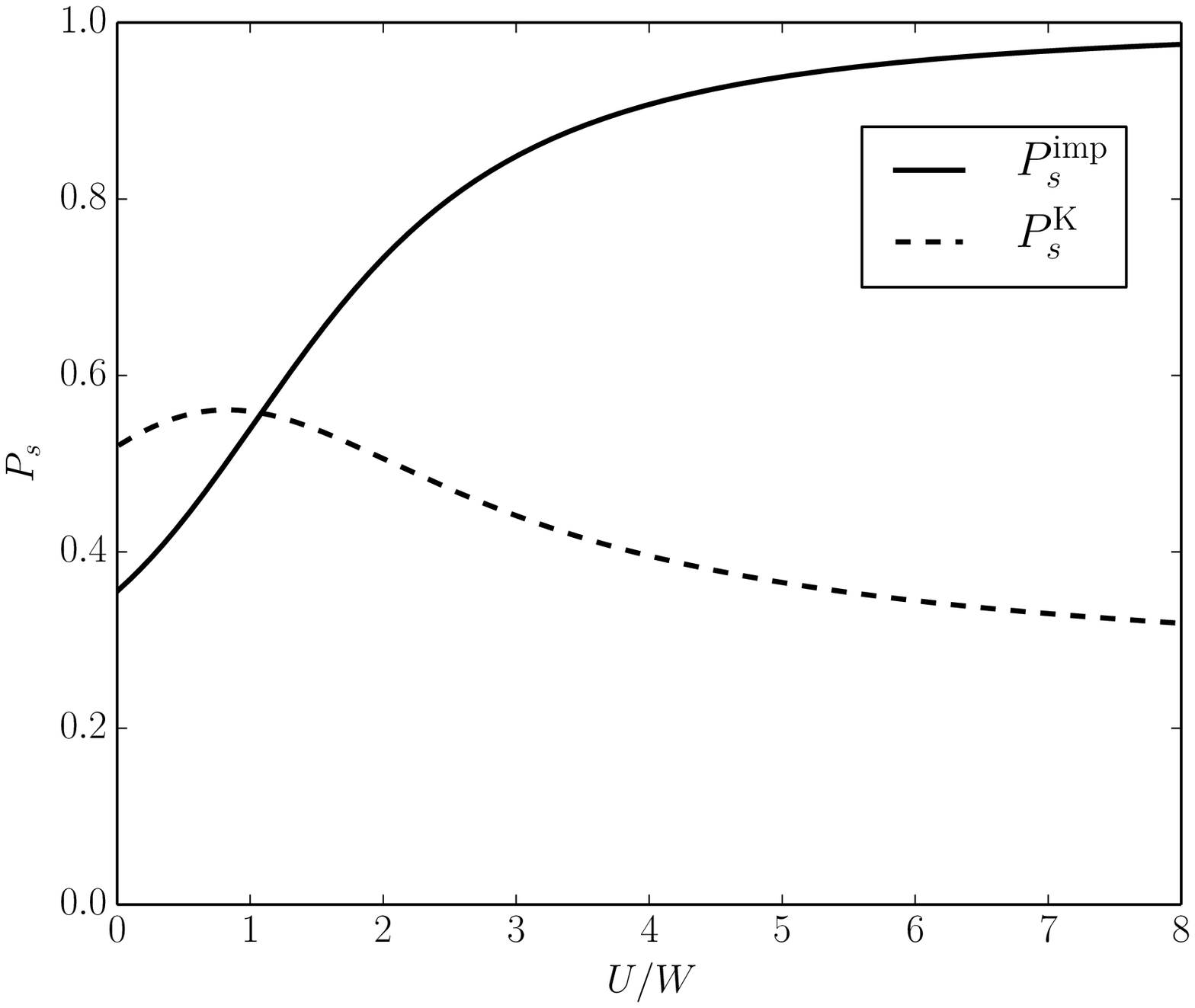}
\caption{Probabilities for a spin singlet between the impurity electrons,
$P_{\rm s}^{\rm imp}(U,V)$, and between impurity and host electrons,
$P_{\rm s}^{\rm K}(U,V)$, as a function of $U$ for $V=0.4$.\label{fig:singletprobtoy}}
\end{figure}

To elucidate the properties of the ground state further, we consider
the probability to find 
a spin singlet formed between the impurities,
\begin{equation}
P_{\rm s}^{\rm imp}(U,V)=\langle \Psi_0 | \frac{1}{4}-\vecS_1\cdot \vecS_2| \Psi_0\rangle
\; ,
\end{equation}
and the probability to find a spin singlet formed between an impurity state and its
local host electron state (`Kondo singlet'),
\begin{equation}
P_{\rm s}^{\rm K}(U,V)=\langle \Psi_0 | \frac{1}{4}-\vecS_l\cdot \vecs_l| \Psi_0\rangle\; ,
\end{equation}
which are equal for site $l=1,2$.
In Fig.~\ref{fig:singletprobtoy} we show both quantities as a function of~$U$
for $V=0.4$.

For small interactions, there is a tendency to form a Kondo spin singlet
because $P_s^{\rm K}$ initially increases as a function of~$U$.
However, $P_s^{\rm K}$ starts to decrease for $U\gtrsim W$. Moreover,
the probability to find a singlet formed by the two impurity spins
dominates for $U\gtrsim W$, $P_s^{\rm imp}>P_s^{\rm K}$. Eventually, 
the impurity spins form a Heisenberg-type singlet pair.


\vspace*{1cm}


\end{document}